\pdfoutput=1
\documentclass[11pt,twoside,a4paper,cmspaper,final,collab]{cms-tdr}
\def\svnVersion{3ffdeb4}\def\svnDate{2026/07/27}\def\cmsCernNoTag{CERN-EP-2026-033}\def\cmsCernDate{\today}\def\cmsMessage{Published in the Journal of High Energy Physics as \href{http://dx.doi.org/10.1007/JHEP07(2026)248}{\doi{10.1007/JHEP07(2026)248}.}}
\begin{document}\cmsNoteHeader{HIG-24-008}

\newcommand{\Likelihood}{\ensuremath{\mathcal{L}}\xspace}
\newcommand{\vtheta}{\ensuremath{\vec{\nu}}\xspace}
\newcommand{\HJMINLO} {{\textsc{HJ-MiNLO}}\xspace}
\newcommand{\mjs}{\ensuremath{m^*_{\text{j}}}\xspace}
\newcommand{\mj}{\ensuremath{m_{\text{j}}}\xspace}
\newcommand{\Hlqq}{\ensuremath{\PH_{\Pell\PQq\PQq}}\xspace}
\newcommand{\Hthreeq}{\ensuremath{\PH_{3\PQq}}\xspace}
\newcommand{\Hfourq}{\ensuremath{\PH_{4\PQq}}\xspace}
\newcommand{\ptja}{\ensuremath{\pt^{\text{j}}}\xspace}
\newcommand{\PNet}{\ensuremath{\textsc{ParticleNet}}\xspace}
\newcommand{\ParT} {{\textsc{ParT}}\xspace}
\newcommand{\ParTfinetuned} {{\textsc{ParT-Finetuned}}\xspace}
\newcommand{\SRone}{\ensuremath{\text{SR}_1}\xspace}
\newcommand{\SRtwo}{\ensuremath{\text{SR}_2}\xspace}
\newcommand{\SRa}{\ensuremath{\text{SR}_\text{a}}\xspace}
\newcommand{\SRb}{\ensuremath{\text{SR}_\text{b}}\xspace}
\newcommand{\SRonea}{\ensuremath{\text{SR}_\text{1a}}\xspace}
\newcommand{\SRoneb}{\ensuremath{\text{SR}_\text{1b}}\xspace}
\newcommand{\SRtwoa}{\ensuremath{\text{SR}_\text{2a}}\xspace}
\newcommand{\SRtwob}{\ensuremath{\text{SR}_\text{2b}}\xspace}
\newcommand{\CRone}{\ensuremath{\text{CR}_{1}}\xspace}
\newcommand{\CRtwo}{\ensuremath{\text{CR}_{2}}\xspace}
\newcommand{\Irel}{\ensuremath{I_\text{rel}}\xspace}
\newcommand{\Imini}{\ensuremath{I_\text{mini}}\xspace}
\newcommand{\TFir}{\ensuremath{\text{TF}_{ir}}\xspace}
\newcommand{\TFonea}{\ensuremath{\text{TF}_{1\text{a}}}\xspace}
\newcommand{\TFoneb}{\ensuremath{\text{TF}_{1\text{b}}}\xspace}
\newcommand{\TFtwoa}{\ensuremath{\text{TF}_{2\text{a}}}\xspace}
\newcommand{\TFtwob}{\ensuremath{\text{TF}_{2\text{b}}}\xspace}
\newcommand{\hgamgam}{\ensuremath{\PH \to \PGg \PGg}\xspace}
\newcommand{\hzz}{\ensuremath{\PH \to \PZ \PZ \to 4\Pell}\xspace}
\newcommand{\hww}{\ensuremath{\PH \to \PW \PW \to \Pe\PGn\PGm\PGn}\xspace}
\newcommand{\htt}{\ensuremath{\PH \to \PGt \PGt}\xspace}
\newcommand{\hbb}{\ensuremath{\PH \to \PQb \PQb}\xspace}
\newcommand{\hwwqq}{\ensuremath{\PH\to\PW\PW\to\Pell\PGn\Pq\Pq/\Pq \Pq\Pq\Pq}\xspace}
\newcommand{\hwwhad}{\ensuremath{\PH\to\PW\PW\to\Pq \Pq\Pq\Pq}\xspace}
\newcommand{\hwwlep}{\ensuremath{\PH\to\PW\PW\to\Pell\PGn\Pq\Pq}\xspace}
\newcommand{\hzzall}{\ensuremath{\PH \to \PZ \PZ}\xspace}
\newcommand{\hwwall}{\ensuremath{\PH \to \PW \PW}\xspace}
\newcommand{\VH}{\PV{}\PH{}\xspace}
\newcommand{\ttH}{\ensuremath{\PQt\PAQt\PH}\xspace}
\newcommand{\ptH}{\ensuremath{\pt^\PH}\xspace}
\newcommand{\fxfx}{{\textsc{FxFx}}\xspace}
\newcommand{\Wjets}{\ensuremath{\PW\text{+jets}}\xspace}
\newcommand{\wjets}{\PW{}+jets\xspace}
\newcommand{\cmenergy}{\ensuremath{\sqrt{s}=13\TeV}\xspace}
\newcommand{\SoverSqrtB}{\ensuremath{S/\sqrt{B}}\xspace}
\newcommand{\wjetslnu}{\ensuremath{\PW(\Pell\PGn)}+jets\xspace}
\newcommand{\mjj}{\ensuremath{m_\text{jj}}\xspace}
\newcommand{\jj}{\ensuremath{\text{jj}}\xspace}
\newlength\cmsTabSkip\setlength{\cmsTabSkip}{1ex}

\cmsNoteHeader{HIG-24-008}

\title{Search for Higgs boson production at high transverse momentum in the \texorpdfstring{$\PW\PW$}{WW} decay channel in proton-proton collisions at \texorpdfstring{$\sqrt{s} = 13\TeV$}{sqrt(s) = 13 TeV}}

\date{\today}

\abstract{
A search for Higgs boson (\PH) production at high transverse momentum (\pt) in the $\PW\PW$ decay channel is presented. The analysis uses proton-proton collisions at $\sqrt{s}=13\TeV$ recorded by the CMS experiment in 2016--2018, corresponding to an integrated luminosity of 138\fbinv. The visible decay products of the Higgs boson are reconstructed as a single large-radius jet with one isolated lepton or none ($1\Pell$ and $0\Pell$, respectively; $\Pell=\Pe,\mu$). The \PH-candidate jets are identified using an advanced transformer-based algorithm and are calibrated with the Lund jet plane reweighting technique. The $1\Pell$ channel is further split into gluon fusion, vector boson fusion, and associated production with hadronically decaying vector boson categories, while the $0\Pell$ channel considers all production processes inclusively. The measured cross section times the \hwwall branching fraction relative to the standard model expectation is $\mu = -0.19^{+0.48}_{-0.46}$, indicating no evidence of a signal above the background. This measurement represents the first dedicated study of highly Lorentz-boosted \hwwall decays, complementing earlier searches for high-\pt Higgs boson in other decay channels.}

\hypersetup{
pdfauthor={CMS Collaboration},
pdftitle={Search for Higgs boson production at high transverse momentum in the WW decay channel in proton-proton collisions at sqrt(s) = 13 TeV}, 
pdfsubject={CMS}, 
pdfkeywords={CMS, boosted, Higgs boson, HWW, WW, LJP, ParT, finetune}
}

\maketitle

\section{Introduction}
\label{Sec:intro}

Since the discovery of a 125\GeV Higgs boson (\PH) by the ATLAS~\cite{ATLAS:2012yve} and CMS~\cite{CMS:2012qbp,cms2013} Collaborations, extensive effort has been devoted to precision measurements of its properties and couplings~\cite{ATLAS:2022vkf,CMS:2022dwd}.
Measuring Higgs boson production at high transverse momentum, \pt, is a core component of this program at the CERN LHC.
The high-\pt regime is especially important because theoretical predictions feature large logarithmic corrections and substantial higher-order effects~\cite{becker2021precise}, making measurements in this regime a powerful test of standard model (SM) calculations.
Additionally, it serves as a sensitive probe for new phenomena, as deviations from the SM predictions could indicate the presence of effects beyond the SM at energy scales not directly accessible by the LHC~\cite{Grojean_2014,Schlaffer_2014,Dawson_2015,Grazzini_2017,Maltoni_2014,Degrande_2017}.

The ATLAS and CMS Collaborations have reported measurements of the differential production cross section of the Higgs boson as a function of its transverse momentum, \ptH, in a number of decay channels: \hbb, \hgamgam, \hww, \htt, and \hzz ($\Pell=\Pe,\mu$),~\cite{CMS:2018gwt,h_gamgam,h_ww,h_tt,h_zz}.
Among those, the sensitivity at the highest \ptH values probed so far is dominated by the \hbb~\cite{Sirunyan_2018,Sirunyan_2020,CMS:2026fsx,Aad_2022,ATLAS:2023jdk,ATLAS:2026iwc} and \htt decay channels~\cite{htt_boosted}, which benefit from large branching fractions. These studies have enabled differential cross section measurements up to $\pt^\PH \approx 1\TeV$~\cite{Aad_2022}, thereby placing constraints on potential deviations from the SM in Lorentz-boosted topologies. 

This paper presents the first search for highly Lorentz-boosted Higgs boson production decaying into a pair of \PW bosons, where at least one \PW boson decays into a pair of quarks, \hwwqq.
The analysis is performed using proton-proton ($\Pp\Pp$) collision data at a center-of-mass energy of \cmenergy, collected with the CMS detector at the LHC in 2016--2018, corresponding to an integrated luminosity of 138\fbinv~\cite{CMS:2021xjt,CMS-PAS-LUM-17-004,CMS-PAS-LUM-18-002}.

The \PW bosons originating from a Higgs boson with $\pt^{\PH} \gtrsim 250 \GeV$ are separated by a small angular distance $\Delta R = \sqrt{(\Delta \eta)^{2} + (\Delta \phi)^{2}} < 0.8$, with $\Delta\eta$ and $\Delta\phi$ denoting pseudorapidity and azimuthal angle difference of the two \PW bosons, respectively.
As a result, their final-state products are merged into a single large-radius jet.
For leptonically decaying \PW bosons, leptons within the jet can still be identified as relatively isolated using a custom isolation variable that depends on the lepton \pt.

\begin{figure}[ht!] \centering
\includegraphics[width=0.75\linewidth]{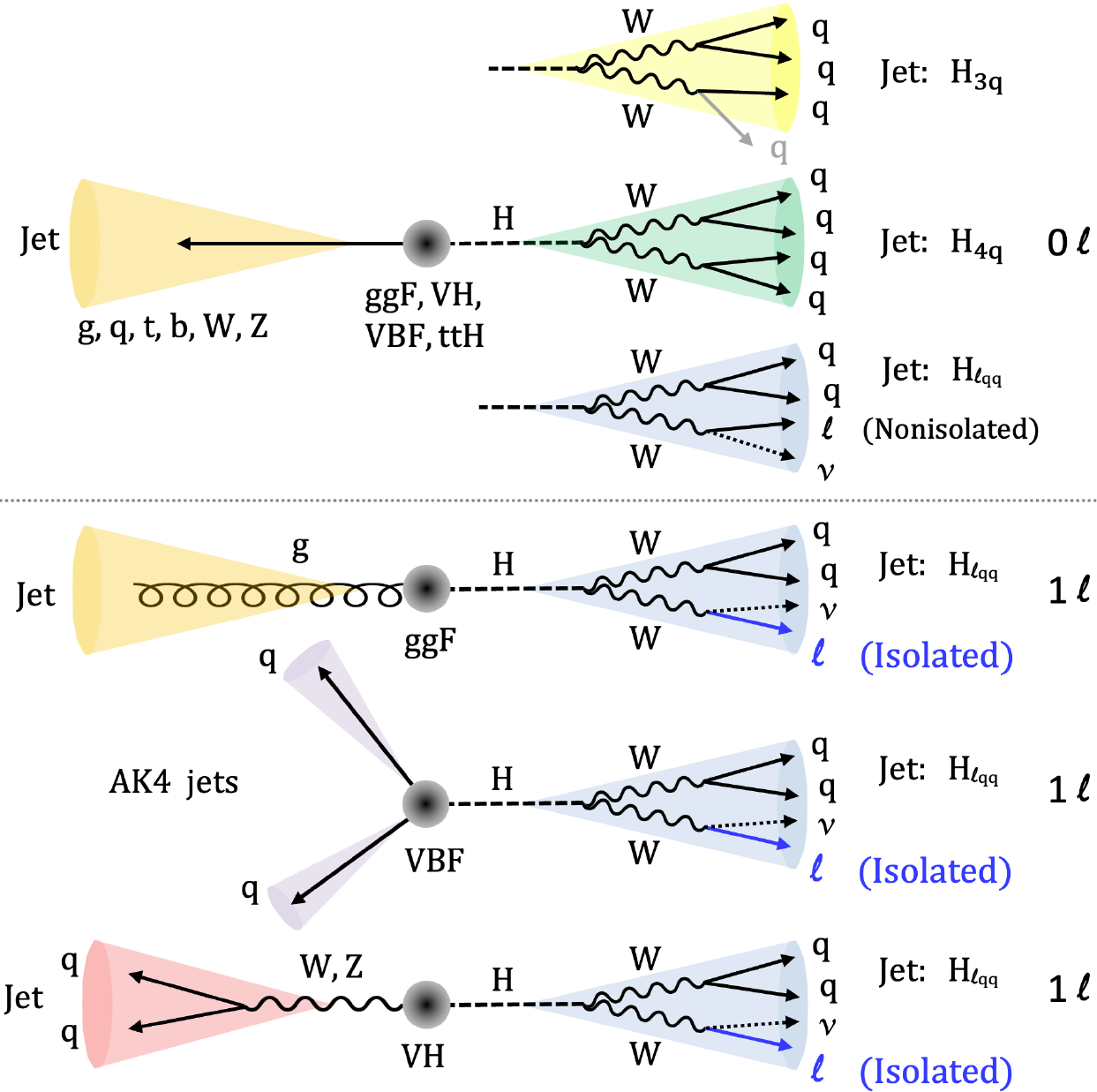}
\caption{Illustration of the event topologies analyzed.  
Right: boosted Higgs boson final states from the \hwwqq decay.  
Left: associated jets corresponding to the different production processes.  
From upper to lower: $0\Pell$ inclusive (all production and decay modes), and the $1\Pell$ ggF, VBF, and \VH production processes.
} \label{fig:Topologies}
\end{figure}

In the analysis, we categorize events into two channels based on whether they have one isolated lepton ($1\Pell$) or none ($0\Pell$), as illustrated in Fig.~\ref{fig:Topologies}.
Two different and complementary approaches are employed for these channels.
The $0\Pell$ channel considers fully-hadronic $\PW\PW$ decays as well as semileptonic $\PW\PW$ decays with nonisolated leptons, while the $1\Pell$ channel exclusively targets $1\Pell+$jets decays with an isolated lepton.
The two channels feature similar baseline event selection, but adopt different Higgs boson identification and background estimation techniques tailored to their distinct background compositions.

While the overall measurement considers all Higgs boson production processes inclusively, the most sensitive channel, $1\Pell$, further categorizes events into the three Higgs boson production processes with the largest cross section at the LHC: gluon fusion (ggF), vector boson fusion (VBF), and associated production with a vector boson (\VH), where the vector boson, \PV (\PW or \PZ), decays hadronically.
This allows for better modeling of the signal, as the relative contributions of these processes to the cumulative Higgs boson production cross section are expected to vary with \ptH~\cite{becker2021precise}. 
The associated production with a top quark-antiquark pair, \ttH, is not explicitly targeted with a dedicated region, although it still contributes as a minor signal component.

The \PH-candidate jet is identified using a jet tagging algorithm known as the particle transformer (\ParT)~\cite{Qu:2022mxj}.
This algorithm uses a self-attention neural network architecture~\cite{Vaswani:2017attention} that dynamically weights jet constituent importance, trained on various large-radius jet topologies with excellent performance for identifying boosted Higgs boson decays to vector bosons, with performance and calibration for \hwwall decays documented in Ref.~\cite{CMS:2026uph}.
The Higgs boson four-momentum is reconstructed from the \PH-candidate jet and the missing \pt vector, \ptvecmiss, for events containing a neutrino in the final state.
In the \VH category, the \PV candidate jet is identified with the \PNet algorithm~\cite{Qu:2019gqs,CMS:2020poo}, while the VBF category requires the presence of two additional well-separated jets.
The signal extraction is performed by fitting the invariant mass distribution of the reconstructed \PH or \PV boson candidates.
Results are also presented as simplified template cross sections (STXS)~\cite{deFlorian:2016spz} following the stage-1.2 binning scheme~\cite{Berger:2922392}, which partitions the production phase space by Higgs boson kinematics.

The paper is organized as follows.
The CMS detector is briefly described in Section~\ref{Sec:cms}, followed by details of the event simulation in Section~\ref{Sec:Sample} and event reconstruction in Section~\ref{Sec:Objects}. 
Section~\ref{Sec:Higgs_identification_ParT_calibration} presents the Higgs boson identification using the \ParT algorithm and its calibration. 
The $0\Pell$ and $1\Pell$ channel analyses are described in Sections~\ref{Sec:0l_channel} and~\ref{Sec:1l_channel}, respectively, including event selections and background estimation methods.
Systematic uncertainties are discussed in Section~\ref{Sec:uncertainties} and results are presented in Section~\ref{Sec:Results}, followed by a summary in Section~\ref{Sec:sum}.
Tabulated results for this analysis are provided in the HEPData record~\cite{hepdata}.

\section{The CMS detector}
\label{Sec:cms}

The CMS apparatus~\cite{Chatrchyan:2008zzk,CMS:2023gfb} is a multipurpose, nearly hermetic detector, designed to trigger on~\cite{CMS:2020cmk,Khachatryan:2016bia,CMS:2024aqx} and identify electrons (\Pe), muons ($\mu$), photons, and (charged and neutral) hadrons~\cite{CMS:2020uim,Sirunyan:2018fpa,CMS:2014pgm}.
Its central feature is a superconducting solenoid of 6\unit{m} internal diameter, providing a magnetic field of 3.8\unit{T}.
Within the solenoid volume are a silicon pixel and strip tracker, a lead tungstate crystal electromagnetic calorimeter (ECAL), and a brass and scintillator hadron calorimeter, each composed of a barrel and two endcap sections.
Forward calorimeters extend the pseudorapidity coverage provided by the barrel and endcap detectors.
Muons are reconstructed using gas-ionization detectors embedded in the steel flux-return yoke outside the solenoid.

Events of interest are selected using a two-tiered trigger system.
The first level, composed of custom hardware processors, uses information from the calorimeters and muon detectors to select events at a rate of around 100\unit{kHz} within a fixed latency of 4\mus~\cite{CMS:2020cmk}.
The second level, known as the high-level trigger, consists of a farm of processors running a version of the full event reconstruction software optimized for fast processing, and reduces the event rate to a few kHz before data storage~\cite{Khachatryan:2016bia,CMS:2024aqx}.

More detailed descriptions of the CMS detector, together with a definition of the coordinate system used and the relevant kinematic variables, can be found in Refs.~\cite{Chatrchyan:2008zzk,CMS:2023gfb}.

\section{Event simulation}
\label{Sec:Sample}

The ggF Higgs boson production process is simulated at next-to-leading order (NLO) accuracy in quantum chromodynamics (QCD) using the \HJMINLO event generator~\cite{Hamilton:2012rf}.
Finite top quark mass effects~\cite{Neumann_2018} are included following the recommendation of Ref.~\cite{becker2021precise}.
The generated Higgs boson is required to have $\pt>200\GeV$.
The \POWHEG generator 2.2~\cite{Nason:2009ai,Luisoni:2013kna,Hartanto:2015uka} is used to simulate Higgs boson production via VBF, \VH, and associated production with a top quark-antiquark pair, \ttH, at NLO accuracy in QCD.
The loop-induced $\Pg\Pg \to \PZ\PH$ process is generated separately at leading order (LO) with \POWHEG~\cite{POWHEG3}.
In all signal simulations, the Higgs boson mass is set to 125\GeV.
The inclusive cross sections for the ggF, VBF, \VH, and \ttH samples are taken from Ref.~\cite{deFlorian:2016spz}.
The decay to a pair of \PW bosons is performed using \textsc{JHUgen} v7.2.7~\cite{Bolognesi:2012mm} for VBF, $\PW\PH$, and quark-induced $\PZ\PH$ samples to properly account for spin correlations and angular distributions in these EW production processes, while \PYTHIA 8.240~\cite{Sjostrand:2014zea} is used for the remaining signal samples where such effects are less critical; all \PW boson decays are considered.
The analysis includes events with \hzzall decays from ggF production, simulated with the \HJMINLO generator, and \htt decays, both of which are treated as background processes with nearly negligible contributions.

Background from jets produced via the strong interaction, commonly referred to as QCD multijet events, is modeled at LO accuracy using the \MGvATNLO~2.6.5 generator~\cite{Alwall_2014}, including up to four final-state partons in the matrix element calculations.
The $\PW(\Pell \PGn)$+jets process for the $1\Pell$ channel is modeled using the \SHERPA Monte Carlo generator \cite{Gleisberg:2008ta,Sherpa:2019gpd}.
In this setup, NLO matrix elements with up to two extra jets and LO matrix elements with up to four extra jets are calculated with \textsc{Comix} 2.2.15 package~\cite{Gleisberg:2008fv}.
For the $0\Pell$ channel, the $\PW(\Pell \PGn)$+jets process is simulated at LO accuracy using \MGvATNLO, in exclusive \HT ranges, where \HT is defined as the scalar sum of the transverse momenta of all generated jets, providing an adequately large event sample.
The same \HT-binning procedure is applied to the $\PV(\Pq\Pq)$+jets process for both the $0\Pell$ and $1\Pell$ channels.
Jets from the matrix element calculations and parton shower description are matched via the MLM scheme~\cite{Alwall:2007fs} for the LO \MGvATNLO samples, while \SHERPA uses CKKW-based merging for $\PW(\Pell\PGn)$+jets, combining NLO (0--2 jets) and LO (3--4 jets) matrix elements~\cite{Hoeche:2012yf}.
The cross section of the $\PV(\Pq\Pq)$+jets process is corrected as a function of the boson \pt for higher-order QCD and electroweak (EW) effects.
The QCD NLO corrections are derived using \MGvATNLO, simulating \PW and \PZ boson production with up to two additional partons and \fxfx jet matching~\cite{Frederix:2012ps}.
The EW NLO corrections are taken from theoretical calculations in Refs.~\cite{kallweit2015nlo,Kallweit_2016,kallweit2015nlo2,Lindert_2017}.

The production of top quark-antiquark pairs (\ttbar) and single top quarks (single \PQt), including $\PQt\PW$ and $t$-channel contributions, are modeled at NLO accuracy using the \POWHEG generator~\cite{POWHEG1,POWHEG2,POWHEG3,Alioli:2011as,Alioli:2009je,Frederix:2012dh}.
Single top quark production in the $s$-channel is simulated with \MGvATNLO at LO accuracy with the MLM jet matching scheme.
The Drell--Yan production of lepton pairs through $\PZ/\gamma^*$ is simulated at NLO accuracy with \MGvATNLO with up to two additional partons, using the \fxfx jet matching scheme.
The EW production of \PW and \PZ bosons with exactly two additional partons is modeled using \MGvATNLO at LO accuracy, corresponding to $\mathcal{O}(\alpS^4)$, 
where $\alpS$ denotes the strong coupling constant.
Diboson processes are modeled at LO accuracy with \PYTHIA~8.226, and the total cross sections are corrected to next-to-NLO accuracy with the \MCFM~7.0 program~\cite{Campbell:2010ff}.
Contributions from other processes are found to be negligible.
Parton showering, fragmentation, and hadronization are modeled with \PYTHIA~8.230 using the underlying event tune CP5~\cite{Sirunyan:2019dfx}, with the exception of the samples produced with \textsc{Sherpa}.
The parton distribution function (PDF) set {NNPDF3.1}~\cite{Ball_2017} at next-to-NLO accuracy is used for all processes.

For all simulated samples, the CMS detector response is modeled by \GEANTfour~\cite{GEANT4,geant4_2}.
Independent samples are generated for each data-taking period using identical generator configurations, but accounting for changes in the accelerator and detector running conditions.
The recorded data samples include additional $\Pp\Pp$ interaction vertices from the same or nearby bunch crossings (pileup), generated with \PYTHIA and added to all simulated events based on the expected pileup distribution. Corrections are applied to the simulated samples to match the pileup multiplicity distribution measured in the recorded data by era.
Parton showering, fragmentation, and hadronization are modeled with \PYTHIA~8.230 using the CP5 underlying event tune~\cite{Sirunyan:2019dfx}, with the exception of the samples produced with \textsc{Sherpa}.

\section{Event reconstruction}
\label{Sec:Objects}

The physics objects (\Pe, \PGm, \PGg, charged and neutral hadrons) are reconstructed using the particle-flow (PF) algorithm~\cite{cmsPF}, which uses an optimized combination of information from the various elements of the CMS detector to reconstruct individual particles (PF candidates).
The primary vertex is taken to be the vertex corresponding to the hardest scattering in the event, evaluated using tracking information alone, as described in Ref.~\cite{CMS-TDR-15-02}.

The PF candidates are clustered into jets using the anti-\kt algorithm~\cite{Cacciari:2008gp, Cacciari:2011ma}.
The clustering algorithm, as implemented by the \textsc{FastJet} package~\cite{Cacciari:2011ma}, is applied twice over the same inputs with distance parameter of 0.4 or 0.8, producing the AK4 or AK8 jet collections, respectively.
The larger radius of the AK8 jet effectively captures the decay products of high-\pt Higgs bosons.
For AK4 jets, the pileup effect is mitigated by excluding tracks that originate from pileup vertices and applying an offset correction to account for remaining contributions.
For AK8 jets, the pileup-per-particle identification algorithm~\cite{Sirunyan:2020foa,Bertolini:2014bba} weights each PF candidate, prior to jet clustering, based on the likelihood of the particle to originate from the hard-scattering vertex.
Jet energy corrections are derived to match the detector response to particle-level jets~\cite{Khachatryan:2016kdb}, and additional selection criteria remove jets dominated by anomalous contributions from instrumental effects or reconstruction failures~\cite{CMS-PAS-JME-16-003}.
The  \ptvecmiss is computed as the negative vector sum of the transverse momenta of all the PF candidates in an event, and its magnitude is denoted as \ptmiss~\cite{CMS:2019ctu}.
The \ptvecmiss is modified to account for corrections to the energy scale of the reconstructed jets in the event.

The AK8 jets must satisfy $\pt > 200\GeV$ and $\abs{\eta} < 2.4$ to be considered in this analysis.
The soft-drop algorithm~\cite{Larkoski:2014wba}, with parameters $\beta=0$ and $z_\text{cut}=0.1$, is applied to mitigate contamination from underlying event and pileup by removing PF candidates consistent with soft and wide-angle radiation from the jet.
For jets originating from the fully hadronic decay of a massive boson, the soft-drop mass (\mj) distribution peaks near the boson mass, while for quark- and gluon-initiated jets, \mj has a smoothly falling spectrum.
Each jet is assigned a discriminant score quantifying its compatibility with an \PH or \PV boson origin, as described in Section~\ref{Sec:Higgs_identification_ParT_calibration}.

Events with AK8 jets compatible with a hadronically decaying \PV boson from the $\PV(\Pq\Pq)\PH$ production in the $1\Pell$ channel are identified using the  \textsc{ParticleNet} algorithm~\cite{Qu:2019gqs,CMS:2020poo} which is based on a graph neural network trained to distinguish large-radius jets originating from scalar particles with two-pronged $\PX\to\PQq\PQq$ decays from QCD jets.
Each prong corresponds to the fragmentation and hadronization of a colored parton from the \PX decay.
To achieve decorrelation from the jet mass, signal resonances between 15 and 250\GeV are used, and jets from signal and background are reweighted to obtain flat \mj and \pt distributions.
Merged $\PV\to\PQq\PQq$ decays are identified using a probability score defined as the sum of three tagger output scores for $\PX\to\PQb\PQb$, $\PQc\PQc$, and $\PQq\PQq$ decays, where $\PQq\in\{\PQu,\PQd,\PQs\}$.

The $\PV\to\PQq\PQq$ jet tagging efficiency is calibrated in semileptonic \ttbar events with quarks using a tag-and-probe method~\cite{CMS-DP-2025-010}.
For the $\PZ\to\PQb\PQb$ tagging, the efficiency is calibrated with the QCD proxy-jet method~\cite{CMS:2025kje}, where a boosted decision tree identifies suitable gluon-splitting, $\Pg\to\PQb\PQb$, jets as proxies for the signal.

The AK4 jets must satisfy $\pt > 30\GeV$.
Jets originating from a bottom quark, denoted as \PQb jets, satisfy $\abs{\eta} < 2.5$ and are identified with the \textsc{DeepJet} algorithm~\cite{Bols:2020bkb, CMS-DP-2018-058}.
The \PQb jet identification requirements in this search use two working points (WPs), ``medium'' and ``tight,'' corresponding to tagging efficiencies of about 75\% and 60\%, and probabilities of misidentifying a light-flavor quark or gluon jet as a \PQb jet of 1\% and 0.1\%, respectively, as determined in a sample of simulated \ttbar events. 
The data-to-simulation differences in \PQb-tagging efficiency are corrected via scale factors (SFs) applied to the simulation~\cite{CMS-DP-2018-058}.

Muons are identified as tracks in the central tracker consistent with either a track or several hits in the muon system, and associated with calorimeter deposits compatible with the muon hypothesis~\cite{Sirunyan:2018fpa}.
Electrons are identified as a primary charged-particle track and potentially many ECAL energy clusters corresponding to this track extrapolation to the ECAL and to possible bremsstrahlung photons emitted along the way through the tracker material.
They are identified with a multivariate discriminant described in Ref.~\cite{Khachatryan:2015hwa}.
Isolation algorithms are designed to measure the amount of energy deposited near an object, such as a lepton.
The relative isolation, \Irel, of \PGm (\Pe) is calculated by summing the \pt of PF candidates in a cone of size $\Delta R < 0.4$ (0.3) centered on the lepton, corrected for neutral pileup contributions, and divided by the \pt of the lepton~\cite{Sirunyan:2018fpa,Khachatryan:2015hwa}.
Leptons emerging from the decay chain \hwwlep of high-\pt Higgs bosons may fail \Irel requirements if they are produced inside or close to the jet from the hadronic \PW boson decay. 
However, they can still be identified using an optimized version of the isolation variable, referred to as ``mini-isolation,'' \Imini~\cite{Rehermann:2010vq}.
The \Imini is defined identically to \Irel, but with a \pt-dependent cone size 
$\Delta R^{\text{mini-iso}}$ of 0.2, $10\GeV/\pt^\ell$, and 0.05 for leptons with 
$\pt^\ell<50\GeV$, $50<\pt^\ell<200\GeV$, and $\pt^\ell>200\GeV$, respectively.

\section{Higgs boson identification and reconstruction}
\label{Sec:Higgs_identification_ParT_calibration}

Higgs boson identification is performed using the \ParT algorithm, a machine-learning model based on self-attention mechanisms and the transformer architecture~\cite{Qu:2022mxj}.
The \ParT algorithm extends the capabilities of \textsc{ParticleNet}~\cite{Qu:2019gqs}, originally developed for distinguishing QCD jets from hadronically decaying resonances into two quarks (``two-pronged'' jets) to a broader range of jet topologies, including those originating from \hwwqq decays.
Tagging these jets is particularly challenging due to the asymmetric kinematics of the intermediate \PW bosons.

For each jet, the \ParT algorithm outputs 37 classification scores, each representing the probability that the jet belongs to a specific category.
These categories distinguish jets by flavor---charm (\PQc), \PQb, light quarks, and hadronically decayed tau ($\tauh$) leptons---as well as by the number of the constituents (quarks or leptons), before parton shower and hadronization.
In particular, \hwwhad jets are subdivided into four-pronged (\Hfourq) and three-pronged (\Hthreeq) classes to account for cases where not all decay products fall within the jet cone.
The complete list of jet classes considered by \ParT is shown in Table~\ref{tab:ParT_Classes}.

The \ParT model is trained on a sample of simulated jets in a kinematic range that satisfies $200 < \pt < 2500\GeV$, $\abs{\eta} < 2.4$, and $20 < \mj < 260\GeV$.
The training set includes jets from scalar resonances decaying into two-quark and $\PW\PW$ channels, and top-like resonances decaying into $\PQb\PW$ channels, as well as jets produced in QCD multijet events.
The generator-level resonance masses range from 15 to 250\GeV.
The training employs mass decorrelation via sample reweighting to achieve uniform \pt and \mj distributions, preventing the tagger from using jet mass as a discriminant and avoiding background mass sculpting effects, following the \textsc{ParticleNet} strategy.
Additionally, in the training samples, the \PH and \PW boson masses are varied together in order to maintain the \PW-to-\PH mass ratio fixed to the SM value of 0.64.
This ensures that one \PW boson remains off-shell throughout the mass range and that the tagger is decorrelated from both the \PH and \PW boson masses simultaneously.
Additional details on the model architecture and the \ParT training are documented in Ref.~\cite{CMS:2026uph}.

\begin{table}[!htbp]\centering
\topcaption{
The 37 \ParT jet classification categories.
The categories are based on the decay modes of \PH and \PV bosons, top quarks, and on QCD processes.
All listed decay products are assumed to be contained within the jet cone, except for neutrinos.
Numbers like $4\PQq$ indicate the multiplicity of the adjacent quark, while those in parentheses indicate the number of \PQc quarks in the preceding quark sequence.
Classes such as $3\Pq$ or $\PQb\Pq$ imply that one quark escapes the jet cone in $\PH\to4\Pq$ or $\PQt\to\PQb\Pq\Pq$ decays, respectively.
Subscripts on \PGt indicate hadronic (h) or leptonic (\Pe, $\mu$) decays.}
\begin{tabular}{l l l}
     & Category and decay mode       & Final state class and substructure                                                             \\ \hline
    \multirow{4}{*}{\rotatebox{90}{Signal}}
     & \hwwall fully hadronic & $3\Pq(0\PQc),\,3\Pq(1\PQc),\,3\Pq(2\PQc),\,4\Pq(0\PQc),\, 4\PQq(1\PQc),\,4\Pq(2\PQc)$          \\[\cmsTabSkip]
     &                               & $\Pe\PGn\Pq{}\Pq(0\PQc),\,\Pe\PGn\Pq\Pq(1\PQc),\,\mu\PGn\Pq\Pq(0\PQc),\,\mu\PGn\Pq\Pq(1\PQc)$          \\
     & \hwwall semileptonic   & $\PGt_{\Pe}\PGn\Pq\Pq(0\PQc),\,\PGt_{\Pe}\PGn\Pq\Pq(1\PQc),\,\PGt_{\PGm}\PGn\Pq\Pq(0\PQc),\,\PGt_{\PGm}\PGn\Pq\Pq(1\PQc),$       \\
     &                               & $\tauh\PGn\Pq\Pq(0\PQc),\,\tauh\PGn\Pq\Pq(1\PQc)$                                              \\[\cmsTabSkip]
    \multirow{5}{*}{\rotatebox{90}{Background}}
     & $\PQt\to\PQb\PW$ hadronic     & $\PQb\Pq(0\PQc),\,\PQb\Pq(1\PQc),\,\PQb\Pq\Pq(0\PQc),\,\PQb\Pq\Pq(1\PQc)$                      \\
     & $\PQt\to\PQb\PW$ leptonic     & $\PQb \Pe\PGn,\,\PQb\mu\PGn,\,\PQb\tauh\PGn,\,\PQb\PGt_{\Pe}\PGn,\,\PQb\PGt_{\PGm}\PGn$ \\
     & $\PH,\PZ,\PW\to\Pq\Pq$        & $\PQb{}\PQb,\,\PQc{}\PQc,\,\PQs{}\PQs,\,\Pq{}\Pq\,(\Pq=\PQu/\PQd)$                             \\
     & $\PH,\PZ\to\PGt\PGt$          & $\tauh\tauh,\,\tauh\PGt_{\Pe},\,\tauh\PGt_{\PGm}$  \\
     & QCD                           & $\PQb,\,\PQb{}\PQb,\,\PQc,\,\PQc{}\PQc,$\,others (light \Pq and gluon)                         \\
  \end{tabular}
  \label{tab:ParT_Classes}
\end{table}

The $0\Pell$ channel defines a discriminant, $P(\PH_{0\Pell})$, from the sum of all 16 \ParT \hwwall scores, including both semileptonic and fully hadronic $\PW\PW$ signal classes, as denoted in the upper part of Table~\ref{tab:ParT_Classes}.
Two exclusive selections based on $P(\PH_{0\Pell})$ are used: $P(\PH_{0\Pell})>0.99$ and $0.92 < P(\PH_{0\Pell}) < 0.99$, defining regions of high and moderate probability of containing $\PH_{0\Pell}$ signal events, respectively. 
These boundaries were chosen to maximize \SoverSqrtB (where $S$ and $B$ denote expected signal and background yields), yielding signal (background) efficiencies of $4.5\%$ ($0.02\%$) and $13.4\%$ ($0.35\%$) for the high and moderate probability regions, respectively.

The $1\Pell$ channel targets semileptonic $\PW\PW$ decays, \Hlqq, by identifying one isolated lepton \emph{inside} the AK8 jet.
This lepton-in-jet topology for background processes constitutes only a small fraction of the \ParT training sample, as lepton-jet overlap is rare in boosted topologies, limiting the tagger's ability to reject \wjetslnu and similar backgrounds in this channel.
To enhance the signal sensitivity and suppress the \wjetslnu background, a dedicated fine-tuning strategy is therefore developed.
Specifically, the activations of the final hidden layer of \ParT are used as input to a shallow neural network (multilayer perceptron) that categorizes jets into four classes: Higgs boson signal, QCD multijet, \ttbar, and \wjetslnu, following the transfer learning approach introduced in Ref.~\cite{Li:2024htp}. 
Fine-tuning is performed using a dedicated simulated data set of AK8 jets originating from these processes, selected such that the jets overlap with the leading lepton within a cone of radius 0.8, reflecting the topology of the single-lepton signal region (SR).

A separate sample of simulated Higgs boson signal events is used to prevent training data leakage into the validation set during the fine-tuning.
The resulting discriminant, $P(\PH_{1\Pell})$, is defined from the Higgs boson probability output of the fine-tuned model, denoted as \ParTfinetuned.

\begin{figure}[htb!]\centering
\includegraphics[width=0.485\linewidth]{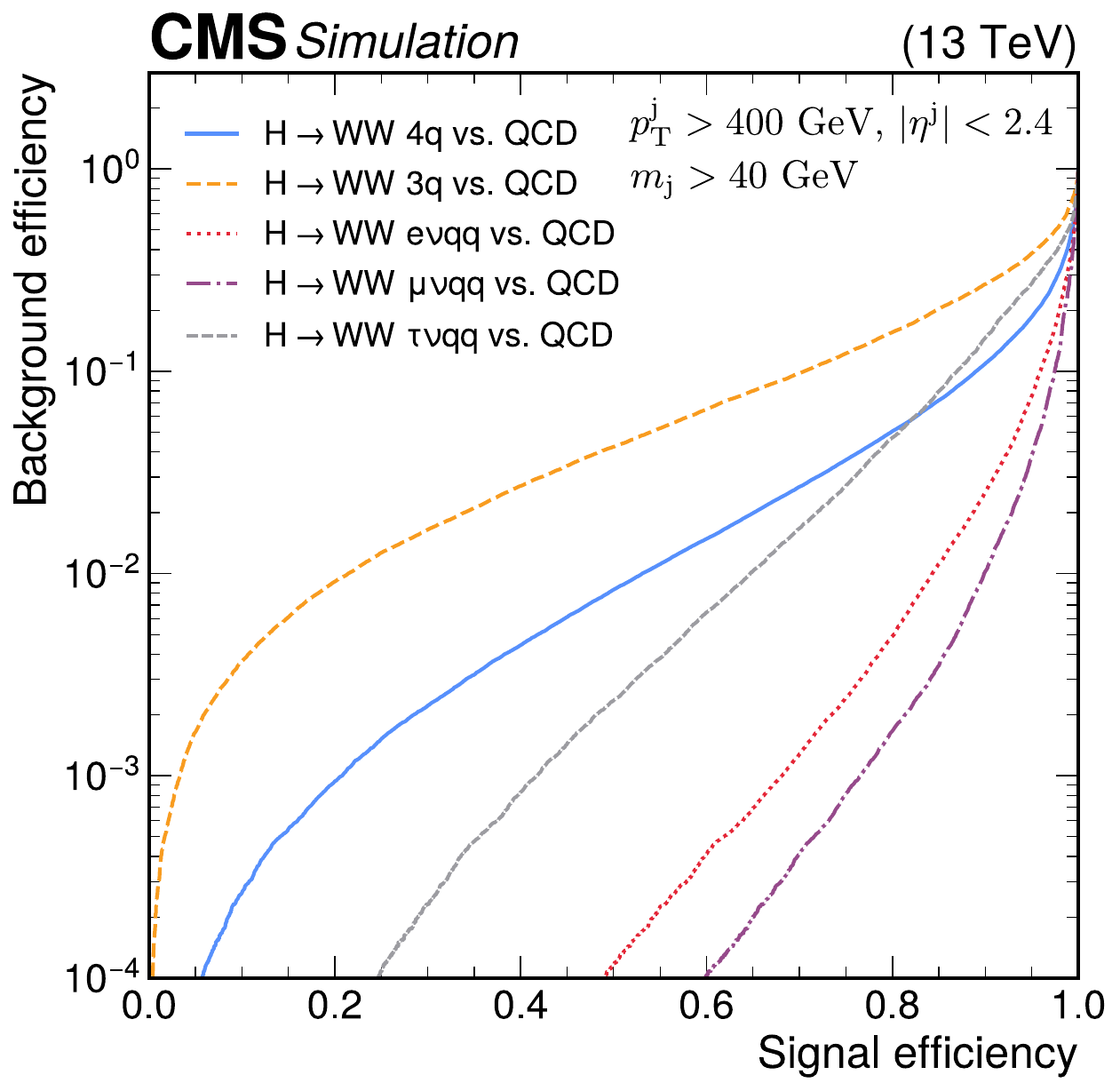}
\includegraphics[width=0.475\linewidth]{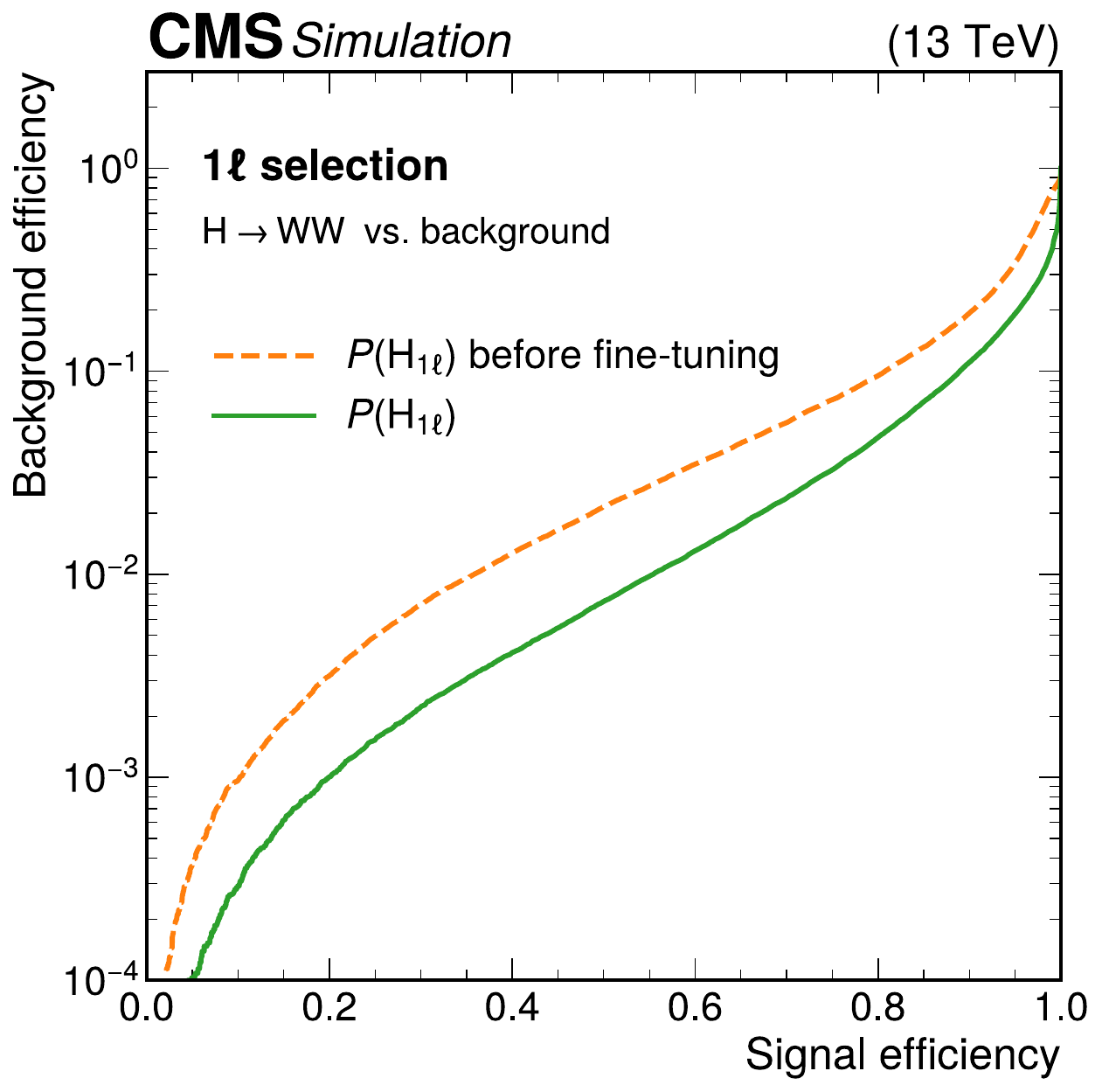}
\caption{Performance curves showing the identification probability of background jets versus \hwwall signal jets for \ParT and \ParTfinetuned. Left: Discrimination performance of the \ParT model for various \hwwall decays against the dominant QCD multijet background. Right: Comparison of $P(\PH_{1\Pell})$ before and after fine-tuning, following the event selection in the $1\Pell$ channel. The background includes jets originating from QCD multijet events, \wjetslnu, and top quark processes.}
\label{fig:ROCs}
\end{figure}

Compared to the original \ParT, the \ParTfinetuned tagger achieves nearly 60\% higher signal efficiency at a background efficiency of 1\% for \hwwlep decays.
This demonstrates the value of transfer learning: training large models on a broad set of classes and datasets, followed by targeted fine-tuning on a smaller subset specific to the topology of interest, as the $1\Pell$ final state here.
A comparison was also made directly using the CMS statistical analysis tool \textsc{Combine}~\cite{CMS:2024onh}, observing similar increase in expected significance. 
Tagger performance curves (background vs. signal efficiency) for simulated jets are shown in Fig.~\ref{fig:ROCs}, using a selection similar to the $0\Pell$ channel: $\ptja>400\GeV$, $\mj>50\GeV$, and $\abs{\eta^\text{j}}<2.4$.

The \Hlqq, \Hfourq, and \Hthreeq simulated signal jets lack a suitable SM counterpart jet for calibration.
Therefore, to calibrate the $P(\PH_{0\Pell})$ and $P(\PH_{1\Pell})$ discriminants, the Lund jet plane (LJP) reweighting technique~\cite{CMS:2025eyd} is applied to simulated jets with a fixed number of quark-initiated prongs.
The method reclusters each jet into subjets, each representing the radiation pattern of an individual quark from the hard interaction.
Corrections for each subjet are derived based on data-to-simulation ratios of LJP densities~\cite{Dreyer:2018nbf}, which provide representations that capture the phase space density of different types of splittings in the two-dimensional space of momentum transfer and angular separation.
These per-splitting corrections, measured in data using subjets from $\PW \to \Pq\Pq$ decays~\cite{CMS:2025eyd}, are combined into a total correction for the simulated jet.
In this analysis, the LJP reweighting is used to correct the signal selection efficiency in the $P(\PH_{0\Pell})$ and $P(\PH_{1\Pell})$ discriminants and to estimate the associated uncertainties.
The corresponding SFs are defined as the ratio of efficiencies after the LJP reweighting to those from the uncorrected simulation, and are validated in \ttbar-enriched control regions as described below.

The simulation is calibrated separately for each dominant signal component: \Hfourq, \Hthreeq, and \Hlqq, corresponding to jets containing four, three, and two quark-originated subjets, respectively.
Jets are matched to their generator-level category and reclustered into the appropriate number of subjets based on their quark content, using the exclusive \kt algorithm~\cite{Catani:1993hr, Ellis:1993tq}.
The splitting pattern is extracted using the clustering history of each subjet.
For \Hlqq jets, reconstructed lepton and photon PF candidates within 
$\Delta R = 0.2$ of the generator-level lepton are removed before 
reclustering, to isolate the hadronic component from the leptonic one.

Individual SFs derived for each signal component are then combined to produce a global SF applicable to all signal jets in each analysis category.
The resulting SFs range from 0.84 to 0.98, with total uncertainties of 9--27\%, as described in Section~\ref{Sec:uncertainties}.

The procedure is validated in regions enriched in \ttbar events, which closely resemble the signal topology in the $0\Pell$ and $1\Pell$ channels and require at least one \PQb-tagged jet.
Generator-level information from the top quark decay is used to classify top quark jets as ``matched'' or ``unmatched,'' with the reweighting applied only to the matched jets using the LJP ratios.
For hadronic top quark decays (which serve as a proxy for jets with four or three quark-originated subjets), the matched component corresponds to $\PQt\to\PQb\PQq\PQq$ decays, where all the \PW boson daughter quarks and the \PQb quark are within the jet radius.
For jets with two quark-originated subjets and one lepton, the matched component comes from \ttbar events where both top quark decays are separated by small angular distances and the visible products from a leptonic \PW boson and a hadronic \PW boson decays are contained within the jet radius; this selection results in a $\PQt\to\Pell\PQq\PQq$ topology.
The LJP-reweighted simulation is found to agree well with the data in the $P(\PH_{0\ell})$ and $P(\PH_{1\ell})$ distributions in a \ttbar-dominated validation sample, confirming the validity of the SF corrections applied to top quark jets.

In this analysis, the Higgs boson is reconstructed as a single large-radius jet corrected for the potential presence of a neutrino.
For signal jets, the \mj distribution exhibits a sharp peak near the nominal Higgs boson mass for \Hfourq jets and a broader peak for the \Hthreeq and \Hlqq components, due to the partial reconstruction of the $\PW\PW$ system.
To improve the mass resolution for the \Hlqq signal, the reconstructed Higgs boson four-momentum is calculated as the sum of the jet four-momentum and the estimated four-momentum of the neutrino.
The four-momentum of any lepton within the AK8 jet cone is included in the \PH-candidate jet momentum, since leptons within the jet cone are treated as PF constituents at jet reconstruction.
The \ptvecmiss is attributed solely to the neutrino, with $\ptvec^\PGn = \ptvecmiss$. Since the \PH-candidate jet is highly boosted, the neutrino is expected to be collinear with the jet axis, thus we assume $\eta^\PGn = \eta^\text{j}$, enabling full four-momentum reconstruction for the neutrino.
Then, the corrected mass of the \PH-candidate jet, referred to as \mjs, is evaluated as the invariant mass of the reconstructed Higgs boson system.
This definition aligns the peak position of the \mjs distribution for \Hlqq signal jets to 125\GeV.

The correction to the jet mass is applied only to events with significant \ptmiss aligned with the candidate jet.
In the $0\Pell$ channel, this is quantified by requiring the ratio $\ptmiss/\ptja$ to be greater than 0.1 and the azimuthal angle between \ptvecmiss and the jet, $\abs{\Delta\phi(\text{j}, \ptvecmiss)}$, to be less than 0.8.
For events that do not meet these conditions, the mass of the \PH-candidate jet is taken as \mj.
For all $1\Pell$ channels, the correction to the jet mass is always applied.

\section{Analysis of the \texorpdfstring{$0\Pell$}{0l} channel}
\label{Sec:0l_channel}

The analysis in the $0\Pell$ channel considers all major Higgs boson production processes (ggF, VBF, \VH, and \ttH) decaying into final states without a lepton (\Hfourq, \Hthreeq) or with a lepton that fails the isolation requirements (\Hlqq).
Four orthogonal SRs are defined, with background predictions derived from a combination of control regions (CRs) in data and simulation.

\subsection{Event selection}
\label{Sec:0l_selection}

Events in the $0\Pell$ channel are selected using a combination of hadronic triggers.
The hadronic triggers impose a minimum threshold on either the $\pt$ of an AK8 jet or the event $\HT$, defined here as the scalar $\pt$ sum of all AK4 jets in the event with $\pt>30\GeV$ and $\abs{\eta}<2.4$.
For AK8 jets used in the trigger selection, a minimum trimmed jet mass of 30 or 50\GeV is also required~\cite{Krohn:2009th}.
The combined hadronic trigger efficiency is 75--80\% for $200<\pt<400\GeV$, 80--95\% for $400<\pt<450\GeV$, and nearly 100\% for $\pt>450\GeV$.
The efficiency is measured using AK8 jets in an independent sample of $\mu$+jets \ttbar events collected with a single-muon trigger.

Offline, events are selected with two or three large-radius jets.
To ensure high trigger efficiency, the highest \pt jet and the highest mass jet are required to satisfy $\pt > 400\GeV$ and $\mj > 50\GeV$, respectively.
The jet with the highest $P(\PH_{0\Pell})$ discriminant value is identified as the \PH-candidate jet and must also satisfy $\mj > 40\GeV$.
Events with isolated leptons, as defined in Section~\ref{Sec:Objects}, are vetoed in the $0\Pell$ channel, but events with nonisolated leptons are included and contribute significantly to the sensitivity of this channel.
No \PQb-tagged jet veto is applied in the $0\ell$ channel, as \ttH production is part of the signal considered.
The 0$\ell$ channel contributes significantly less than the 1$\ell$ due to (a) the overwhelming QCD multijet background, (b) the $\ell$ isolation requirements, and (c) reduced trigger efficiency for $\ptH< 400 \GeV$.

In signal events, the dominant contribution originates from \Hfourq jets ($\approx$28\%), 
followed by \Hthreeq ($\approx$14\%) and \Hlqq ($\approx$13\%) jets, accounting for 
$\approx$55\% of signal events in total. The remaining $\approx$45\%, designated as 
``Other,'' arises because the \PH-candidate jet \pt threshold of 200\GeV allows for 
partially boosted topologies, where resolved \PW bosons from \PH decays or jets from 
associated production processes are misidentified as the \PH-candidate. This category 
includes jets matched to a \PV boson ($\approx$18\%), a gluon or light-flavor quark 
($\approx$10\%), a top quark from \ttH production ($\approx$5\%), rare \PH decay 
topologies ($\approx$3\%), and jets unmatched to any of the above objects ($\approx$9\%).

\begin{figure}[!htbp]\centering
\includegraphics[width=0.490\linewidth]{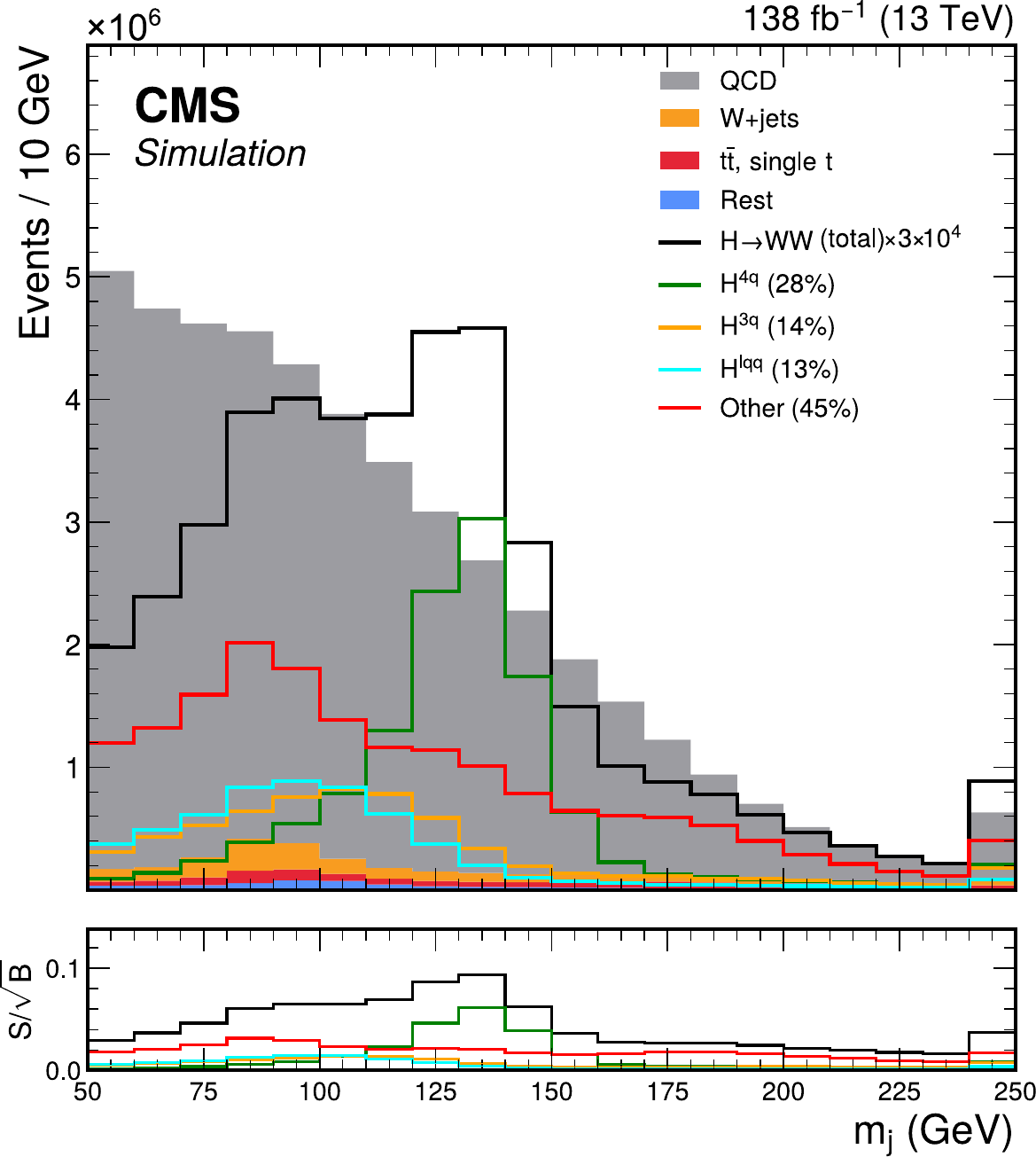}
\includegraphics[width=0.497\linewidth]{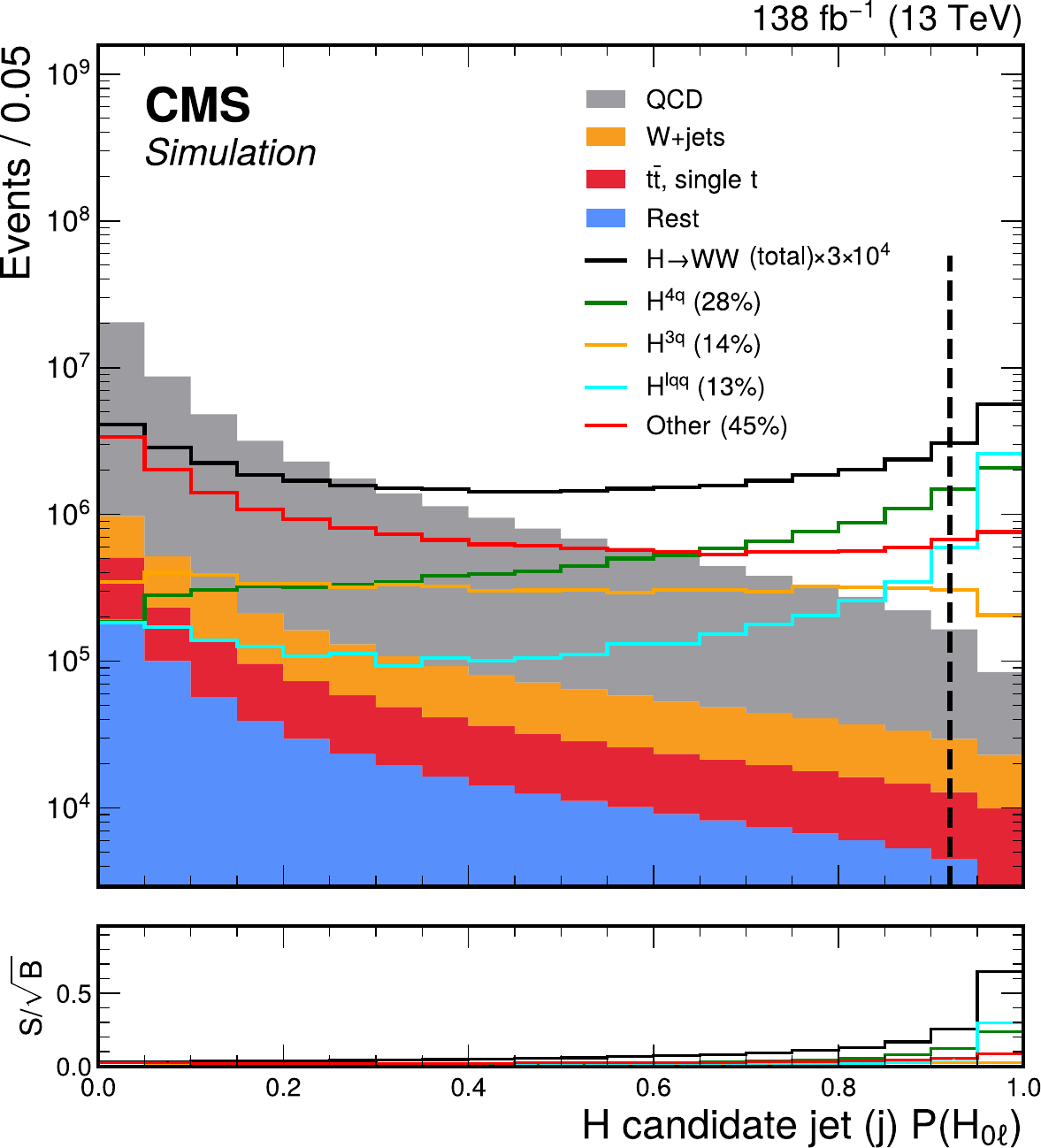}
\includegraphics[width=0.497\linewidth]{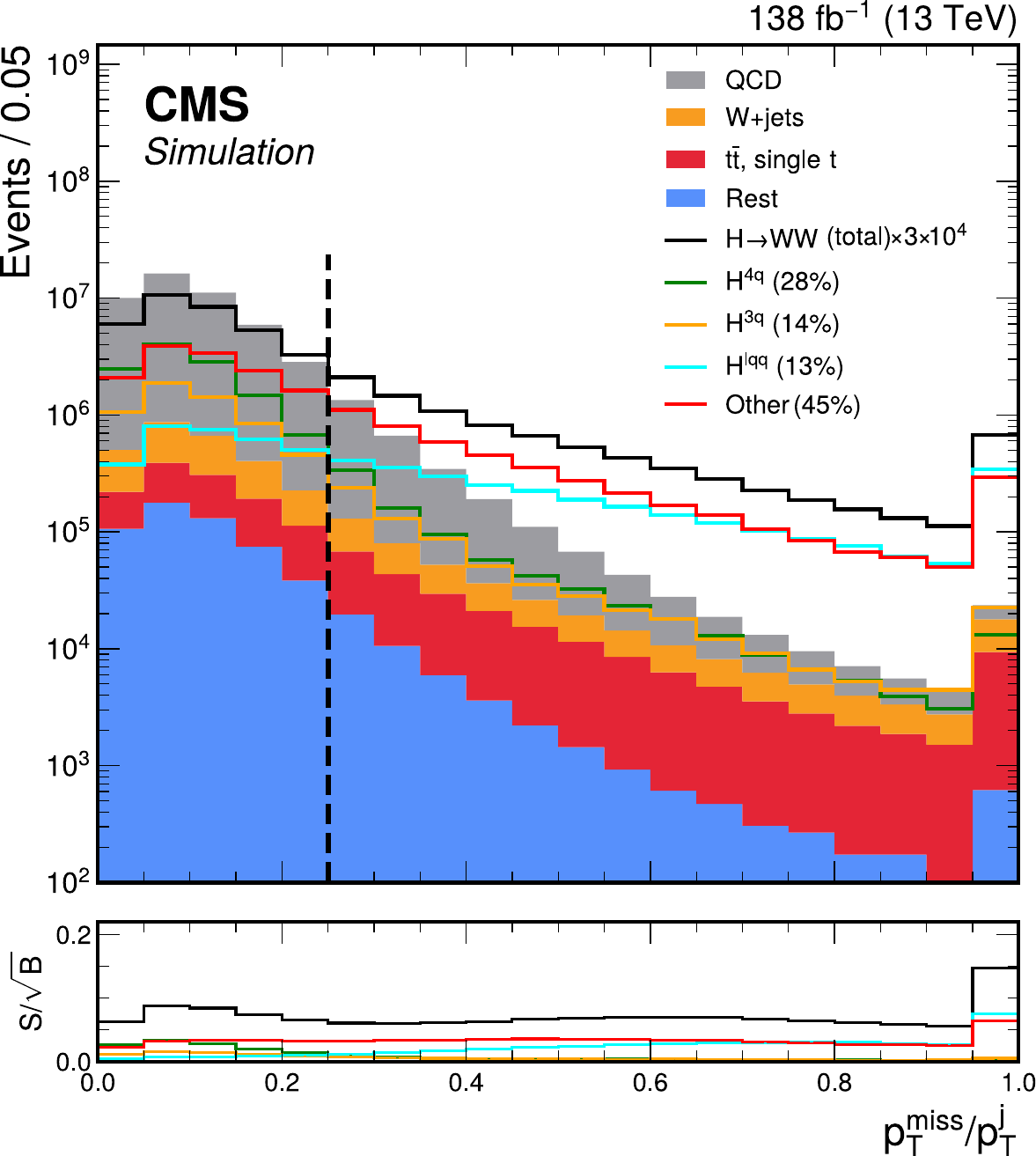}
\includegraphics[width=0.485\linewidth]{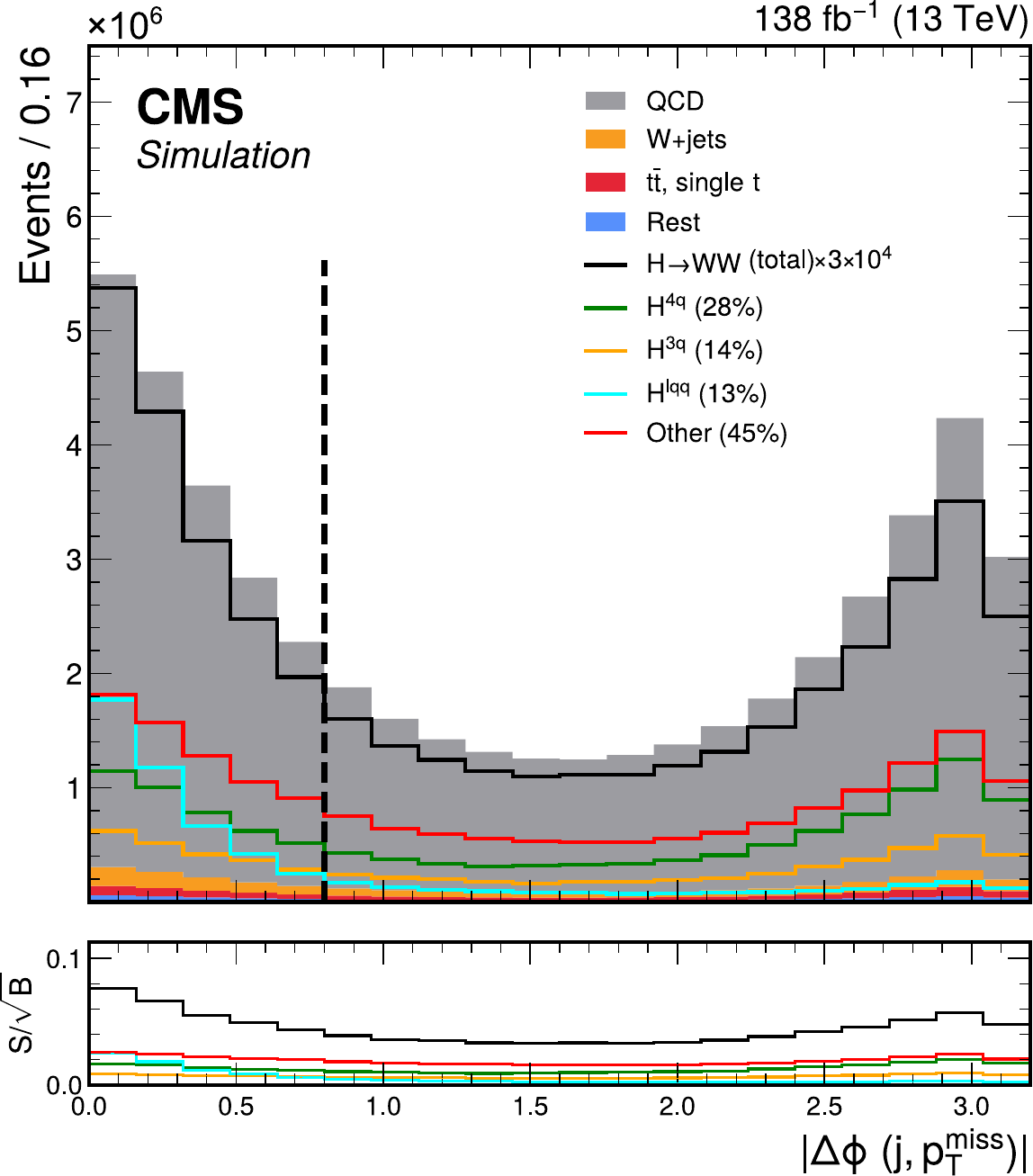}
\caption{Distributions of the total simulated background and total signal 
(scaled by a factor of $3\times10^{4}$ for visibility) passing event 
selection in the $0\Pell$ channel, prior to the requirements on 
$P(\PH_{0\Pell})$, $\ptmiss/\ptja$, and $|\Delta\phi(\text{j},\ptvecmiss)|$ 
used to define the SR.
The signal is split into classes as defined in the text.
The upper left and upper right panels show the soft-drop mass and \ParT score 
distributions for the \PH-candidate jet (j) $P(\PH_{0\ell})$, respectively.
The lower left and lower right panels display the $\ptmiss/\ptja$ ratio and the 
angle $\abs{\Delta\phi(\text{j}, \ptvecmiss)}$, respectively.
Vertical lines indicate the selection conditions imposed to define the SRs.}
\label{fig:0l_Preselection}
\end{figure}

Figure~\ref{fig:0l_Preselection} shows the distributions of the \PH-candidate jet mass $\mj$, $P(\PH_{0\Pell})$ discriminant, $\ptmiss/\ptja$, and $\abs{\Delta\phi(\text{j}, \ptmiss)}$, in simulated background and signal events, where j corresponds to the \PH-candidate jet.
The \Hlqq signal, which is associated with energetic neutrinos aligned with the jet axis, can be isolated from the other signal components and the multijet background by selecting on $\ptmiss/\ptja$ and $\abs{\Delta\phi(\text{j},\ptmiss)}$.

The SR is defined by $P(\PH_{0\Pell}) > 0.92$.
It is first divided into two main categories based on the ratio of \ptmiss to the \PH-candidate jet $\pt$: \SRone with $\ptmiss/\ptja<0.25$, which includes contributions from all three major signal components, and \SRtwo with $\ptmiss/\ptja>0.25$, where an additional requirement of $\abs{\Delta\phi(\text{j},\ptvecmiss)}<0.8$ ensures alignment between the reconstructed neutrino and the \PH-candidate jet in the transverse plane.
Each category is further divided by signal purity into \SRa $(P(\PH_{0\Pell})>0.99$) for high purity and \SRb ($0.92<P(\PH_{0\Pell})<0.99$) for medium purity, resulting in four statistically independent signal regions labeled \SRonea, \SRoneb, \SRtwoa, and \SRtwob.

Because of the high-\ptmiss requirement, \SRtwoa and \SRtwob are dominated by the \Hlqq component, accounting for 75 and 93\% of the total signal events, respectively.
Despite the low-\ptmiss selection, the composition of signal events in \SRonea is also dominated by \Hlqq at 63\% because of the greater efficiency at high $P(\PH_{0\Pell})$ (Fig.~\ref{fig:ROCs}, left; Fig.~\ref{fig:0l_Preselection}, upper right) compared to \Hfourq, which follows at 27\%.
The \SRoneb is populated by \Hfourq at 57\%, followed by \Hlqq at 19\%.
The \Hthreeq contribution ranges from 1 to 8\%.
Therefore, the sensitivity to Higgs boson production in the 0$\Pell$ channel is driven by the semileptonic $\PW\PW$ decays with a nonisolated lepton.
The presence of \Hlqq across all SRs necessitates a specialized Higgs boson mass reconstruction, incorporating information about the neutrino via \ptmiss and $\abs{\Delta\phi(\text{j}, \ptvecmiss)}$ even in the low-\ptmiss regions \SRonea and \SRoneb, as described at the end of Section~\ref{Sec:Higgs_identification_ParT_calibration}.
All kinematic requirements defining the four SRs and two CRs (described in Section~\ref{Sec:0l_prediction}) are summarized in Table~\ref{tab:0l_selection}. 
All conditions and selection boundaries in Table~\ref{tab:0l_selection} were chosen to maximize \SoverSqrtB.

The dominant \Hlqq signal exhibits a broad \mj distribution peaking at masses lower than the Higgs boson mass due to the missing neutrino.
The rightmost column of Table~\ref{tab:0l_selection} specifies conditions defining the phase space where \mj is corrected for neutrino reconstruction and replaced by \mjs; elsewhere, $\mjs=\mj$.
These conditions are designed to shift the mean of the \Hlqq mass distribution toward the Higgs boson mass and to minimize its width.
This replacement shifts the mass distributions for all events in the affected regions to higher values.

\begin{table}[!htbp]\centering
\topcaption{Kinematic requirements used to define the SRs and CRs in the $0\Pell$ channel. The rightmost columns list the conditions, combined with logical ``AND'', under which \mj is replaced by the corrected \mjs mass. The $\abs{\Delta\phi}$ denotes the azimuthal angle difference between \ptvecmiss and the Higgs boson candidate \pt vector.} 
\label{tab:0l_selection}
\begin{tabular}{c c ccc c c}
Region && $\ptmiss/\ptja$ & $P(\PH_{0\Pell})$  & $\abs{\Delta\phi(\text{j}, \ptvecmiss)}$ && Apply \ptmiss correction if \\ \hline
 \SRonea && $<$0.25 & $>$0.99 & any && $\abs{\Delta\phi}<0.8,\,\ptmiss/\ptja>0.1$ \\
 \SRoneb && $<$0.25 & $0.92$--$0.99$ & any && $\abs{\Delta\phi}<0.8,\,\ptmiss/\ptja>0.1$ \\
 \SRtwoa && $>$0.25 & $>$0.99 & $\abs{\Delta\phi}<0.8$ && always \\
 \SRtwob && $>$0.25 & $0.92$--$0.99$ & $\abs{\Delta\phi}<0.8$ && always \\[\cmsTabSkip]
 \CRone  && $<$0.25 & $<$0.92 & any && $\abs{\Delta\phi}<0.8,\,\ptmiss/\ptja>0.1$ \\
 \CRtwo  && $>$0.25 & $<$0.92 & $\abs{\Delta\phi}<0.8$ && always \\
 \end{tabular}
\end{table}
\subsection{Background estimation}
\label{Sec:0l_prediction}

The statistical analysis is performed over the \mjs spectra in the four SRs, with background predictions derived separately for QCD multijet and other processes.
The QCD multijet production dominates \SRonea, \SRoneb, and \SRtwob ($>$90\%) and contributes subdominantly to \SRtwoa (30\%); it is estimated using a data-driven method.
The \wjetslnu and top quark backgrounds, which contribute significantly to \SRtwoa, as well as diboson processes, are estimated from simulation and validated in data.

\subsubsection{The QCD multijet background}
\label{Sec:0l_pred_QCD}

The QCD multijet process is the dominant background in all the SRs, except for \SRtwoa.
We estimate this background using a data-driven method implementing two CRs, the \CRone and \CRtwo, defined by requiring $P(\PH_{0\Pell})<0.92$, but still fulfilling the $\ptmiss/\ptja$ condition of \SRone and \SRtwo, respectively.

The QCD multijet background in each $\text{SR}_{ir}$ ($i=1,2$; $r=\text{a},\text{b}$) is estimated from the data distribution in the corresponding CR, multiplied by a polynomial transfer function (\TFir) to account for the shape differences between the SR and CR.
These differences arise from the kinematic correlation between \mjs and the variables defining the SR, as well as residual differences in \ParT efficiency between data and simulation for background processes.

The prediction of the QCD background yield in $\text{SR}_{ir}$ takes the form
\begin{equation}
  N_{\text{SR}_{ir}}^{\text{QCD},k} = N_{\text{CR}_i}^{\text{QCD},k} \;
  \TFir({\mjs}^{k}),
  \label{eq:pred_0l}
\end{equation}
where ${\mjs}^{k}$ is the center of the $k^\text{th}$ \mjs bin, $N_{\text{SR}_i}^{\text{QCD},k}$ is the estimated number of QCD background events in bin $k$ of $\text{SR}_{ir}$, and $N_{\text{CR}_i}^{\text{QCD},k}$ is the number of data events minus the number of predicted non-QCD background events in bin $k$ of $\text{CR}_i$.

\begin{figure}[!htbp]
  \centering
  \includegraphics[width=0.47\linewidth]{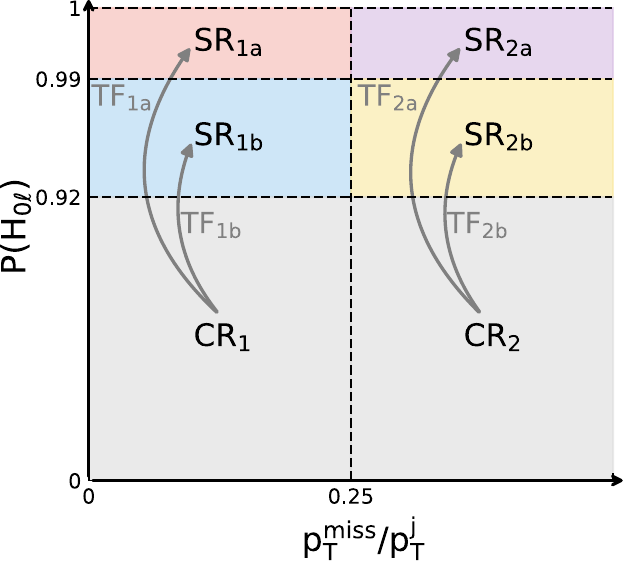}
  \includegraphics[width=0.52\linewidth]{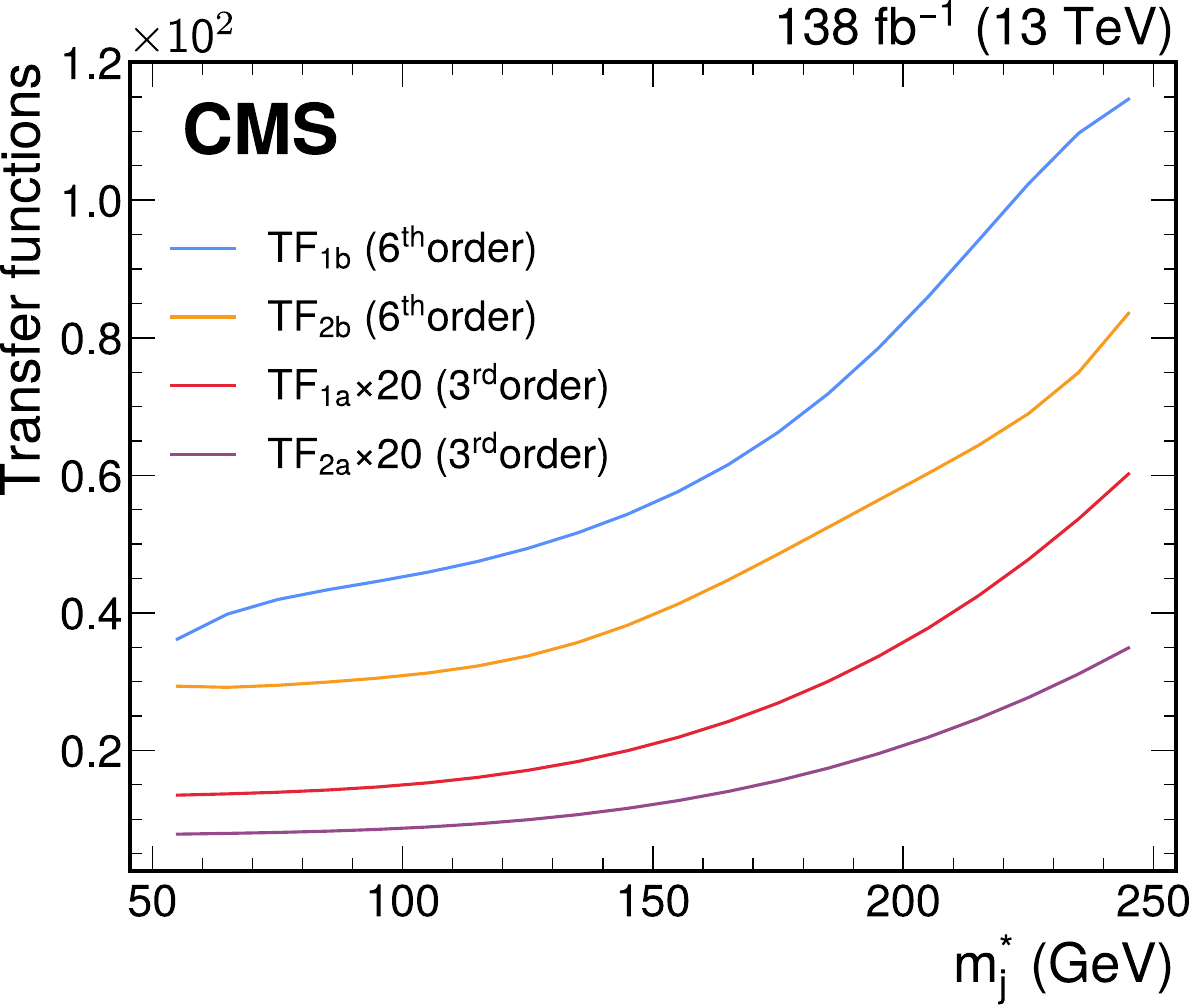}
  \caption{
    Illustration of the SRs and CRs, and the TFs used to relate the QCD background in the different regions (left).
    The TFs used to predict the QCD process in the four SRs as a function of the \mjs (right).
  }
  \label{fig:0l_TFs}
\end{figure}

The QCD background in \SRonea and \SRtwoa is estimated from \CRone and \CRtwo using two transfer functions, \TFonea and \TFtwoa, 
which share the same shape (determined by a simultaneous fit across all four regions) and have independent normalizations.
For \SRoneb and \SRtwob, the QCD background is estimated with two independent transfer functions, \TFoneb and \TFtwob, each having its shape and normalization determined by a fit to its corresponding CR and SR pair. 
The resulting transfer functions are shown in Fig.~\ref{fig:0l_TFs} (right).

The polynomial transfer function, \TFir, is defined using the Bernstein basis:
\begin{equation}
  \TFir(\mjs) = \sum_{l=0}^{n} a_{l} b_{l,n}(\mjs) \label{eq:TF}
\end{equation}
where $a_{l}$ are fitted coefficients, and $b_{l,n}$ are the \mjs-dependent one-dimensional Bernstein basis polynomials of degree $n$, properly transformed to the \mjs range probed.
The optimal order of the transfer function, $\TFir(\mjs)$, is determined to be 3 for $\text{TF}_{i\text{a}}$ and 6 for $\text{TF}_{i\text{b}}$, based on the results of a Fisher F-test~\cite{ftest}, performed independently for regions ``a'' and ``b''.
These high-degree polynomials capture the smoothly varying kinematic differences between the SRs and CRs without introducing spurious structures.

The parameters of the polynomial function are treated as unconstrained in the fit to data.
The post-fit results of the four TFs are shown in Fig.~\ref{fig:0l_TFs} (right).
They demonstrate monotonic behavior and a smooth shape variation near the 125\GeV region, where the signal is most abundant.
The predicted spectra for the QCD process are presented and discussed in Section~\ref{Sec:Results}.

\subsubsection{Other background processes}
\label{Sec:0l_pred_Wjets_Top}

The background events from \wjetslnu and top quark production are predicted by simulation.
Their normalization is validated in data in two background-enriched samples within a kinematic phase space consistent with \SRtwoa.
The first sample is the subset of \SRtwoa defined by requiring $\mjs> 160\GeV$, while the second sample is defined as the \SRtwoa but inverting the angular requirement to $\abs{\Delta\phi(\text{j},\ptvecmiss)}>0.8$. 
These samples receive equal contributions from top quark and \wjetslnu production processes. 
The simulated events match the data \mjs distribution at a 5\% level in both normalization and shape within statistical uncertainties.
Furthermore, this consistency has been verified under the following conditions: (a) applying a stricter than the SR $\ptmiss/\ptja>0.25$ requirement, which further suppresses QCD contributions, (b) excluding narrow \PQb-tagged jets, thereby enhancing \wjets fraction, and (c) requiring at least one \PQb-tagged jet, increasing the presence of the top quark background.

Based on the observed data-to-simulation agreement in the aforementioned validation samples (all of which require a high \ParT score) and in relevant samples in the $1\Pell$ channel, a 5\% rate uncertainty is assigned to each of the \wjets and top quark processes. 
This uncertainty collectively accounts for the limited simulation sample size, residual QCD contamination, and potential cross section mismodeling.

Events originating from the $\hzzall\to4\PQq$ process are suppressed relative to $\hwwall\to4\PQq$ decays because of the smaller branching fraction and reduced \ParT tagger efficiency, by factors of 10 and 3, respectively.
This small fraction of background events entering \SRone is estimated using simulation.

\section{Analysis in the \texorpdfstring{$1\Pell$}{1l} channel}
\label{Sec:1l_channel}

The analysis in the $1\Pell$ channel focuses on events with one isolated lepton, targeting the \Hlqq final state.
Events are partitioned into three mutually exclusive categories, targeting the ggF, VBF, and \VH production processes.

\subsection{Event selection}
\label{Sec:1l_selection}

Online, events are collected using a combination of single-lepton (\Pgm or \Pe) triggers.
For \PGm (\Pe), these triggers require a $\pt$ threshold of 50 (115)\GeV and no isolation requirement, or a $\pt$ above 24--27 (27--35)\GeV and a \Irel requirement.
The range indicates the difference in triggers across the different data-taking periods.
At high \pt, single-electron triggers are complemented by a single photon trigger with thresholds of $\pt > 175\GeV$ for the 2016 data-taking period and $\pt > 200\GeV$ for the 2017 and 2018 periods.
The isolated-lepton triggers allow for a lower $\pt^\ell$ threshold, even if they are less efficient for the signal.
This motivates the following definition of an ``isolated lepton'' in the offline analysis.
Muons are required to have $\Imini<0.2$ for $\pt>55\GeV$ and $\Irel<0.2$ for $30<\pt<55\GeV$.
Electrons are required to have $\Irel<0.15$ for $38<\pt<120\GeV$, with no \Imini requirement for $\pt>120\GeV$.

Offline, events are required to contain at least one large-radius AK8 jet with $\pt > 250\GeV$ and exactly one isolated \PGm or \Pe.
The \PH-candidate jet is defined as the AK8 jet satisfying $P(\PH_{1\Pell})>0.75$ and $\mj > 40\GeV$, with an isolated lepton located in its vicinity, requiring the $\Delta R$ distance between the lepton and the jet axis to lie within $0.03 < \Delta R < 0.8$. Therefore, the lepton is one of the PF candidates of the jet, contributing to its mass and momentum.

The VBF category ($1\Pell$ VBF) selects events with two additional AK4 jets accompanying the \PH-candidate jet, with the requirement that the jets be well separated and do not geometrically overlap.
The two AK4 jets are required to have a pseudorapidity separation of $\abs{\Delta\eta_{\jj}}>3.5$ and an invariant mass of $\mjj>1\TeV$; the selection criteria were chosen to maximize the sensitivity to VBF production.
If an event contains more than two AK4 jets, the two highest-\pt jets are used to compute $\Delta\eta_\jj$ and $\mjj$.
Events containing fewer than two AK4 jets, or where $\abs{\Delta\eta_{\jj}}<3.5$ or $\mjj<1\TeV$, fall into the \VH or ggF categories.
Events with an additional AK8 jet tagged by the \textsc{ParticleNet} algorithm as originating from a \PV boson ($P(\PV)>0.9$), satisfying $\pt>250\GeV$ and $\mj>40\GeV$, are assigned to the \VH category.
This \PV boson identification corresponds to selection efficiencies of approximately 33 and 0.5\% for signal and background jets, respectively. 
Events without additional jets satisfying the above criteria are assigned to the ggF category.  
The \PH-candidate jet must also satisfy tagger requirements specific to each category: $P(\PH_{1\Pell}) > 0.93$ for the ggF category, and $P(\PH_{1\Pell}) > 0.905$ for the VBF and \VH categories.
This selection condition has been verified to preserve the background \mjs peak location and shape, with no artificial enhancement observed near 125\GeV across different tagger score selections.

For the ggF and VBF categories, we require $\ptmiss > 20\GeV$ and, since signal events feature \ptvecmiss aligned with the \PH-candidate jet, we apply $\abs{\Delta\phi(\text{j},\ptvecmiss)} < \pi/2$.
To suppress \ttbar and single \PQt backgrounds, events must not contain any \PQb-tagged AK4 jets with the tight WP in the hemisphere opposite to the \PH-candidate jet. 
In the \VH category, we reject events with a \PQb-tagged AK4 jet that does not overlap with \PV- or \PH-tagged large-radius jet.
Furthermore, events are required to have $\ptmiss>30\GeV$, but there is no requirement on $\abs{\Delta\phi(\mathrm{j},\ptvecmiss)}$ in this category.  

The ggF category is subdivided into three bins based on the \pt of the reconstructed \PH-candidate jet ($\pt^\text{j}$), with ranges of 250--350, 350--500, and ${>}500\GeV$.
These bins are chosen to maximize the diagonal elements of the migration matrix ($M_{\text{mig}}$), which is formed using $\pt^\text{j}$ at the reconstructed level and the corresponding \ptH at the generator level.
The corresponding \ptH bins are 200--300, 300--450, and ${>}450\GeV$, consistent with the recommendations of the simplified template cross section (STXS) framework version 1.2~\cite{deFlorian:2016spz,Berger:2922392}. This choice facilitates future combined measurements of the differential Higgs boson production cross section as a function of \ptH.
The migration between generator-level STXS bins and the reconstruction-level categories used in the $1\ell$ channel is given by the matrix $M_{\rm mig}$:
\begin{equation*}
M_\text{mig} = \begin{pmatrix}
0.72 & 0.05 & 0.14 & 0.09 \\
0.02 & 0.78 & 0.18 & 0.02 \\
0.05 & 0.20 & 0.70 & 0.05 \\
0.05 & 0.04 & 0.26 & 0.65
\end{pmatrix},
\end{equation*}
where each row presents the fraction of generated signal events in the 
corresponding STXS bin reconstructed in each analysis category. 
The rows correspond to the VBF ($\mjj^\text{gen} > 1\TeV$, where 
$\mjj^\text{gen}$ is the generator-level dijet invariant mass) and 
ggF ($\ptH \in [200,300],~[300,450],~>450\GeV$) STXS bins, and the 
columns correspond to the VBF and ggF analysis categories at 
reconstruction level ($\ptja \in [250,350],~[350,500],~>500\GeV$). 
The dominant diagonal indicates good correspondence between the 
generator and the reconstruction binning.

\begin{figure}[!htbp]
  \centering
  \includegraphics[width=0.55\linewidth]{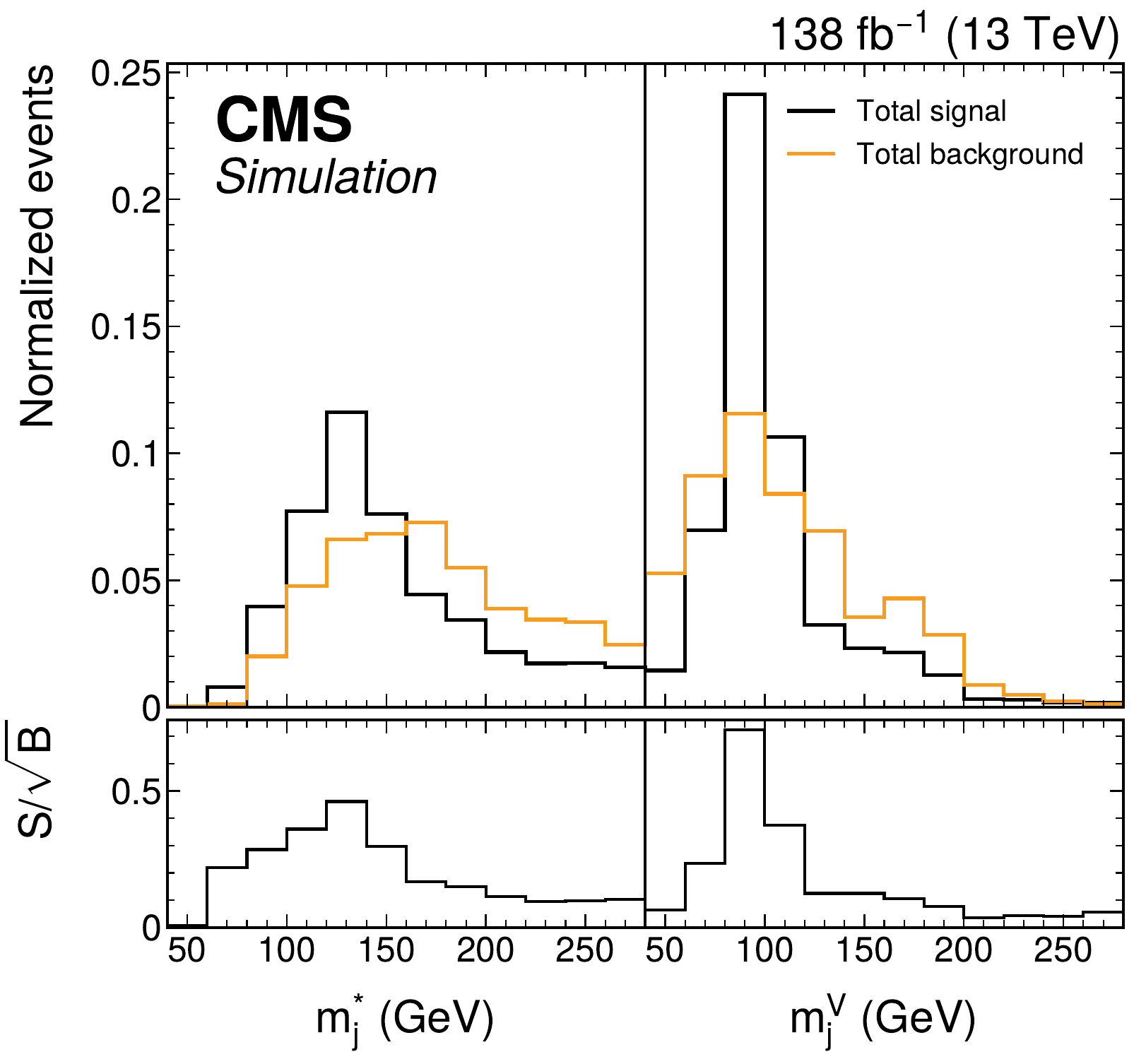}
\caption{Distributions of \mjs (left) and the $m_\text{j}^\PV$ (right) in the \VH SR for simulated signal and background, normalized to unit area. 
The lower panels show the bin-by-bin \SoverSqrtB. 
The superior signal-to-background separation of $m_\text{j}^\PV$, owing to its 
narrower resonant peak, motivates its use as the \VH discriminant.}
  \label{fig:vh_mass_choice}
\end{figure}

The signal extraction is performed with a binned maximum likelihood fit, using the \mjs in the ggF and VBF channels, and the \PV-candidate jet soft-drop mass, $m_\text{j}^\PV$, in the \VH channel.
The \mjs distribution exhibits a broad peak, with the resolution limited by the presence of one or more neutrinos from decays of \PW bosons or \PGt leptons.
For the \VH channel, the $m_\text{j}^\PV$ distribution exhibits a sharper peak and more distinct shape differences from key backgrounds than \mjs, as shown in Fig.~\ref{fig:vh_mass_choice}, making it the preferred discriminant for this channel.
A coarse binning is applied to the $m_\text{j}^\PV$ distribution, concentrating most signal events in a single bin ($70<m_\text{j}^\PV<110\GeV$) that captures events from both $\PW\PH$ and $\PZ\PH$ production processes.
Limited event yields prevent a two-dimensional fit using both resonance masses. 

\subsection{Background estimation}
\label{Sec:1l_prediction}

The dominant sources of background are the \wjetslnu and \ttbar processes, where a high-\pt lepton from a \PW boson decay is merged within a jet.
Both contributions are estimated from simulation with corrections derived from data CRs.
The background from events with misidentified leptons or leptons from heavy-flavor hadron decays (nonprompt leptons) is suppressed by the identification and isolation requirements imposed on the \PGm and \Pe candidates, as well as the \ParT tagger selection, while the remaining contribution is estimated directly from data.
Other subdominant sources of background originate from the \PZ{}+jets and diboson processes, and are estimated from simulation.

\subsubsection{Nonprompt-lepton background}
\label{subSec:fake_prediction}

Background events containing nonprompt leptons misidentified as prompt, isolated leptons could be selected for the SR.
This background includes the QCD multijet process.
Using the data-driven method described below, its contribution to the total background yield in the signal and control regions is estimated to be less than 2\%.
The contribution of this background is obtained by reweighting events from a control sample containing lepton candidates that pass looser selection criteria than those of the SR, but fail the SR requirements.
This sample has a minor contribution from EW processes containing genuine prompt leptons, which is subtracted from data using the simulation-based estimate.
The weights, called misidentified-lepton factors, are measured in QCD-enriched 
samples, as $\epsilon_{\text{misID}}/(1 - \epsilon_{\text{misID}})$, where 
$\epsilon_{\text{misID}}$ is the probability of a nonprompt lepton satisfying the less stringent criteria to also pass the SR criteria.
The QCD-enriched samples follow the SR selection, except for requiring exactly 
one loose (rather than tight and isolated) lepton, no \PQb-tagged AK4 jets at the loose \textsc{DeepJet} WP, and $\ptmiss < 20\GeV$ to suppress prompt-lepton contamination from \wjets and ensure orthogonality to the SR.

The misidentified-lepton factors are parameterized as a function of the \pt and $\eta$ of the leptons.
The uncertainty relative to the subtraction of the prompt-lepton background is also propagated into the misidentified-lepton factors.
This method is described in detail in Ref.~\cite{CMS:2018zzl}.

\subsubsection{The \texorpdfstring{\wjets}{W+jets} and top quark backgrounds}
\label{subSec:1lep_prediction_wjets}

The background contributions from $\PW(\Pell\PGn)$+jets and \ttbar processes are estimated from simulation, with their rates constrained using data from CRs enriched in these processes.
The CRs are defined by minimally modifying the selections of each SR to ensure orthogonality while preserving the same kinematic features.

To estimate the $\PW(\Pell\PGn)$+jets contribution, a single ``\Wjets CR'' is defined as a sideband in the \ParT  discriminant, $P(\PH_{1\Pell})\in[0.75,0.90]$.
In the combined fit, this CR simultaneously constrains the $\PW(\Pell\PGn)$+jets background in the ggF, VBF, and \VH SRs, yielding a normalization SF of $0.89^{+0.13}_{-0.12}$.
The correlated treatment is motivated by the kinematic similarity between the CR and SRs. Fitting each SR independently yields stable SFs and negligible impact on the result, indicating that acceptance differences among these regions are negligible. 
In addition, theoretical uncertainties on the \Wjets process are treated as uncorrelated across the three SRs, as described in Section~\ref{Sec:uncertainties}, providing further freedom to absorb residual per-SR effects.

The \ttbar contribution rate is estimated using two CRs: the ``ggF/VBF Top CR'' for the ggF and VBF categories collectively, and the ``\VH Top CR'' specific to the \VH category.
These two CRs select distinct regions of phase space to account for possible mismodeling of the selection efficiency $P(\PV)$, which is applied only in the \VH category, allowing independent normalization in the fit. 
The ggF/VBF Top CR requires $P(\PH_{1\Pell})>0.90$ and at least one \PQb-tagged jet at the tight WP.
The \VH Top CR requires $P(\PH_{1\Pell})>0.75$, at least one \PQb-tagged jet at the medium WP, and a \PV-tagged AK8 jet.
In the combined fit of all channels, the \ttbar normalization scale factors are $0.97^{+0.17}_{-0.14}$ for the ggF and VBF categories, and $2.01^{+0.38}_{-0.31}$ for the \VH category.
It has been verified that the larger SF arises from mismodeled selection efficiency specific to the requirement of two jets tagged for a signature that includes a hadronic \PV boson decay, which is common to the \PNet and \ParT tagging requirements.  
In the \ttbar events passing the $1\Pell$ selection, there is only 
one hadronic \PW boson decay. Therefore, passing all tagging 
requirements relies on another jet in the event, with extra radiation 
mimicking a two-prong structure; a topology not captured by the 
tagging efficiency scale factors.
This situation is common to \VH top CR and SR for \ttbar events, but does not affect the signal, which has two genuine hadronic \PV boson decays.

We studied the potential mismodeling of the \ttbar acceptance between the inclusive ggF/VBF Top CR and the VBF SR, where the latter probes a different kinematic phase space requiring two forward AK4 jets. Comparing the data-to-simulation agreement in the Top CR under its nominal selection and with additional VBF-like forward-jet requirements, accounting for the statistical limitations of the VBF category, yields a rate difference of $20\%$. Assigning this as an additional rate uncertainty exclusively to the \ttbar normalization in the VBF SR was found to have a negligible impact on the result.

\section{Systematic uncertainties}
\label{Sec:uncertainties}

We consider several sources of systematic uncertainties of experimental and theoretical nature, which are summarized in Table~\ref{tab:SysUnc_com}.
They are included as nuisance parameters in the signal extraction procedure treated according to the frequentist paradigm~\cite{CMS-NOTE-2011-005}.

\begin{table}[!t]
  \topcaption{Systematic uncertainty sources considered in the analysis.
    Left to right columns: the sources, the channels, whether the uncertainty affects signal (S) or background (B), its influence on shape (s) or rate (r), and whether the nuisances are (un)correlated (u or $\checkmark$) among different process models (P) or among the data-taking years (Y).}
  \centering
\begin{tabular}{c l c c c c c}
      & & Channel & \multicolumn{2}{c}{Effect} & \multicolumn{2}{c}{Corr.} \\
      & {Source of uncertainty} &  & \begin{tabular}[c]{@{}c@{}}S and/or B\end{tabular} & \begin{tabular}[c]{@{}c@{}}s or r\end{tabular} & P & Y \\
      \hline
      \multirow{23}{*}{\rotatebox{90}{Experimental}} & Integrated luminosity & $0\Pell\,\&\,1\Pell$ & S \& B & r & $\checkmark$ & $\checkmark$ \\
      & Pileup                                         & $0\Pell\,\&\,1\Pell$ & S \& B & s & $\checkmark$ & u \\
      & Trigger                                        & $0\Pell\,\&\,1\Pell$ & S \& B & r & $\checkmark$ & $\checkmark$ \\
      & Jet energy scale                               & $0\Pell\,\&\,1\Pell$ & S \& B & s & $\checkmark$ & $\checkmark$ \\
      & Jet energy resolution                          & $0\Pell\,\&\,1\Pell$ & S \& B & s & $\checkmark$ & u \\
      & Jet mass scale \& resolution                   & $0\Pell\,\&\,1\Pell$ & S \& B & s & $\checkmark$ & u \\
      & Prefiring                                      & $0\Pell\,\&\,1\Pell$ & S \& B & s & $\checkmark$ & u \\
      & Unclustered energy                             & $0\Pell\,\&\,1\Pell$ & S \& B & s & $\checkmark$ & $\checkmark$ \\
      & Simulated sample size                         & $0\Pell\,\&\,1\Pell$ & S \& B & s & u & u \\
      & \ParT tagging efficiency                       & $0\Pell\,\&\,1\Pell$ & S & r & u & $\checkmark$ \\
      & {\PGm}/{\Pe} rejection                            & $0\Pell$ & S & s & \NA & $\checkmark$ \\
      & QCD background estimate TFs                    & $0\Pell$ & B & s & \NA & $\checkmark$ \\
      & QCD estimate statistical uncertainty            & $0\Pell$ & B & s & \NA & u \\
      & Top quark \& \wjets bkg. normalization         & $0\Pell$ & B & r & \NA & $\checkmark$ \\
      & \PNet tagging efficiency                       & $1\Pell$ & S & r & u & $\checkmark$ \\
      & $\PQb$ tagging efficiency                      & $1\Pell$ & S \& B & r & $\checkmark$ & $\checkmark$ \\
      & {\PGm}/{\Pe} reconstruction, \& isolation & $1\Pell$ & S \& B & r & $\checkmark$ & $\checkmark$ \\
      & \ttbar \& single \PQt floating normalization   & $1\Pell$ & B & r & $\checkmark$ & $\checkmark$ \\
      & \Wjets floating normalization                  & $1\Pell$ & B & r & $\checkmark$ & $\checkmark$ \\
      & Misid. $\Pell$ rate, stat. uncertainty             & $1\Pell$ & B & s & \NA & u \\
      & Misid. $\Pell$ rate, EW SF stat. uncertainty       & $1\Pell$ & B & s & $\checkmark$ & $\checkmark$ \\
      & Misid. $\Pell$ rate, SF flavor \& normalization            & $1\Pell$ & B & r & $\checkmark$ & $\checkmark$ \\[\cmsTabSkip]
      \multirow{8}{*}{\rotatebox{90}{Theoretical}} & Branching fraction $\mathcal{B}(\PH\to\PW\PW)$ & $0\Pell\,\&\,1\Pell$ & S & r & $\checkmark$ & $\checkmark$ \\
      & $\alpS$                                       & $0\Pell\,\&\,1\Pell$ & S \& B & r & $\checkmark$ & $\checkmark$ \\
      & Parton shower model                           & $0\Pell\,\&\,1\Pell$ & S \& B & s & u & $\checkmark$ \\
      & Ren. \& fact. scales, rate \& acceptance & $0\Pell\,\&\,1\Pell$ & S \& B & s & u & $\checkmark$ \\
      & PDF                                           & $0\Pell\,\&\,1\Pell$ & S \& B & s & u & $\checkmark$ \\
      & PDF acceptance                                & $0\Pell\,\&\,1\Pell$ & S \& B & s & u & $\checkmark$ \\
      & QCD LO/NLO correction \PV{}+jets              & $1\Pell$ & B & s & u & $\checkmark$ \\
      & EW NLO correction \PV{}+jets                  & $1\Pell$ & B & r & u & $\checkmark$ \\
    \end{tabular}
  \label{tab:SysUnc_com}
\end{table}

Common experimental uncertainties between channels, affecting all simulation samples (both signal and background), include the uncertainty in the integrated luminosity, which varies between 1.2 and 2.5\% for individual data-taking periods and amounts to 1.6\% overall~\cite{CMS:2021xjt,CMS-PAS-LUM-17-004,CMS-PAS-LUM-18-002}, and the uncertainty in the modeling of pileup interactions, evaluated by varying the total pp inelastic cross section by $\pm4.5\%$~\cite{CMS:2018mlc}.
The jet energy scale and resolution corrections, as well as those of unclustered particles, are varied within their uncertainty as a function of jet \pt and $\eta$, and are propagated to the acceptance and mass observables of SRs.
These uncertainties range from 0.5 to 8.0\%.

Uncertainties in the trigger efficiency for the combination of single-lepton triggers, and the efficiencies for lepton reconstruction, identification, and isolation are evaluated as functions of lepton \pt and $\eta$.
The overall magnitude of these uncertainties is 0.5\% for the trigger efficiency uncertainty and 3--5\% for uncertainties in the reconstruction and identification efficiency for \PGm and \Pe.

The jet mass scale and resolution are corrected using data-to-simulation SFs close to unity (0.98--1.00 and 0.80--1.20, respectively), indicating good agreement between data and simulation~\cite{CMS:2025kje}. 
The corresponding uncertainties are propagated to the analysis observables, \mjs and $m_\text{j}^\PV$, for signal processes and background \PW boson matched jets, ranging from 0.5 to 7.0\%.

During 2016--2017, mistiming in the ECAL endcaps (${2.5<\abs{\eta}<3.0}$) caused partial trigger efficiency loss (``prefiring'')~\cite{CMS:2020cmk}, corrected in simulation using control samples in data. The associated shape and rate uncertainties were found to have a minor impact but are included for all simulation-based samples.

The $\hwwall$ efficiency corrections for $P(\PH_{0\Pell})$ and $P(\PH_{1\Pell})$ are applied exclusively to signal processes, as only these processes contain such a genuine \PH-candidate jet.
Uncertainties from the LJP reweighting method are propagated to these corrections and have a relative impact ranging from 5 to 15\%.
Three main sources of uncertainty are considered: (1) statistical, due to the limited event count used to derive the data-to-simulation ratio of LJP densities; (2) systematic, from modeling effects impacting the simulation in the ratio's denominator; and (3) extrapolation, related to applying the correction to subjets with higher \pt than those used in the control sample.
The dominant uncertainty stems from the LJP method's assumption that jets can be reclustered into a prescribed number of subjets, each matching a quark.
This uncertainty is decomposed into three components: (3a) ``number of prongs,'' corresponding to the ambiguity in choosing the number of subjets and assessed by varying the number of subjets used in reclustering; (3b) ``unclustered,'' corresponding to the presence of a generator-level quark failing to match any subjet even after varying the number of subjets, evaluated by scaling the weights for such events by factors of 5 and 1/5; and (3c) ``distortion,'' arising from limitations in the reclustering procedure, such as misassignment of PF candidates to subjets, which can distort the reconstructed splitting tree. The latter component, which is the dominant one, is estimated by comparing LJP densities from the original \Wjets sample and simulated signal, with their ratio providing the corresponding uncertainty. Details are given in Ref.~\cite{CMS:2025eyd}.

The LJP-based correction to the signal efficiency for the $P(\PH_{0\Pell})$ discriminant, combining \Hlqq, \Hthreeq, and \Hfourq final states, is $0.84^{+0.14}_{-0.22}$ for $P(\PH_{0\Pell})>0.99$ and $0.95^{+0.13}_{-0.15}$ for $0.92<P(\PH_{0\Pell})<0.99$. 
For the $P(\PH_{1\Pell})$ discriminant, considering the \Hlqq final state, the correction is $0.98^{+0.10}_{-0.24}$ for $P(\PH_{1\Pell})>0.905$ and $0.95^{+0.10}_{-0.28}$ for  $P(\PH_{1\Pell})>0.93$. 
Unclustered quarks dominate the uncertainty, with rate effects ranging from 11 to 16\% for $P(\PH_{0\Pell})$ and from 9 to 27\% for $P(\PH_{1\Pell})$.

In the \VH regions, the signal and top quark processes are assigned \PV-candidate jet tagging efficiency uncertainties of 3--7\%, determined from calibration in semileptonic \ttbar events for \PW boson decays~\cite{CMS-DP-2025-010} and from the QCD proxy method for \PZ boson decays~\cite{CMS:2025kje}, respectively.

There are also experimental uncertainties in the estimation of the nonprompt-lepton background in the $1\Pell$ channel.
This background is affected by the statistical uncertainty in the data sample used to derive the misidentified-lepton factors and by the uncertainty in the estimate of the prompt lepton contamination from EW processes that is subtracted from that sample.
An overall 25\% normalization uncertainty is assigned to the nonprompt-lepton background template based on a closure test in a QCD multijet-enriched sample, accounting for systematic uncertainties in the $\epsilon_{\text{misID}}$ determination.

The \wjetslnu and \ttbar background yields are determined by floating normalization parameters constrained by the corresponding CRs as detailed in Section~\ref{subSec:1lep_prediction_wjets}.
The resulting uncertainties in the fitted normalizations, corresponding to the statistical uncertainties, are about 15\% for \wjetslnu and 20\% for \ttbar.

In addition to experimental uncertainties, theoretical uncertainties are included in the final fit to account for inaccuracies in the modeling of SM processes.
Uncertainties in the $\PV(\Pq\Pq)$+jets processes account for missing higher-order QCD and mixed QCD-EW effects beyond those described in Section~\ref{Sec:Sample}, following the prescription of Ref.~\cite{Lindert_2017}.
The QCD NLO correction is implemented as a shape nuisance, propagating to the fit observables through the boson \pt dependence, while the mixed QCD-EW correction is implemented as a rate nuisance.
The uncertainties due to the renormalization and factorization scales chosen for the simulated Higgs boson samples are propagated to the total expected yield of the Higgs boson signal according to the prescription recommended in Ref.~\cite{becker2021precise}.

Uncertainties in the event yields due to initial- and final-state radiation are also calculated for all Higgs boson production processes by varying the renormalization scale and non-singular term using the \PYTHIA8 showering algorithm~\cite{Sjostrand:2014zea}.
The PDFs and $\alpS$ uncertainties are further split between the cross section normalization uncertainties computed in Ref.~\cite{deFlorian:2016spz} for the Higgs boson signal and their effect on the acceptance.
These uncertainties are neglected for other processes estimated from data.

Finally, the uncertainties due to the limited number of events in the simulated samples are included independently for each mass distribution bin in each category. 
These uncertainties are modeled with a single nuisance parameter per bin, following the Barlow--Beeston procedure~\cite{Barlow:1993dm} using the simplifying approximation from Ref.~\cite{Conway:2011in}.

The dominant systematic uncertainty in the $0\Pell$ channel arises from the modeling of the QCD multijet background through the fitted coefficients of the transfer function.
In the $1\Pell$ channel, the dominant uncertainty arises from the asymmetric uncertainty in the \ParT tagger efficiency and the size of simulated signal samples in the VBF category.

\section{Results}
\label{Sec:Results}

The signal is extracted from a binned maximum likelihood fit to the \mjs (or $m_\text{j}^\PV$) distribution using the sum of the signal and background contributions.
In the $0\Pell$ channel, the \mjs distribution spans the 50--250\GeV range with 10\GeV binning.
In the $1\Pell$ channel, it covers the 75--235\GeV range with 20\GeV binning.
In the \VH channel, the \PV-candidate jet soft-drop mass covers 40--180\GeV with a variable bin width.
The test statistic chosen to determine the signal yield and the associated confidence intervals is based on the profile likelihood ratio~\cite{CMS-NOTE-2011-005},
\begin{equation}
t_{\mu} = -2\ln\left(\frac{\Likelihood(\mu,\hat{\hat{\vtheta}}({\mu}))}{\Likelihood(\hat{\mu},\hat{\vtheta})}\right),
\end{equation}
where $\Likelihood$ denotes the combined likelihood function, and $\mu$ is the signal strength, defined as the ratio of the measured cross section to the SM expectation, taken as the parameter of interest.
The $\hat{\mu}$ is its best fit value, while $\hat{\hat{\vtheta}}({\mu})$ and $\hat{\vtheta}$ represent the conditional and global maximum likelihood estimators of the nuisance parameters, respectively.
The following results have been determined using \textsc{Combine} tool~\cite{CMS:2024onh}, which is based on the \textsc{roofit}~\cite{roofit} and \textsc{roostats}~\cite{roostats} frameworks.

\begin{figure}\centering
  \includegraphics[width=0.49\linewidth]{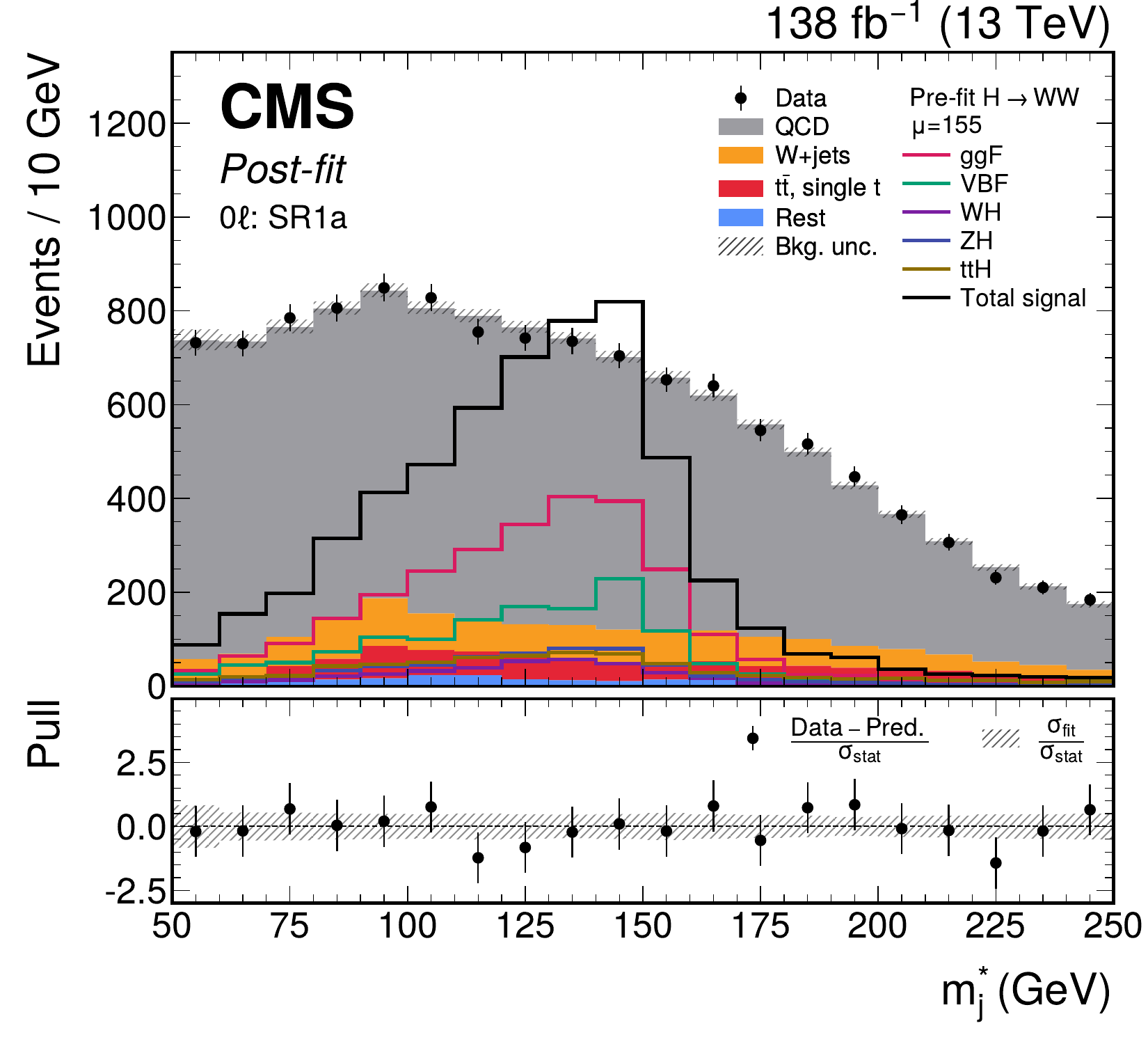}
  \includegraphics[width=0.49 \linewidth]{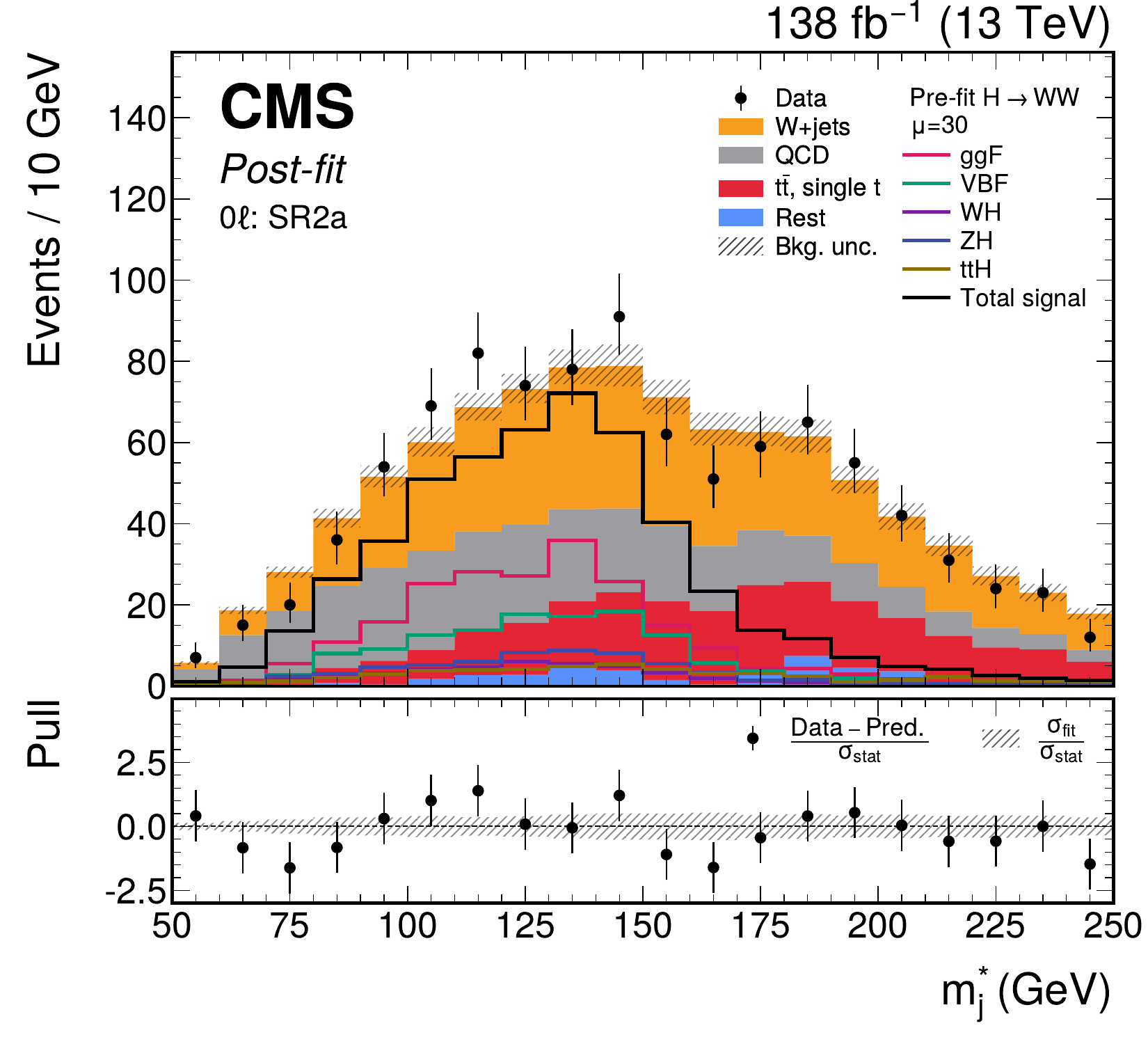}
  \includegraphics[width=0.49\linewidth]{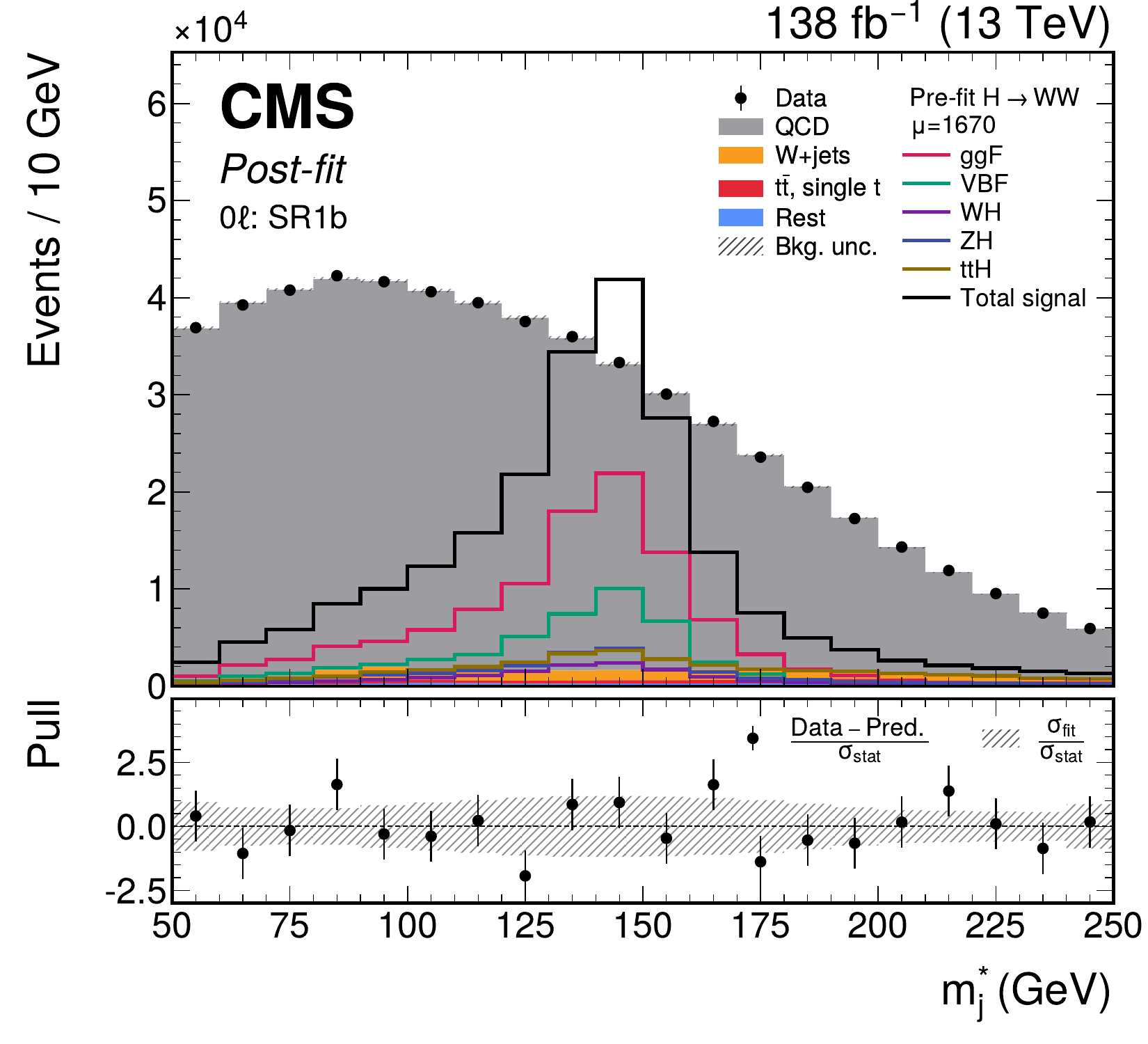}
  \includegraphics[width=0.49 \linewidth]{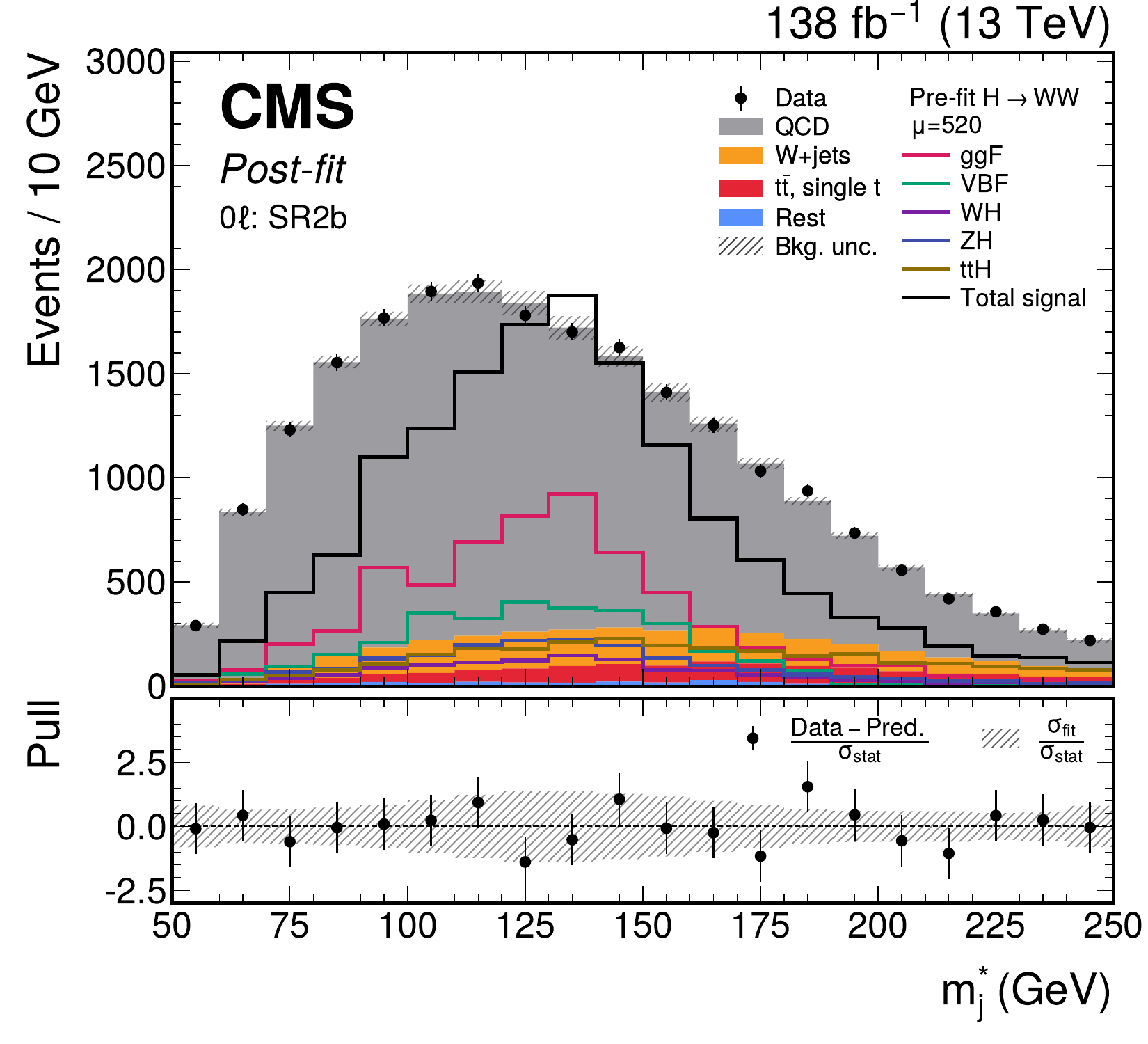}
  \caption{
Post-fit \mjs distributions in the $0\Pell$ channel, showing the predicted background with total uncertainty, observed data, and the expected pre-fit signal scaled by the labeled strength $\mu$.
From left to right, upper to lower, the plots correspond to \SRonea, \SRtwoa, \SRoneb, and \SRtwob.
The lower panel of each plot presents the pull distribution, as well as the $\sigma_\text{fit}$ normalized to the $\sigma_\text{stat}$. }
  \label{fig:0l_postfit}
\end{figure}

\begin{figure}[htbp]  \centering
\includegraphics[width=0.47\textwidth]{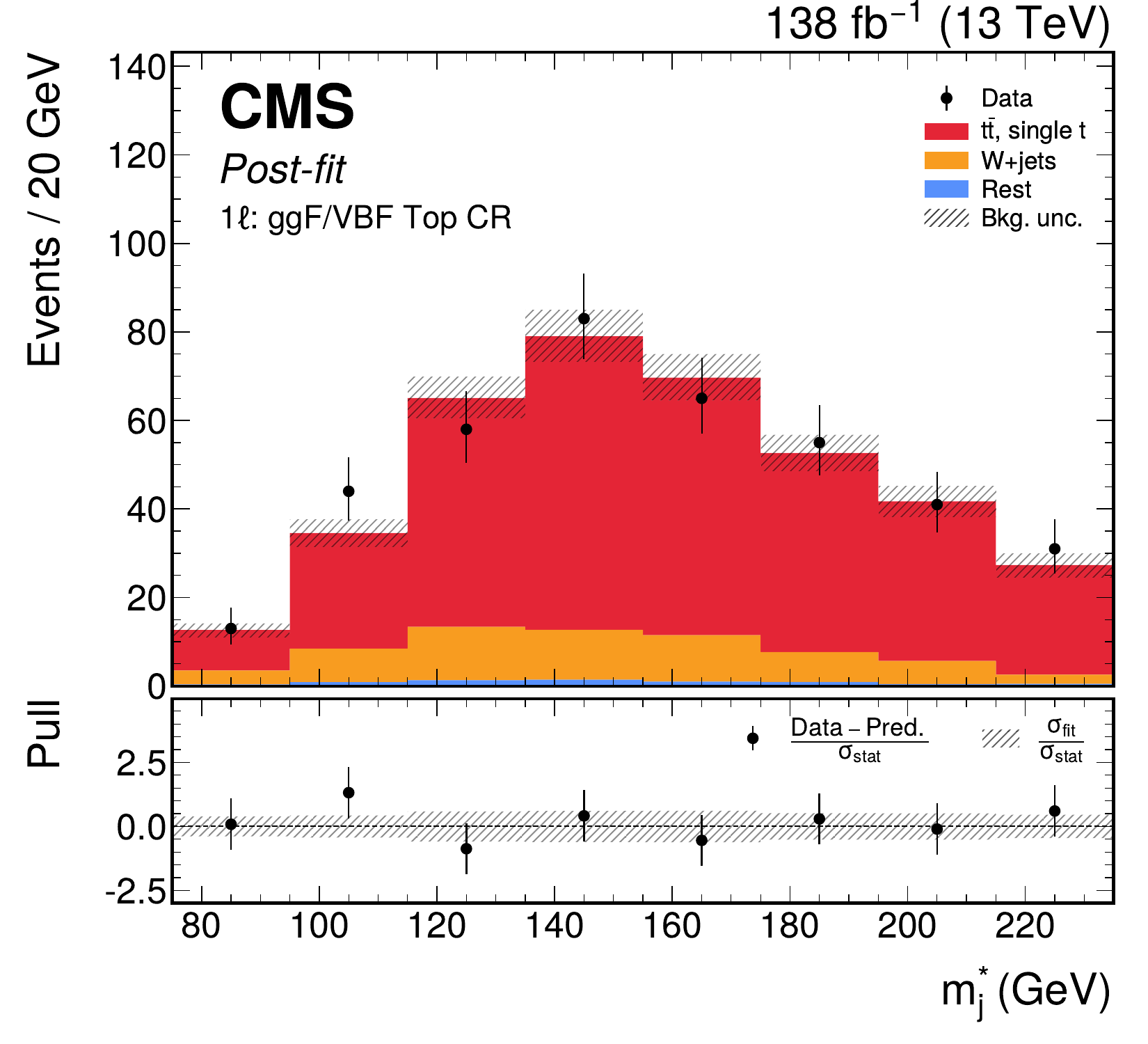} 
\includegraphics[width=0.47\textwidth]{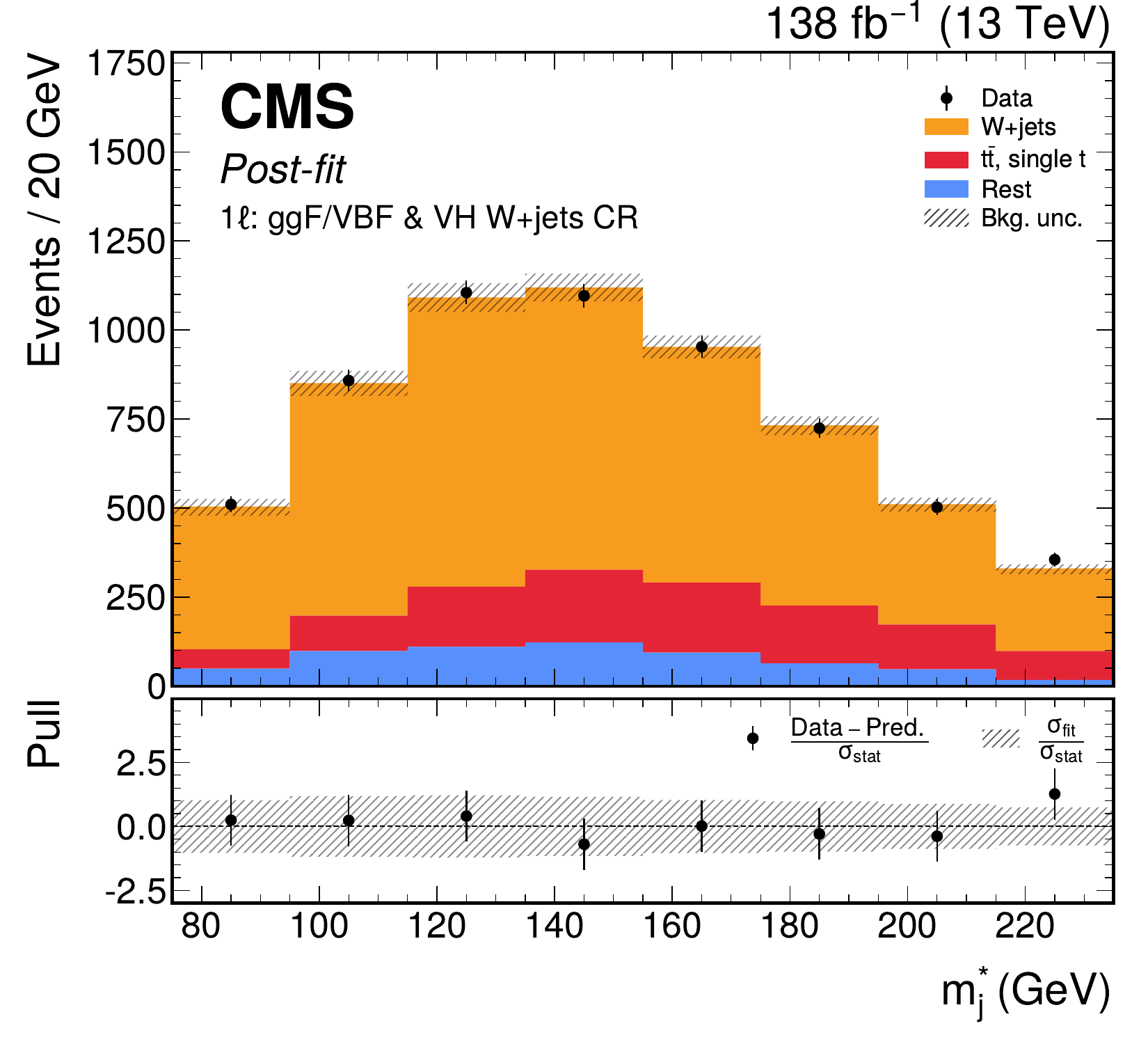}
\includegraphics[width=0.47\textwidth]{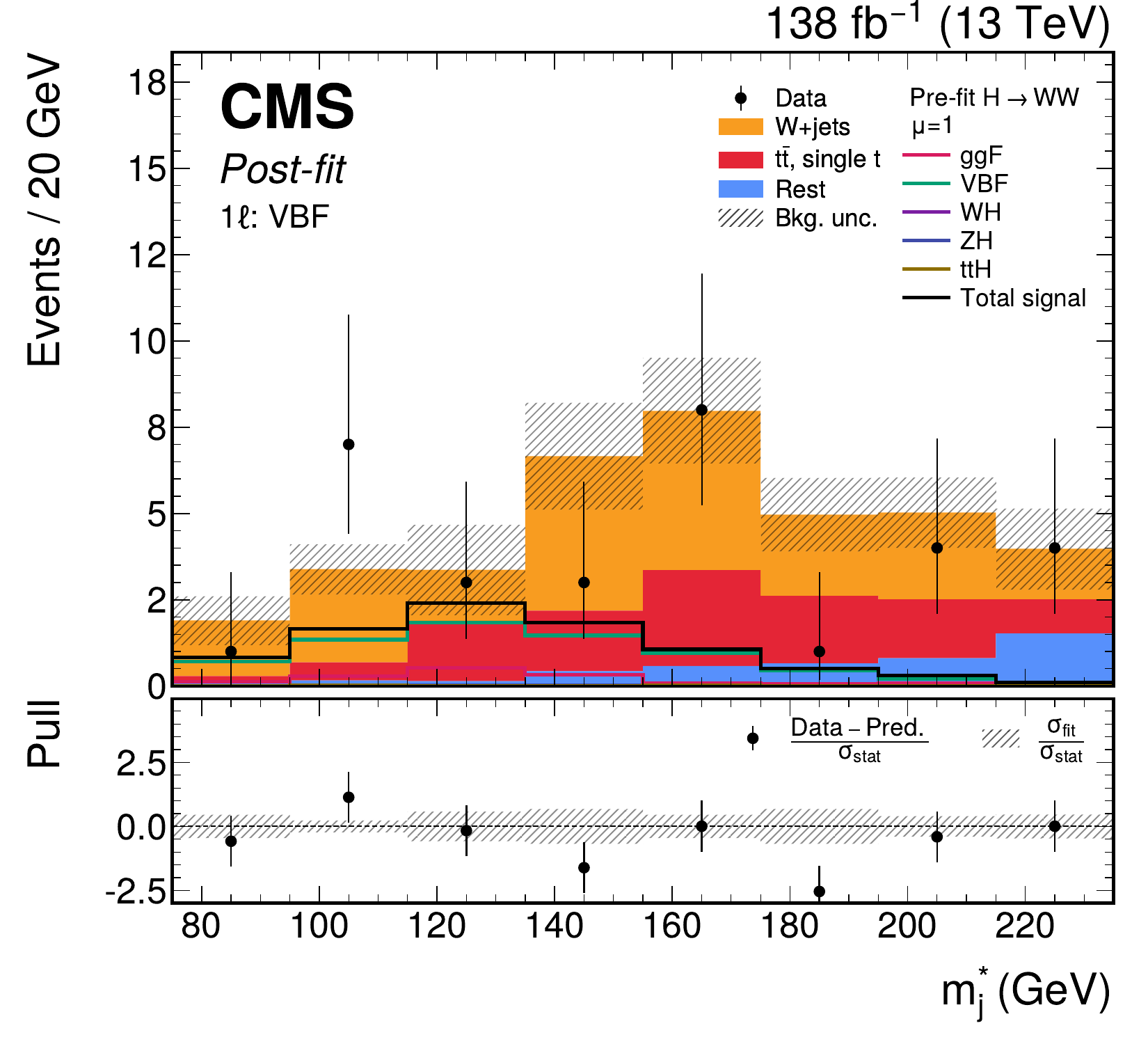} 
\includegraphics[width=0.47\textwidth]{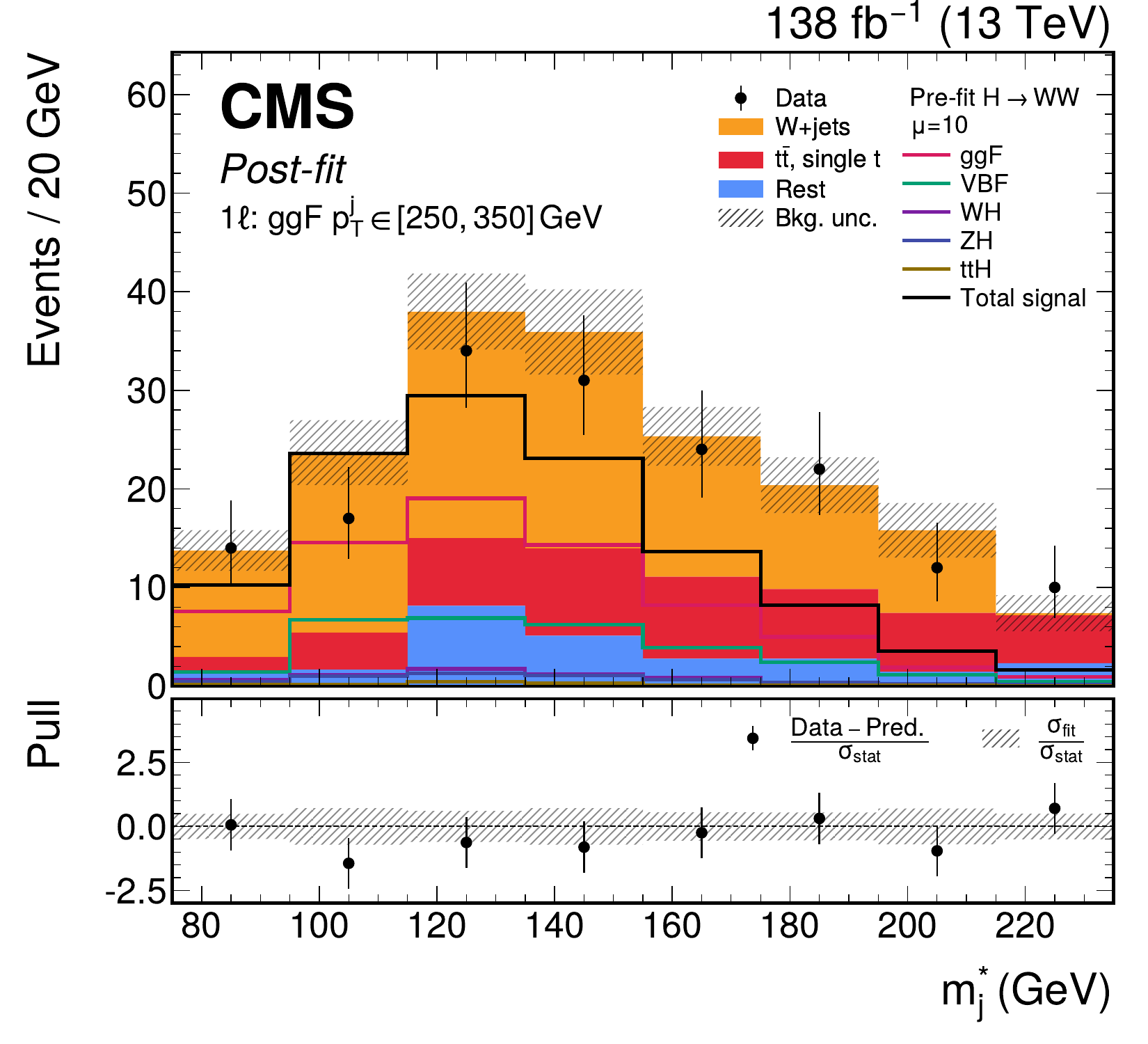}
\includegraphics[width=0.47\textwidth]{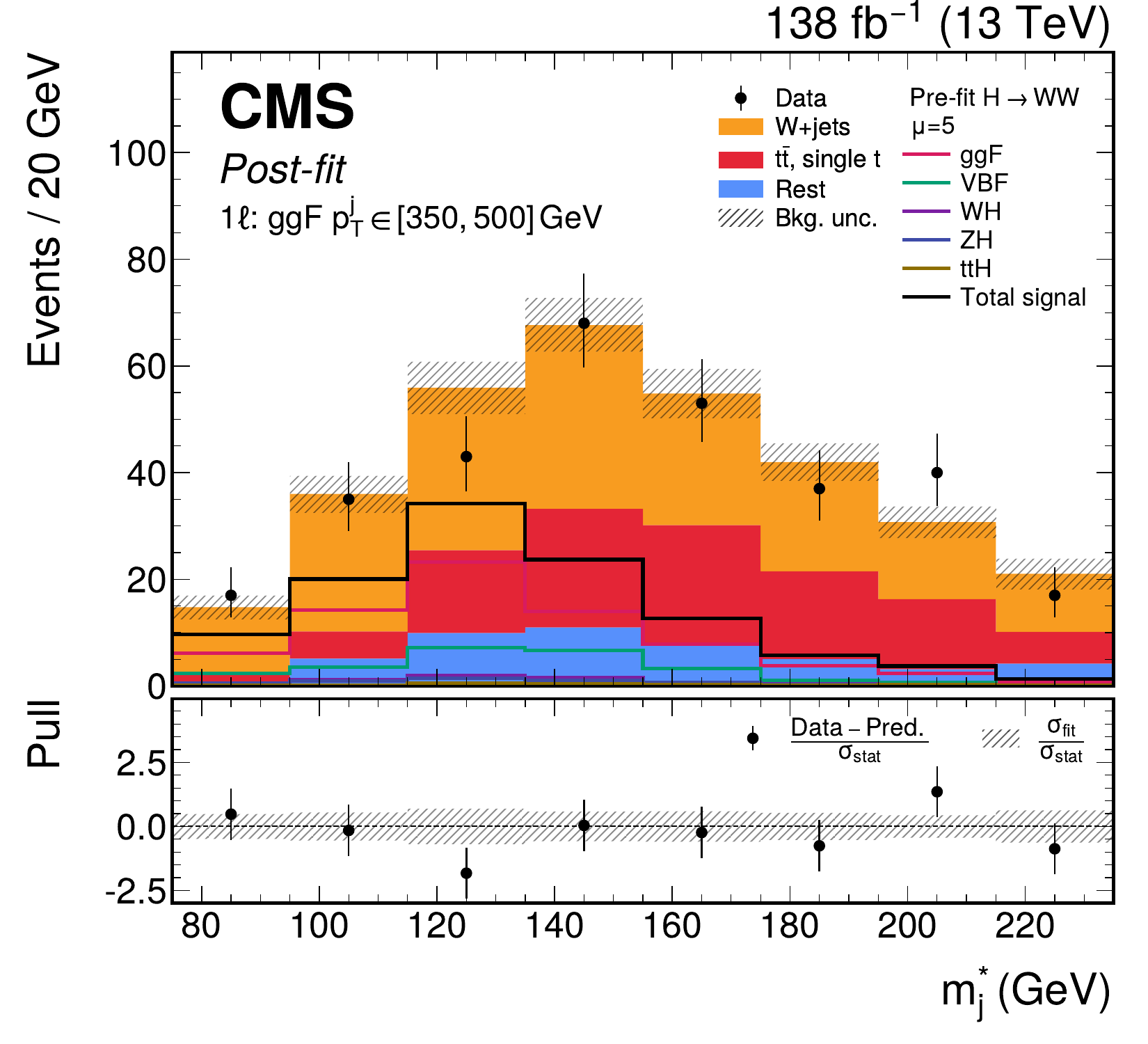}
\includegraphics[width=0.47\textwidth]{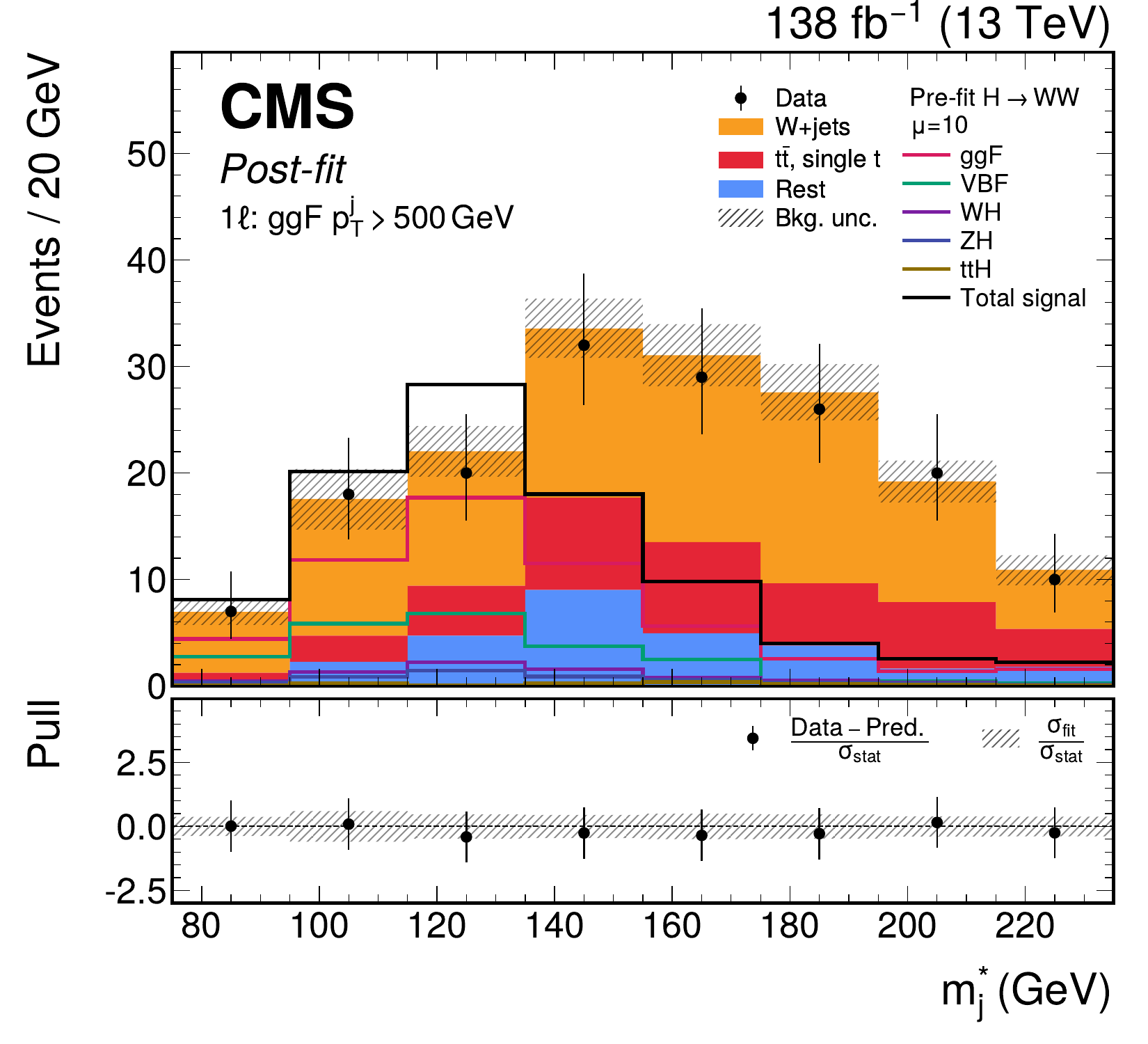} 
\caption{
Post-fit \mjs distributions in the $1\Pell$ channel, showing the predicted background with total uncertainty, observed data, and the expected pre-fit signal scaled by the labeled strength $\mu$.
Left to right and upper to lower: Top CR, \Wjets CR, VBF SR, and the ggF SRs binned in \pt as $[250,350)$, $[350,500)$, and $[500,+\infty)\GeV$, respectively.
The lower panel of each plot presents the pull distribution, as well as $\sigma_\text{fit}$ normalized to the $\sigma_\text{stat}$.}
\label{fig:1lep_postfit}
\end{figure}

\begin{figure}[htbp]  \centering
\includegraphics[width=0.495\textwidth]{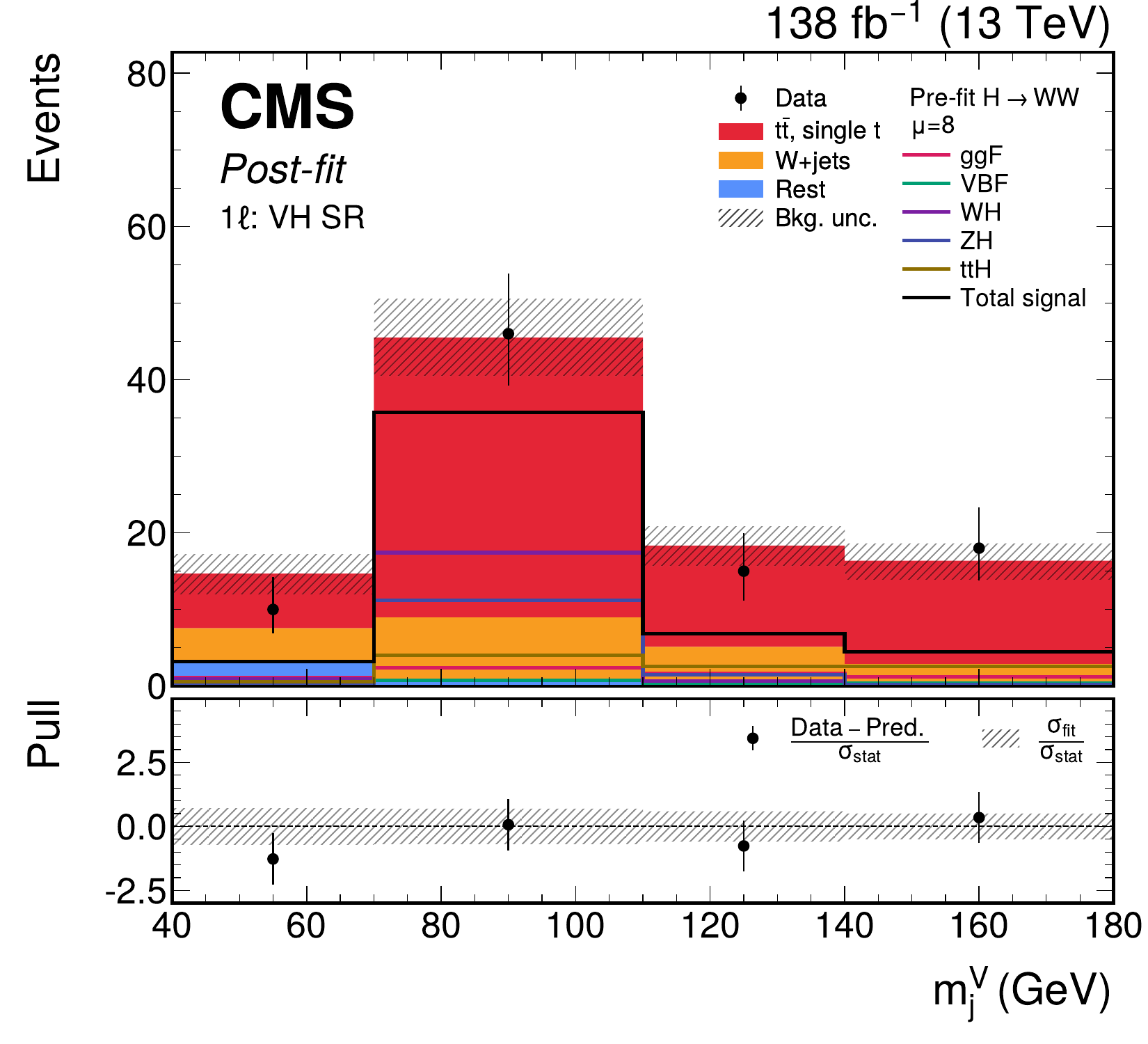}
\includegraphics[width=0.495\textwidth]{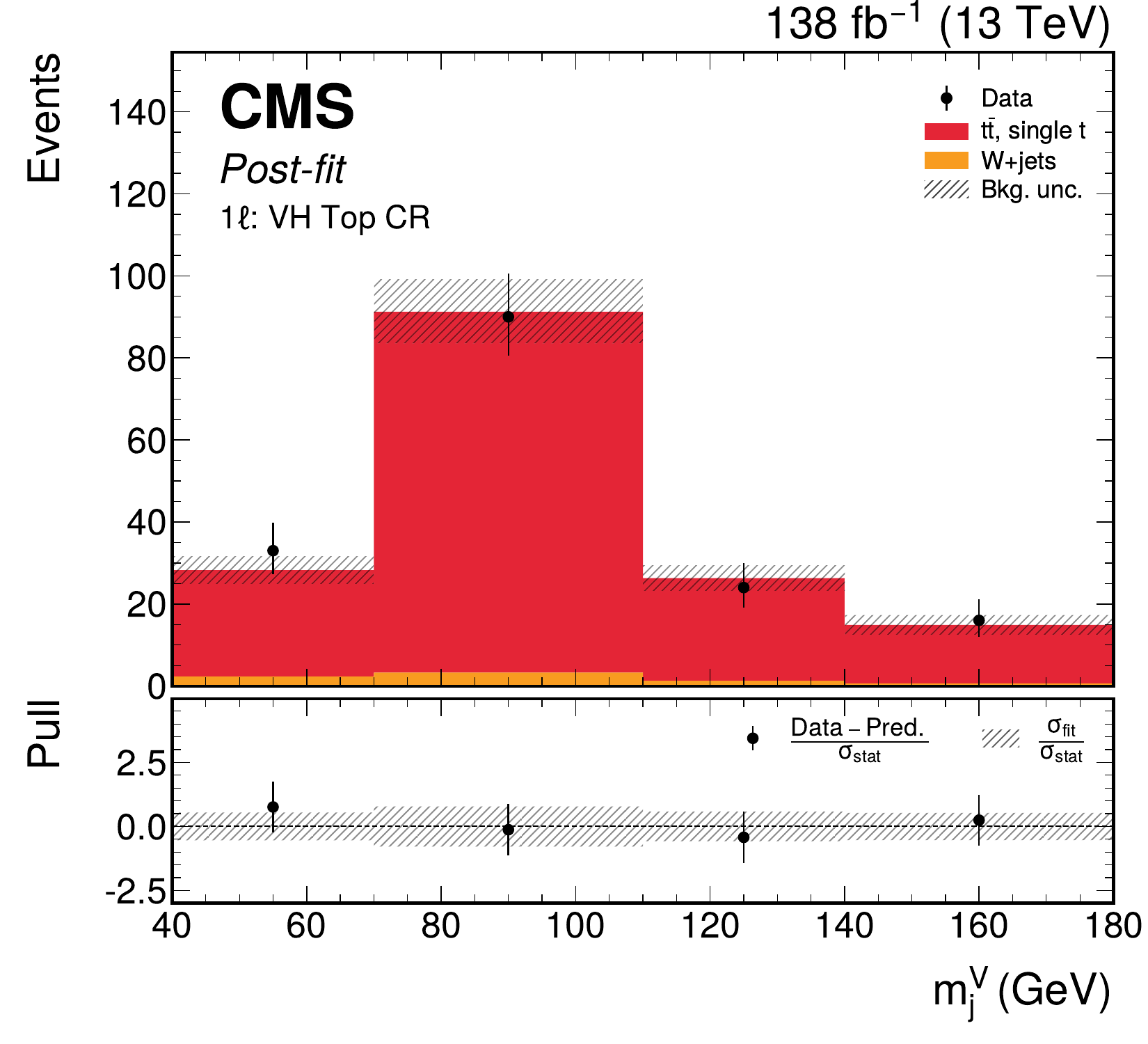}
\caption{
Post-fit $m_\text{j}^\PV$ distributions in the \VH channel, showing the predicted background with total uncertainty, observed data, and expected signal, split by production process.
Left to right: \VH SR and \VH Top CR.
The lower panel of each plot presents the pull distribution, as well as $\sigma_\text{fit}$ normalized to the $\sigma_\text{stat}$.
The predicted pre-fit signal is scaled for visibility.
}
\label{fig:vh_postfit}
\end{figure}

Separate fits are first performed for each production process and final state: the six regions of the $0\Pell$ channel (two CRs and four SRs) and the eight regions in the $1\Pell$ channel (three CRs and five SRs), to extract individual results.
A simultaneous fit of all the regions is then performed to obtain the combined result.
The observed data, post-fit background, and pre-fit signal distributions are shown in Figs.~\ref{fig:0l_postfit}, \ref{fig:1lep_postfit}, and \ref{fig:vh_postfit}. These correspond, respectively, to the $0\Pell$ SRs, the $1\Pell$ (ggF and VBF) SRs and CRs, and the $1\Pell$ \VH SR and Top CR.
The lower panels display the pull distributions, defined as the difference between the observed and post-fit predicted background event yields divided by the statistical uncertainty in the data, $\sigma_{\text{stat}}$. 
The total fit uncertainty $\sigma_\text{fit}$ (also normalized to $\sigma_\text{stat}$) is overlaid.
The data and fitted distributions are summed for all data-taking periods.
Each signal production process is shown together with its sum, using pre-fit predictions scaled to the signal strength indicated in the legend, to allow direct comparison with the background.

\begin{figure}[!htbp]
  \centering
  \includegraphics[width=0.55\linewidth]{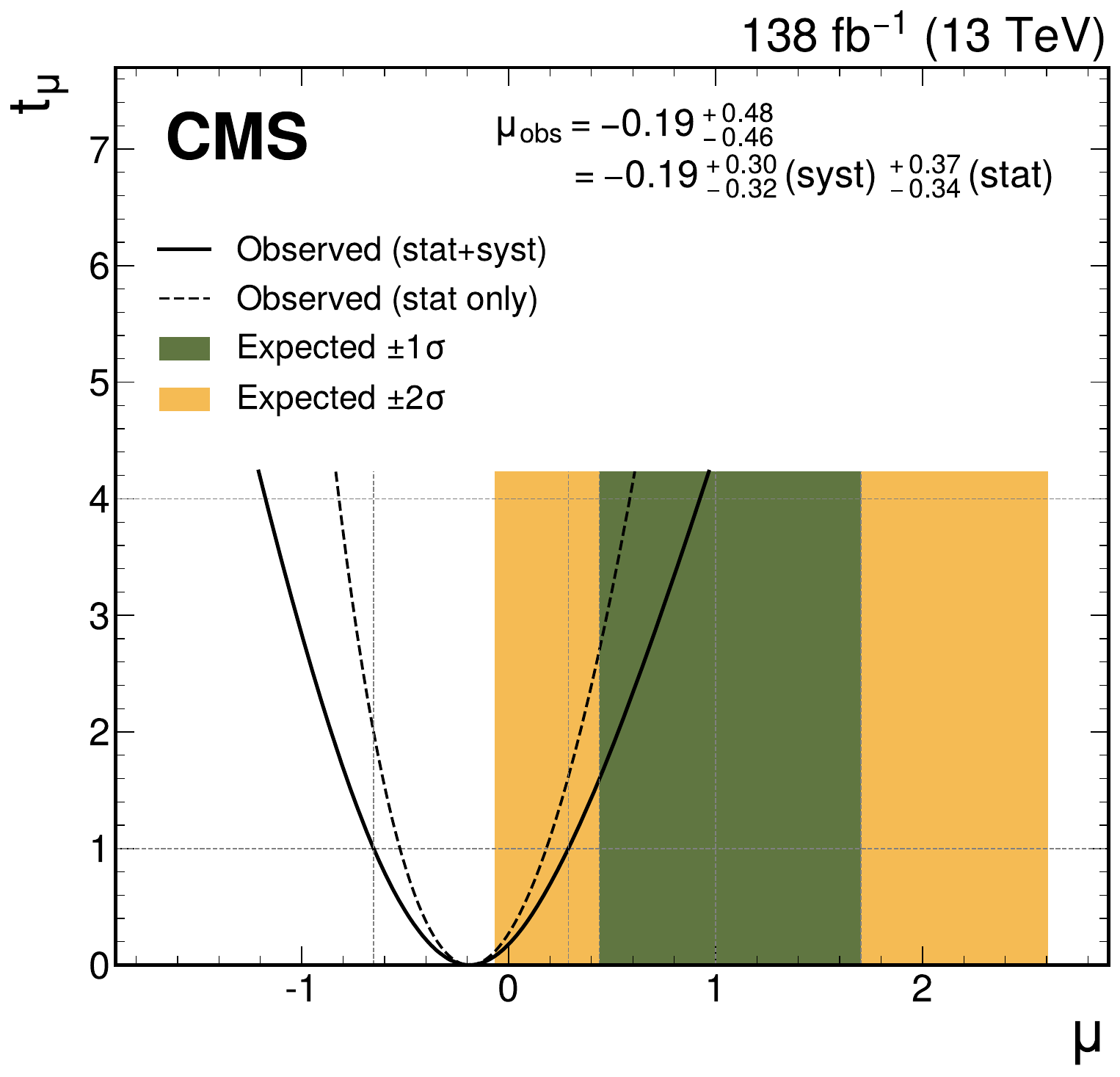}
  \caption{
    Observed scan of the profile likelihood test statistic $t_\mu$ as a function of the signal strength $\mu$ for the combination of all the channels.
    The solid lines correspond to profiling all statistical and systematic uncertainties, while the dashed lines correspond to profiling only the statistical uncertainties.}
  \label{fig:nll_scan}
\end{figure}

The best fit value of the signal strength $\mu$, defined as the measured cross section times the \hwwall branching fraction relative to the SM expectation, and a 68\% confidence level (\CL) interval are extracted as $\mu = -0.19^{+0.48}_{-0.46}$ following Ref.~\cite{CMS:2014fzn}.
Figure~\ref{fig:nll_scan} shows the observed scan of the test statistic $t_\mu$ as a function of $\mu$ for the combination of all channels. The result is decomposed into statistical and systematic uncertainties, yielding $\mu = -0.19^{+0.48}_{-0.46} = -0.19^{+0.30}_{-0.32}\syst^{+0.37}_{-0.34}\stat$.

To estimate the signal significance, we use the test statistic $q_0$ for the discovery of a positive signal,
\begin{equation}
  q_{0} =  \begin{cases} -2\ln\left(\frac{\Likelihood(0,\hat{\hat{\vtheta}}({0}))}{\Likelihood(\hat{\mu},\hat{\vtheta})}\right) & \text{if } \hat{\mu}>0 \\
              0                                                                                                      & \text{otherwise}.
  \end{cases}
\end{equation}
The expected and observed significances of the signal are estimated using the asymptotic formulae, $Z_0=\sqrt{q_0}$~\cite{Cowan:2010js}. The expected significance is 1.86$\sigma$, while the observed significance is found to be 0.0$\sigma$ as the best fit value of the signal strength is negative. The observed result lies below the SM expectation, $\mu=1$, by 2.1 standard deviations when allowing for negative signal strength based on the uncapped test statistic $t_0$~\cite{Cowan:2010js}.

The results are summarized in Table~\ref{tab:results}. The signal strength scales all the signal processes by the same factor.

\begin{table}[!htbp]
  \centering
  \topcaption{Observed and expected signal strength $\mu$ (second column) and significance $\sigma$ (third column) for $\hwwall$ in the $0\Pell$ and $1\Pell$ channels, followed by the combined results.}
  \renewcommand{\arraystretch}{1.3}
  \begin{tabular}{c cc cc}
    & \multicolumn{2}{c}{Signal strength $\mu$}
    & \multicolumn{2}{c}{Significance $\sigma$} \\
    Channel  & Observed & Expected & Observed & Expected \\ \hline
    \rule{0pt}{.5ex}$0\Pell$ (Inclusive) & $+3.61^{+3.11}_{-2.73}$ & $+1.00^{+2.86}_{-2.79}$ & 1.32 & 0.36 \\
    $1\Pell$ (ggF\,+\,VBF) & $-0.30^{+0.49}_{-0.50}$ & $+1.00^{+0.80}_{-0.62}$ & 0.00 & 1.68 \\
    \rule{0pt}{1ex}$1\Pell$ \VH & $+0.03^{+1.92}_{-1.77}$ & $+1.00^{+1.75}_{-1.52}$ & 0.01 & 0.65 \\
    \rule{0pt}{1ex}Combination & $-0.19^{+0.48}_{-0.46}$ & $+1.00^{+0.70}_{-0.56}$ & 0.00 & 1.86 \\
  \end{tabular}
  \label{tab:results}
\end{table}

\begin{figure}[!htbp]  \centering
\includegraphics[width=1.\linewidth]{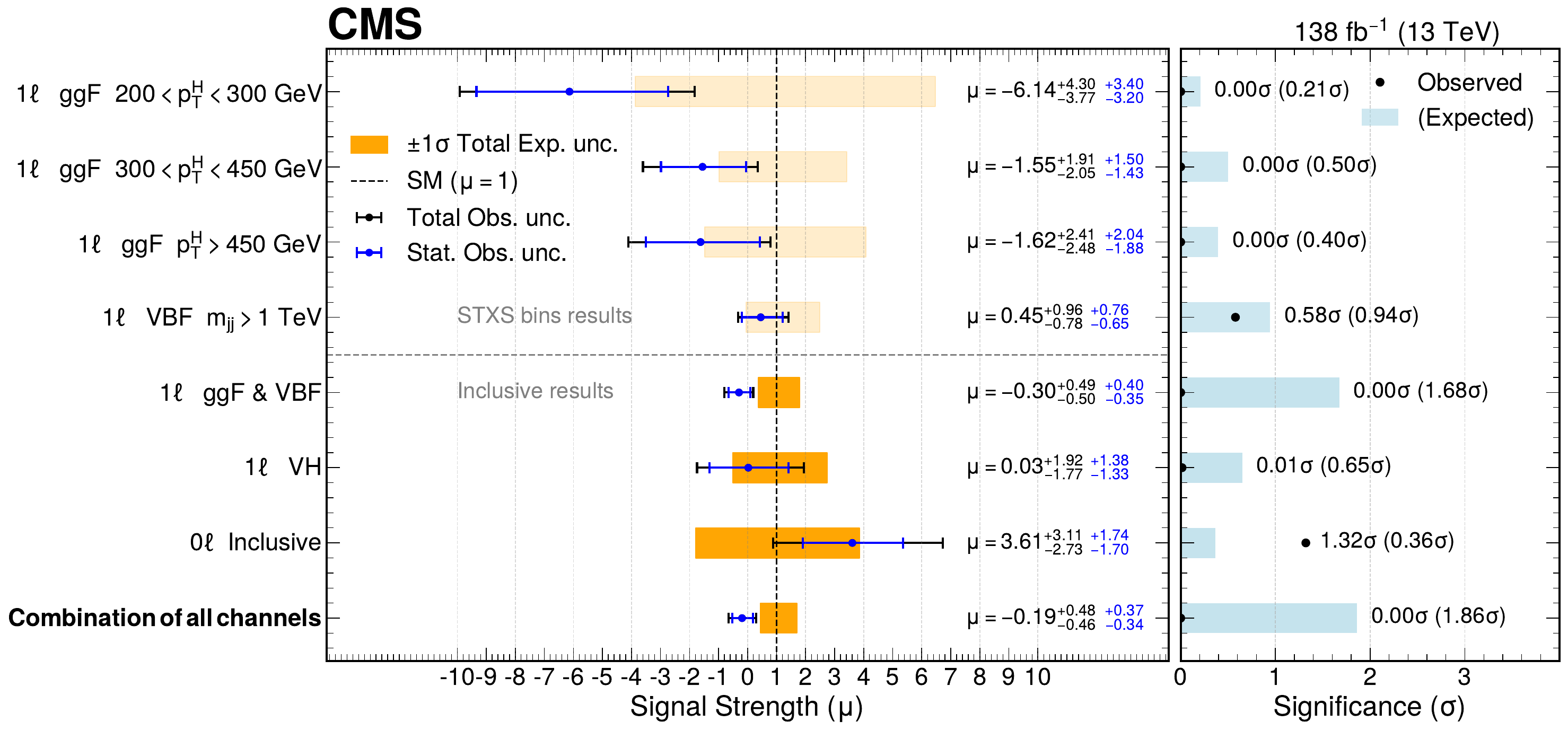}
\caption{
Observed and expected signal strength (left) and significance (right) for $\hwwall$ in all the channels using the full data set. 
Combined results are presented alongside individual contributions from each channel. 
Total expected uncertainties are indicated by yellow bands, while signal strength significances are shown with light blue bars.
Blue and black lines represent statistical and total observed uncertainties, respectively.}
\label{fig:sig_strength}
\end{figure}

\begin{figure}[!htbp]\centering
\includegraphics[width=0.8\linewidth]{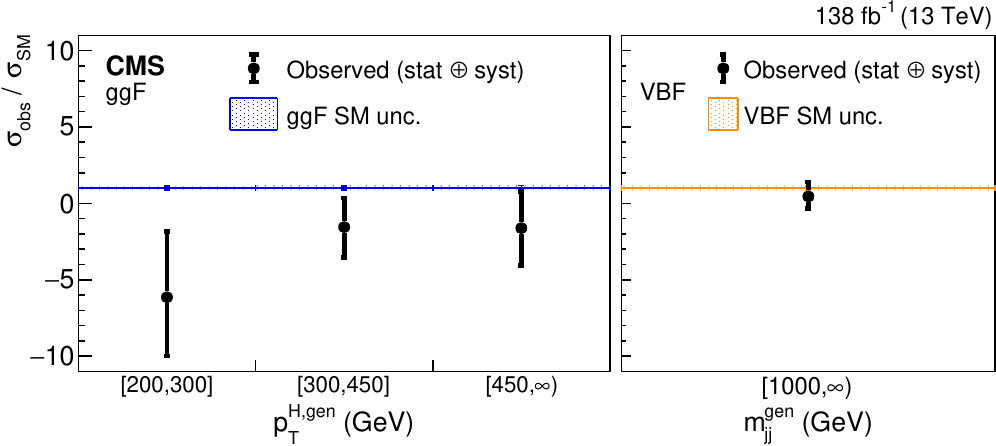}
\caption{
Unfolded measurement of the STXS cross sections in generator-level bins for three bins of Higgs boson $\pt$ and one bin of $\mjj$ in the $1\Pell$ channel. Measured cross sections are divided by standard-model expectations. 
Blue and orange uncertainty bands include theoretical uncertainties affecting the signal acceptance.
}
\label{fig:stxs}
\end{figure}

Figure~\ref{fig:sig_strength} shows the observed and expected signal strength (left) and significance (right) for $\hwwall$ in all the channels for the full data set.
Combined results are shown alongside individual contributions from each channel.
For the $1\Pell$ channel, results are also shown for an alternative measurement using four independent signal strengths for the ggF and VBF processes, corresponding to different STXS generator-level \ptH bins introduced in Section~\ref{Sec:1l_selection}.
For the ggF process, a signal strength is assigned to each bin of \ptH: $[200, 300)$, $[300, 450)$, and $[450,+\infty)\GeV$. 
For the VBF process, an additional signal strength is assigned to the contribution satisfying $\mjj>1\TeV$ for two forward generator-level jets.

The unfolded fiducial cross sections for the alternative measurement in the $1\Pell$ channel are shown in Fig.~\ref{fig:stxs}, together with predictions from the signal event generators.
The cross section is extracted simultaneously in generator-level ggF and VBF bins following the STXS stage 1.2 scheme, similar to Ref.~\cite{CMS:2024ddc}, as described in Section~\ref{Sec:1l_selection}, to enable future combined differential cross section measurements in \ptH bins.
A maximum likelihood unfolding technique is employed to correct for detector acceptance and resolution effects in the measured production cross section.
Each of the three cross section bins is modeled as a separate process with an independent signal strength parameter in the likelihood fit.
Only theoretical uncertainties that impact the signal acceptance in the analysis selection are considered.
The best fit cross sections and their uncertainties are obtained by scaling each fitted signal strength and its uncertainty by the corresponding simulated SM cross section.

A breakdown of the uncertainty in $\mu$, denoted $\Delta\mu$, is provided 
in Table~\ref{tab:unc_breakdown}, obtained from profile-likelihood scans 
when each group of nuisance parameters is fixed to its post-fit value 
while the remaining parameters are profiled.
Finally, the results of recent CMS searches for high-\pt \PH production are summarized in Table~\ref{tab:complementarity}.

\begin{table}[htbp]\centering
\topcaption{Relative contributions to $\Delta\mu$ from each uncertainty source, symmetrized as $(\Delta\mu_{+}+\Delta\mu_{-})/2$ and normalized to the total uncertainty. As nuisance-parameter groups are not statistically orthogonal, 
their sum may exceed 100\%.}
\resizebox{\textwidth}{!}{
\begin{tabular}{l c}
    Uncertainty source & \begin{tabular}[c]{@{}c@{}}Relative\\ contribution to $\Delta\mu$\end{tabular} \\ \hline
    Total statistical uncertainty                                                                          & 75\% \\
    Total systematic uncertainty                                                                           & 66\% \\[\cmsTabSkip]
    Simulated sample size                                                                                  & 44\% \\
    \ParT tagging efficiency                                                                               & 35\% \\
    $1\Pell$ background normalizations (\ttbar, \wjets)                                                   & 34\% \\
    $0\Pell$ QCD background model (TFs \& statistics)                                                     & 28\% \\
    $\PQb$ tagging efficiency                                                                              & 26\% \\
    Background theory (NLO effects, parton shower, \alpS)                                                  & 21\% \\
    Other sources (pileup, prefiring, unclustered energy, misid. $\Pell$, \PNet tagger eff.)               & 17\% \\
    Signal theory (renormalization \& factorization scales, PDF, parton shower, $\alpS$, $\mathcal{B}$)    & 15\% \\
    Jet energy scale and resolution                                                                        & 13\% \\
    Triggers                                                                                               & 11\% \\
    {\PGm}/{\Pe} reconstruction, isolation \& identification                                               & 9\%  \\
    Jet mass scale \& resolution                                                                           & 4\%  \\
    Integrated luminosity                                                                                  & 3\%  \\
\end{tabular}}
\label{tab:unc_breakdown}
\end{table}

\begin{table}[!htbp]\centering
\topcaption{Summary and comparison of the results with complementary CMS boosted $\PH$ searches~\cite{CMS:2024ddc,CMS:2026fsx,htt_boosted,Sirunyan_2020}.
For $\PH\to\PQb\PQb$ ($\PH\to\PW\PW$) VBF, the bins are on $m_{\text{jj}}^{\rm gen}$: $1.0$--$1.5$ and $>1.5\,\TeV$ ($>1\,\TeV$).}
\vspace{-0.3cm}
\resizebox{\textwidth}{!}{
\begin{tabular}{c c c c c c c c}
& & & & & & \multicolumn{2}{c}{Significance} \\[-2pt]
Decay & Production & Ref. & $\ptH$ [\GeV] \& bins & $\mu_{\rm obs}$ total & $\mu_{\rm obs}$ per $\ptH$ bin & Obs. & Exp. \\
\hline
\rule{0pt}{3.5ex}\multirow{4}{*}[-1.2ex]{$\PH\to\PQb\PQb$} & Inclusive & \cite{Sirunyan_2020} & $>450$ & $3.7^{+1.6}_{-1.5}$ & --- & 2.5 & 0.7 \\[6pt]
                & ggF             & \multirow{2}{*}{\cite{CMS:2024ddc}} & 300--450--650--$\infty$ & $1.6^{+1.7}_{-1.5}$ & $-2.1^{+4.8}_{-5.0}$ \hfill $1.9^{+1.8}_{-1.9}$ \hfill $2.0^{+3.0}_{-2.5}$ & \multirow{2}{*}{4.0} & \multirow{2}{*}{---} \\
                & VBF             &  & $>200$ & $4.9^{+1.9}_{-1.6}$ & $7.7^{+6.5}_{-6.1}$ ~~~~~~$2.9^{+1.7}_{-1.5}$ & & \\[6pt]
                & $\PV(\Pq\Pq)\PH$ & \cite{CMS:2026fsx} & $>450$ & $0.72^{+0.75}_{-0.71}$ & --- & 1.0 & 1.64 \\[15pt]
$\PH\to\PGt\PGt$ & Inclusive       & \cite{htt_boosted} & $>250$ & $1.64^{+0.68}_{-0.54}$ & --- & 3.5 & 2.2 \\[15pt]
\multirow{3}{*}{$\PH\to\PW\PW$} & ggF+VBF & \multirow{3}{*}{\begin{tabular}{@{}c@{}} This \\ analysis \end{tabular}} & 200--300--450--$\infty$ & $-0.30^{+0.49}_{-0.50}$ & $-6.15^{+4.32}_{-3.85}$ \hfill $-1.56^{+1.91}_{-2.01}$ \hfill $-1.62^{+2.42}_{-2.45}$ & 0.00 & 1.68 \\
                & VH              &  & $>250$ & $0.03^{+1.92}_{-1.77}$ & --- & 0.01 & 0.65 \\
                & Combination     &  & $>250$ & $-0.19^{+0.48}_{-0.46}$ & --- & 0.00 & 1.86 \\
\end{tabular}}
\label{tab:complementarity}
\end{table}

\section{Summary} \label{Sec:sum}

A search for Higgs boson (\PH) production at high transverse momentum is presented in the decay channel \hwwall.
The analysis uses proton-proton collision data collected at $\sqrt{s} = 13\TeV$ with the CMS experiment, corresponding to an integrated luminosity of 138\fbinv, and focuses on $\PW\PW$ decays with one or no isolated lepton ($1\Pell$ and $0\Pell$, respectively; $\Pell=\Pe,\mu$) in the final state.
The final states are characterized by a single large-radius jet containing the hadronic decay products of the \PW bosons, utilizing the jet substructure resulting from the Lorentz-boosted topology of the Higgs boson decay.
The $1\Pell$ channel categorizes events by the dominant Higgs boson production mechanisms: gluon fusion, vector boson fusion, and vector boson associated production, while the $0\Pell$ channel remains inclusive across all production processes.
The particle transformer algorithm leverages advanced machine-learning techniques to identify \PH-candidate jets with intricate substructure, missing transverse momentum aligned with the jet, or leptons inside the jet.
It is calibrated with the Lund jet plane reweighting method and fine-tuned to optimize the expected signal significance in the $1\Pell$ channel, achieving 60\% higher signal efficiency than the baseline tagger.
The invariant mass of the candidate jet \PH or vector boson is used for signal extraction.
The expected signal significance is 1.86 standard deviations, while the observed signal strength relative to the standard model expectation is $\mu = -0.19^{+0.48}_{-0.46}$, indicating no evidence of a signal above the background.
These measurements represent the first dedicated study of highly Lorentz-boosted \hwwall decays, complementing earlier searches for high transverse momentum Higgs boson production in other decay channels and production processes.

\begin{acknowledgments}
We congratulate our colleagues in the CERN accelerator departments for the excellent performance of the LHC and thank the technical and administrative staffs at CERN and at other CMS institutes for their contributions to the success of the CMS effort. In addition, we gratefully acknowledge the computing centers and personnel of the Worldwide LHC Computing Grid and other centers for delivering so effectively the computing infrastructure essential to our analyses. Finally, we acknowledge the enduring support for the construction and operation of the LHC, the CMS detector, and the supporting computing infrastructure provided by the following funding agencies: SC (Armenia), BMBWF and FWF (Austria); FNRS and FWO (Belgium); CNPq, CAPES, FAPERJ, FAPERGS, and FAPESP (Brazil); MES and BNSF (Bulgaria); CERN; CAS, MoST, and NSFC (China); MINCIENCIAS (Colombia); MSES and CSF (Croatia); RIF (Cyprus); SENESCYT (Ecuador); ERC PRG and PSG, TARISTU24-TK10 and MoER TK202 (Estonia); Academy of Finland, MEC, and HIP (Finland); CEA and CNRS/IN2P3 (France); SRNSF (Georgia); BMFTR, DFG, and HGF (Germany); GSRI (Greece); NKFIH (Hungary); DAE and DST (India); IPM (Iran); SFI (Ireland); INFN (Italy); MSIT and NRF (Republic of Korea); MES (Latvia); LMTLT (Lithuania); MOE and UM (Malaysia); BUAP, CINVESTAV, CONACYT, LNS, SEP, and UASLP-FAI (Mexico); MOS (Montenegro); MBIE (New Zealand); PAEC (Pakistan); MES, NSC, and NAWA (Poland); FCT (Portugal); MESTD (Serbia); MICIU/AEI and PCTI (Spain); MOSTR (Sri Lanka); Swiss Funding Agencies (Switzerland); MST (Taipei); MHESI (Thailand); TUBITAK and TENMAK (T\"{u}rkiye); NASU (Ukraine); STFC (United Kingdom); DOE and NSF (USA).

\hyphenation{Rachada-pisek} Individuals have received support from the Marie-Curie program and the European Research Council and Horizon 2020 Grant, contract Nos.\ 675440, 724704, 752730, 758316, 765710, 824093, 101115353, 101002207, 101001205, and COST Action CA16108 (European Union); the Leventis Foundation; the Alfred P.\ Sloan Foundation; the Alexander von Humboldt Foundation; the Science Committee, project no. 22rl-037 (Armenia); the Fonds pour la Formation \`a la Recherche dans l'Industrie et dans l'Agriculture (FRIA) and Fonds voor Wetenschappelijk Onderzoek contract No. 1228724N (Belgium); the Beijing Municipal Science \& Technology Commission, No. Z191100007219010, the Fundamental Research Funds for the Central Universities, the Ministry of Science and Technology of China under Grant No. 2023YFA1605804, the Natural Science Foundation of China under Grant No. 12535004, and USTC Research Funds of the Double First-Class Initiative No.\ YD2030002017 (China); the Ministry of Education, Youth and Sports (MEYS) of the Czech Republic; the Shota Rustaveli National Science Foundation, grant FR-22-985 (Georgia); the Deutsche Forschungsgemeinschaft (DFG), among others, under Germany's Excellence Strategy -- EXC 2121 ``Quantum Universe" -- 390833306, and under project number 400140256 - GRK2497; the Hellenic Foundation for Research and Innovation (HFRI), Project Number 2288 (Greece); the Hungarian Academy of Sciences, the New National Excellence Program - \'UNKP, the NKFIH research grants K 131991, K 133046, K 138136, K 143460, K 143477, K 146913, K 146914, K 147048, 2020-2.2.1-ED-2021-00181, TKP2021-NKTA-64, and 2025-1.1.5-NEMZ\_KI-2025-00004 (Hungary); the Council of Science and Industrial Research, India; ICSC -- National Research Center for High Performance Computing, Big Data and Quantum Computing, FAIR -- Future Artificial Intelligence Research, and CUP I53D23001070006 (Mission 4 Component 1), funded by the NextGenerationEU program, the Italian Ministry of University and Research (MUR) under Bando PRIN 2022 -- CUP I53C24002390006, PRIN PRIMULA 2022RBYK7T (Italy); the Latvian Council of Science; the Ministry of Education and Science, project no. 2022/WK/14, and the National Science Center, contracts Opus 2021/41/B/ST2/01369, 2021/43/B/ST2/01552, 2023/49/B/ST2/03273, and the NAWA contract BPN/PPO/2021/1/00011 (Poland); the Funda\c{c}\~ao para a Ci\^encia e a Tecnologia (Portugal); the National Priorities Research Program by Qatar National Research Fund; MICIU/AEI/10.13039/501100011033, ERDF/EU, ``European Union NextGenerationEU/PRTR", projects PID2022-142604OB-C21, PID2022-139519OB-C21, PID2023-147706NB-I00, PID2023-148896NB-I00, PID2023-146983NB-I00, PID2023-147115NB-I00, PID2023-148418NB-C41, PID2023-148418NB-C42, PID2023-148418NB-C43, PID2023-148418NB-C44, PID2024-158190NB-C22, RYC2021-033305-I, RYC2024-048719-I, CNS2023-144781, CNS2024-154769 and Plan de Ciencia, Tecnolog{\'i}a e Innovaci{\'o}n de Asturias, Spain; the Chulalongkorn Academic into Its 2nd Century Project Advancement Project, the National Science, Research and Innovation Fund program IND\_FF\_68\_369\_2300\_097, and the Program Management Unit for Human Resources \& Institutional Development, Research and Innovation, grant B39G680009 (Thailand); the Eric \& Wendy Schmidt Fund for Strategic Innovation through the CERN Next Generation Triggers project under grant agreement number SIF-2023-004; the Kavli Foundation; the Nvidia Corporation; the SuperMicro Corporation; the Welch Foundation, contract C-1845; and the Weston Havens Foundation (USA).
\end{acknowledgments}

\bibliography{auto_generated}\cleardoublepage \appendix\section{The CMS Collaboration \label{app:collab}}\begin{sloppypar}\hyphenpenalty=5000\widowpenalty=500\clubpenalty=5000

\providecommand{\href}[2]{#2}\begingroup\raggedright\begin{thebibliography}{100}%
\makeatletter
\providecommand{\hrefCMSnoop }[0]{\@secondoftwo}%
\makeatother
\providecommand{\doi}{\texttt{doi:}\begingroup \urlstyle{tt}\Url}

\bibitem{ATLAS:2012yve}
\hrefCMSnoop {}{{ATLAS Collaboration}, ``Observation of a new particle in the
  search for the standard model {Higgs} boson with the {ATLAS} detector at the
  {LHC}'',} \textit{ Phys. Lett. B} \textbf{ 716} (2012) 1,
  \href{http://dx.doi.org/10.1016/j.physletb.2012.08.020}{\doi{10.1016/j.physletb.2012.08.020}},
  \href{http://www.arXiv.org/abs/1207.7214}{\texttt{arXiv:1207.7214}}.

\bibitem{CMS:2012qbp}
\hrefCMSnoop {}{{CMS Collaboration}, ``Observation of a new boson at a mass of
  {125\GeV} with the {CMS} experiment at the {LHC}'',} \textit{ Phys. Lett. B}
  \textbf{ 716} (2012) 30,
  \href{http://dx.doi.org/10.1016/j.physletb.2012.08.021}{\doi{10.1016/j.physletb.2012.08.021}},
  \href{http://www.arXiv.org/abs/1207.7235}{\texttt{arXiv:1207.7235}}.

\bibitem{cms2013}
\hrefCMSnoop {}{{CMS Collaboration}, ``{Observation of a new boson with mass
  near 125\GeV in $\Pp\Pp$ collisions at $\sqrt{s} = 7$ and 8\TeV}'',} \textit{
  JHEP} \textbf{ 06} (2013) 081,
  \href{http://dx.doi.org/10.1007/JHEP06(2013)081}{\doi{10.1007/JHEP06(2013)081}},
  \href{http://www.arXiv.org/abs/1303.4571}{\texttt{arXiv:1303.4571}}.

\bibitem{ATLAS:2022vkf}
\hrefCMSnoop {}{G.~Aad { et~al.}, ``{A detailed map of Higgs boson interactions
  by the ATLAS experiment ten years after the discovery}'',} \textit{ Nature}
  \textbf{ 607} (2022) 52,
  \href{http://dx.doi.org/10.1038/s41586-022-04893-w}{\doi{10.1038/s41586-022-04893-w}},
  \href{http://www.arXiv.org/abs/2207.00092}{\texttt{arXiv:2207.00092}}.
  [Erratum: \DOI{10.1038/s41586-023-06248-5}].

\bibitem{CMS:2022dwd}
\hrefCMSnoop {}{{CMS Collaboration}, ``{A portrait of the Higgs boson by the
  CMS experiment ten years after the discovery}'',} \textit{ Nature} \textbf{
  607} (2022) 60,
  \href{http://dx.doi.org/10.1038/s41586-022-04892-x}{\doi{10.1038/s41586-022-04892-x}},
  \href{http://www.arXiv.org/abs/2207.00043}{\texttt{arXiv:2207.00043}}.
  [Corrigendum: \DOI{10.1038/s41586-023-06164-8}].

\bibitem{becker2021precise}
\hrefCMSnoop {}{K.~Becker { et~al.}, ``{Precise predictions for boosted Higgs
  production}'',} \textit{ SciPost Phys. Core} \textbf{ 7} (2024) 001,
  \href{http://dx.doi.org/10.21468/SciPostPhysCore.7.1.001}{\doi{10.21468/SciPostPhysCore.7.1.001}},
  \href{http://www.arXiv.org/abs/2005.07762}{\texttt{arXiv:2005.07762}}.

\bibitem{Grojean_2014}
\hrefCMSnoop {}{C.~Grojean, E.~Salvioni, M.~Schlaffer, and A.~Weiler, ``Very
  boosted {Higgs} in gluon fusion'',} \textit{ JHEP} \textbf{ 05} (2014) 22,
  \href{http://dx.doi.org/10.1007/jhep05(2014)022}{\doi{10.1007/jhep05(2014)022}},
  \href{http://www.arXiv.org/abs/1312.3317}{\texttt{arXiv:1312.3317}}.

\bibitem{Schlaffer_2014}
M.~Schlaffer\hrefCMSnoop {}{ { et~al.}, ``Boosted {Higgs} shapes'',} \textit{
  Eur. Phys. J. C} \textbf{ 74} (2014) 3120,
  \href{http://dx.doi.org/10.1140/epjc/s10052-014-3120-z}{\doi{10.1140/epjc/s10052-014-3120-z}},
  \href{http://www.arXiv.org/abs/1405.4295}{\texttt{arXiv:1405.4295}}.

\bibitem{Dawson_2015}
\hrefCMSnoop {}{S.~Dawson, I.~M. Lewis, and M.~Zeng, ``Usefulness of effective
  field theory for boosted {Higgs} production'',} \textit{ Phys. Rev. D}
  \textbf{ 91} (2015) 074012,
  \href{http://dx.doi.org/10.1103/physrevd.91.074012}{\doi{10.1103/physrevd.91.074012}},
  \href{http://www.arXiv.org/abs/1501.04103}{\texttt{arXiv:1501.04103}}.

\bibitem{Grazzini_2017}
\hrefCMSnoop {}{M.~Grazzini, A.~Ilnicka, M.~Spira, and M.~Wiesemann, ``Modeling
  {BSM} effects on the {Higgs} transverse-momentum spectrum in an {EFT}
  approach'',} \textit{ JHEP} \textbf{ 03} (2017) 115,
  \href{http://dx.doi.org/10.1007/jhep03(2017)115}{\doi{10.1007/jhep03(2017)115}},
  \href{http://www.arXiv.org/abs/1612.00283}{\texttt{arXiv:1612.00283}}.

\bibitem{Maltoni_2014}
\hrefCMSnoop {}{F.~Maltoni, K.~Mawatari, and M.~Zaro, ``{Higgs characterisation
  via vector-boson fusion and associated production: NLO and parton-shower
  effects}'',} \textit{ Eur. Phys. J. C} \textbf{ 74} (2014) 2710,
  \href{http://dx.doi.org/10.1140/epjc/s10052-013-2710-5}{\doi{10.1140/epjc/s10052-013-2710-5}},
  \href{http://www.arXiv.org/abs/1311.1829}{\texttt{arXiv:1311.1829}}.

\bibitem{Degrande_2017}
C.~Degrande\hrefCMSnoop {}{ { et~al.}, ``Electroweak {Higgs} boson production
  in the standard model effective field theory beyond leading order in
  {QCD}'',} \textit{ Eur. Phys. J. C} \textbf{ 77} (2017) 262,
  \href{http://dx.doi.org/10.1140/epjc/s10052-017-4793-x}{\doi{10.1140/epjc/s10052-017-4793-x}},
  \href{http://www.arXiv.org/abs/1609.04833}{\texttt{arXiv:1609.04833}}.

\bibitem{CMS:2018gwt}
\hrefCMSnoop {}{{CMS Collaboration}, ``{Measurement and interpretation of
  differential cross sections for Higgs boson production at $\sqrt{s} =
  13\TeV$}'',} \textit{ Phys. Lett. B} \textbf{ 792} (2019) 369,
  \href{http://dx.doi.org/10.1016/j.physletb.2019.03.059}{\doi{10.1016/j.physletb.2019.03.059}},
  \href{http://www.arXiv.org/abs/1812.06504}{\texttt{arXiv:1812.06504}}.

\bibitem{h_gamgam}
\hrefCMSnoop {}{{CMS Collaboration}, ``{Measurement of the Higgs boson
  inclusive and differential fiducial production cross sections in the diphoton
  decay channel with $\Pp\Pp$ collisions at $ \sqrt{s} = 13\TeV$}'',} \textit{
  JHEP} \textbf{ 07} (2023) 091,
  \href{http://dx.doi.org/10.1007/JHEP07(2023)091}{\doi{10.1007/JHEP07(2023)091}},
  \href{http://www.arXiv.org/abs/2208.12279}{\texttt{arXiv:2208.12279}}.

\bibitem{h_ww}
\hrefCMSnoop {}{{CMS Collaboration}, ``{Measurement of the inclusive and
  differential Higgs boson production cross sections in the leptonic $\PW\PW$
  decay mode at $\sqrt{s} = 13\TeV$}'',} \textit{ JHEP} \textbf{ 03} (2021)
  003,
  \href{http://dx.doi.org/10.1007/JHEP03(2021)003}{\doi{10.1007/JHEP03(2021)003}},
  \href{http://www.arXiv.org/abs/2007.01984}{\texttt{arXiv:2007.01984}}.

\bibitem{h_tt}
\hrefCMSnoop {}{{CMS Collaboration}, ``{Measurement of the inclusive and
  differential Higgs boson production cross sections in the decay mode to a
  pair of $\tau$ leptons in pp collisions at $\sqrt{s} = 13\TeV$}'',} \textit{
  Phys. Rev. Lett.} \textbf{ 128} (2022) 081805,
  \href{http://dx.doi.org/10.1103/PhysRevLett.128.081805}{\doi{10.1103/PhysRevLett.128.081805}},
  \href{http://www.arXiv.org/abs/2107.11486}{\texttt{arXiv:2107.11486}}.

\bibitem{h_zz}
\hrefCMSnoop {}{{CMS Collaboration}, ``{Measurements of inclusive and
  differential cross sections for the Higgs boson production and decay to
  four-leptons in proton-proton collisions at $ \sqrt{s} = 13\TeV$}'',}
  \textit{ JHEP} \textbf{ 08} (2023) 040,
  \href{http://dx.doi.org/10.1007/JHEP08(2023)040}{\doi{10.1007/JHEP08(2023)040}},
  \href{http://www.arXiv.org/abs/2305.07532}{\texttt{arXiv:2305.07532}}.

\bibitem{Sirunyan_2018}
\hrefCMSnoop {}{{CMS Collaboration}, ``Inclusive search for a highly boosted
  {Higgs} boson decaying to a bottom quark-antiquark pair'',} \textit{ Phys.
  Rev. Lett.} \textbf{ 120} (2018) 071802,
  \href{http://dx.doi.org/10.1103/physrevlett.120.071802}{\doi{10.1103/physrevlett.120.071802}},
  \href{http://www.arXiv.org/abs/1709.05543}{\texttt{arXiv:1709.05543}}.

\bibitem{Sirunyan_2020}
\hrefCMSnoop {}{{CMS Collaboration}, ``Inclusive search for highly boosted
  {Higgs} bosons decaying to bottom quark-antiquark pairs in proton-proton
  collisions at {$\sqrt{s}=13\TeV$}'',} \textit{ JHEP} \textbf{ 12} (2020) 085,
  \href{http://dx.doi.org/10.1007/jhep12(2020)085}{\doi{10.1007/jhep12(2020)085}},
  \href{http://www.arXiv.org/abs/2006.13251}{\texttt{arXiv:2006.13251}}.

\bibitem{CMS:2026fsx}
\hrefCMSnoop {}{{CMS Collaboration}, ``{Search for a boosted Higgs boson
  decaying to bottom quark pairs in association with a \PW or \PZ boson in
  proton-proton collisions at $\sqrt{s} = 13\TeV$}'',} 2026.
  \href{http://www.arXiv.org/abs/2601.05362}{\texttt{arXiv:2601.05362}}.
  Submitted to \emph{Phys. Lett. B}.

\bibitem{Aad_2022}
\hrefCMSnoop {}{{ATLAS Collaboration}, ``{Constraints on Higgs boson production
  with large transverse momentum using $\PH\to\bbbar$ decays in the ATLAS
  detector}'',} \textit{ Phys. Rev. D} \textbf{ 105} (2022) 092003,
  \href{http://dx.doi.org/10.1103/PhysRevD.105.092003}{\doi{10.1103/PhysRevD.105.092003}},
  \href{http://www.arXiv.org/abs/2111.08340}{\texttt{arXiv:2111.08340}}.

\bibitem{ATLAS:2023jdk}
\hrefCMSnoop {}{{ATLAS Collaboration}, ``Study of high-transverse-momentum
  {Higgs} boson production in association with a vector boson in the
  {$\Pq\Pq\PQb\PQb$} final state with the {ATLAS} detector'',} \textit{ Phys.
  Rev. Lett.} \textbf{ 132} (2024) 131802,
  \href{http://dx.doi.org/10.1103/PhysRevLett.132.131802}{\doi{10.1103/PhysRevLett.132.131802}},
  \href{http://www.arXiv.org/abs/2312.07605}{\texttt{arXiv:2312.07605}}.

\bibitem{ATLAS:2026iwc}
\hrefCMSnoop {}{{ATLAS Collaboration}, ``{Evidence of Higgs boson inclusive
  production at high transverse momentum decaying to a pair of \PQb-quarks with
  the ATLAS detector}'',} 2026.
  \href{http://www.arXiv.org/abs/2603.19369}{\texttt{arXiv:2603.19369}}.
  Submitted to \emph{Phys. Rev. Lett.}

\bibitem{htt_boosted}
\hrefCMSnoop {}{{CMS Collaboration}, ``{Measurement of the production cross
  section of a Higgs boson with large transverse momentum in its decays to a
  pair of \PGt leptons in proton-proton collisions at $\sqrt{s}=13\TeV$}'',}
  \textit{ Phys. Lett. B} \textbf{ 857} (2024) 138964,
  \href{http://dx.doi.org/10.1016/j.physletb.2024.138964}{\doi{10.1016/j.physletb.2024.138964}},
  \href{http://www.arXiv.org/abs/2403.20201}{\texttt{arXiv:2403.20201}}.

\bibitem{CMS:2021xjt}
\hrefCMSnoop {}{{CMS Collaboration}, ``{Precision luminosity measurement in
  proton-proton collisions at $\sqrt{s} = 13\TeV$ in 2015 and 2016 at CMS}'',}
  \textit{ Eur. Phys. J. C} \textbf{ 81} (2021) 800,
  \href{http://dx.doi.org/10.1140/epjc/s10052-021-09538-2}{\doi{10.1140/epjc/s10052-021-09538-2}},
  \href{http://www.arXiv.org/abs/2104.01927}{\texttt{arXiv:2104.01927}}.

\bibitem{CMS-PAS-LUM-17-004}
\href {https://cds.cern.ch/record/2621960}{{CMS Collaboration}, ``{CMS}
  luminosity measurement for the 2017 data-taking period at {$\sqrt{s} =
  13\TeV$}'',} {CMS Physics Analysis Summary} CMS-PAS-LUM-17-004, 2018.

\bibitem{CMS-PAS-LUM-18-002}
\href {https://cds.cern.ch/record/2676164/}{{CMS Collaboration}, ``{CMS}
  luminosity measurement for the 2018 data-taking period at {$\sqrt{s} =
  13\TeV$}'',} CMS Physics Analysis Summary CMS-PAS-LUM-18-002, 2019.

\bibitem{Qu:2022mxj}
\href {https://proceedings.mlr.press/v162/qu22b.html}{H.~Qu, C.~Li, and
  S.~Qian, ``Particle transformer for jet tagging'',} in \textit{ Proc. 39th
  Int. Conf. on Machine Learning}, K.~Chaudhuri { et~al.}, eds., volume 162,
  p.~18281.
\newblock 2022.
\newblock
  \href{http://www.arXiv.org/abs/2202.03772}{\texttt{arXiv:2202.03772}}.

\bibitem{Vaswani:2017attention}
A.~Vaswani\href
  {https://proceedings.neurips.cc/paper_files/paper/2017/file/3f5ee243547dee91fbd053c1c4a845aa-Paper.pdf}{
  { et~al.}, ``{Attention is all you need}'',} in \textit{ Advances in Neural
  Information Processing Systems}, I.~Guyon { et~al.}, eds., volume~30.
\newblock 2017.
\newblock
  \href{http://www.arXiv.org/abs/1706.03762}{\texttt{arXiv:1706.03762}}.

\bibitem{CMS:2026uph}
\hrefCMSnoop {}{{CMS Collaboration}, ``{Particle transformers for identifying
  Lorentz-boosted Higgs bosons decaying to a pair of \PW bosons}'',} 2026.
  \href{http://www.arXiv.org/abs/2604.09809}{\texttt{arXiv:2604.09809}}.
  Submitted to \emph{JHEP}.

\bibitem{Qu:2019gqs}
\hrefCMSnoop {}{H.~Qu and L.~Gouskos, ``{ParticleNet}: {J}et tagging via
  particle clouds'',} \textit{ Phys. Rev. D} \textbf{ 101} (2020) 056019,
  \href{http://dx.doi.org/10.1103/PhysRevD.101.056019}{\doi{10.1103/PhysRevD.101.056019}},
  \href{http://www.arXiv.org/abs/1902.08570}{\texttt{arXiv:1902.08570}}.

\bibitem{CMS:2020poo}
\hrefCMSnoop {}{{CMS Collaboration}, ``{Identification of heavy, energetic,
  hadronically decaying particles using machine-learning techniques}'',}
  \textit{ JINST} \textbf{ 15} (2020) P06005,
  \href{http://dx.doi.org/10.1088/1748-0221/15/06/P06005}{\doi{10.1088/1748-0221/15/06/P06005}},
  \href{http://www.arXiv.org/abs/2004.08262}{\texttt{arXiv:2004.08262}}.

\bibitem{deFlorian:2016spz}
\hrefCMSnoop {}{{LHC Higgs Cross Section Working Group}, ``{Handbook of LHC
  Higgs Cross Sections: 4. Deciphering the nature of the Higgs sector}'',}
  \textit{ CERN Yellow Rep. Monogr.} \textbf{ 2} (2017)
  \href{http://dx.doi.org/10.23731/CYRM-2017-002}{\doi{10.23731/CYRM-2017-002}},
  \href{http://www.arXiv.org/abs/1610.07922}{\texttt{arXiv:1610.07922}}.

\bibitem{Berger:2922392}
\href {https://cds.cern.ch/record/2922392}{{LHC Higgs Cross Section Working
  Group}, ``Simplified template cross sections --- {Stage} 1.1 and 1.2'',}
  {Public Note} LHCHWG-INT-2025-001, 2025.

\bibitem{hepdata}
\hrefCMSnoop {}{``{HEPD}ata record for this analysis'',} 2026.
\newblock
  \href{http://dx.doi.org/10.17182/hepdata.167419}{\doi{10.17182/hepdata.167419}}.

\bibitem{Chatrchyan:2008zzk}
\hrefCMSnoop {}{{CMS Collaboration}, ``The {CMS} experiment at the {CERN}
  {LHC}'',} \textit{ JINST} \textbf{ 3} (2008) S08004,
\href{http://dx.doi.org/10.1088/1748-0221/3/08/S08004}{\doi{10.1088/1748-0221/3/08/S08004}}.

\bibitem{CMS:2023gfb}
\hrefCMSnoop {}{{CMS Collaboration}, ``Development of the {CMS} detector for
  the {CERN LHC Run 3}'',} \textit{ JINST} \textbf{ 19} (2024) P05064,
  \href{http://dx.doi.org/10.1088/1748-0221/19/05/P05064}{\doi{10.1088/1748-0221/19/05/P05064}},
  \href{http://www.arXiv.org/abs/2309.05466}{\texttt{arXiv:2309.05466}}.

\bibitem{CMS:2020cmk}
\hrefCMSnoop {}{{CMS Collaboration}, ``{Performance of the CMS Level-1 trigger
  in proton-proton collisions at $\sqrt{s} = 13\TeV$}'',} \textit{ JINST}
  \textbf{ 15} (2020) P10017,
  \href{http://dx.doi.org/10.1088/1748-0221/15/10/P10017}{\doi{10.1088/1748-0221/15/10/P10017}},
  \href{http://www.arXiv.org/abs/2006.10165}{\texttt{arXiv:2006.10165}}.

\bibitem{Khachatryan:2016bia}
\hrefCMSnoop {}{{CMS Collaboration}, ``{The CMS trigger system}'',} \textit{
  JINST} \textbf{ 12} (2017) P01020,
  \href{http://dx.doi.org/10.1088/1748-0221/12/01/P01020}{\doi{10.1088/1748-0221/12/01/P01020}},
\href{http://www.arXiv.org/abs/1609.02366}{\texttt{arXiv:1609.02366}}.

\bibitem{CMS:2024aqx}
\hrefCMSnoop {}{{CMS Collaboration}, ``Performance of the {CMS} high-level
  trigger during {LHC Run 2}'',} \textit{ JINST} \textbf{ 19} (2024) P11021,
  \href{http://dx.doi.org/10.1088/1748-0221/19/11/P11021}{\doi{10.1088/1748-0221/19/11/P11021}},
  \href{http://www.arXiv.org/abs/2410.17038}{\texttt{arXiv:2410.17038}}.

\bibitem{CMS:2020uim}
\hrefCMSnoop {}{{CMS Collaboration}, ``{Electron and photon reconstruction and
  identification with the CMS experiment at the CERN LHC}'',} \textit{ JINST}
  \textbf{ 16} (2021) P05014,
  \href{http://dx.doi.org/10.1088/1748-0221/16/05/P05014}{\doi{10.1088/1748-0221/16/05/P05014}},
  \href{http://www.arXiv.org/abs/2012.06888}{\texttt{arXiv:2012.06888}}.

\bibitem{Sirunyan:2018fpa}
\hrefCMSnoop {}{{CMS Collaboration}, ``Performance of the {CMS} muon detector
  and muon reconstruction with proton-proton collisions at
  ${\sqrt{s}=13\TeV}$'',} \textit{ JINST} \textbf{ 13} (2018) P06015,
  \href{http://dx.doi.org/10.1088/1748-0221/13/06/P06015}{\doi{10.1088/1748-0221/13/06/P06015}},
\href{http://www.arXiv.org/abs/1804.04528}{\texttt{arXiv:1804.04528}}.

\bibitem{CMS:2014pgm}
\hrefCMSnoop {}{{CMS Collaboration}, ``{Description and performance of track
  and primary-vertex reconstruction with the CMS tracker}'',} \textit{ JINST}
  \textbf{ 9} (2014) P10009,
  \href{http://dx.doi.org/10.1088/1748-0221/9/10/P10009}{\doi{10.1088/1748-0221/9/10/P10009}},
  \href{http://www.arXiv.org/abs/1405.6569}{\texttt{arXiv:1405.6569}}.

\bibitem{Hamilton:2012rf}
\hrefCMSnoop {}{K.~Hamilton, P.~Nason, C.~Oleari, and G.~Zanderighi, ``{Merging
  \PH/\PW/\PZ + 0 and 1 jet at NLO with no merging scale: a path to parton
  shower + NNLO matching}'',} \textit{ JHEP} \textbf{ 05} (2013) 082,
  \href{http://dx.doi.org/10.1007/JHEP05(2013)082}{\doi{10.1007/JHEP05(2013)082}},
  \href{http://www.arXiv.org/abs/1212.4504}{\texttt{arXiv:1212.4504}}.

\bibitem{Neumann_2018}
\hrefCMSnoop {}{T.~Neumann, ``{NLO} {Higgs}+jet production at large transverse
  momenta including top quark mass effects'',} \textit{ J. Phys. Commun.}
  \textbf{ 2} (2018) 095017,
  \href{http://dx.doi.org/10.1088/2399-6528/aadfbf}{\doi{10.1088/2399-6528/aadfbf}},
  \href{http://www.arXiv.org/abs/1802.02981}{\texttt{arXiv:1802.02981}}.

\bibitem{Nason:2009ai}
\hrefCMSnoop {}{P.~Nason and C.~Oleari, ``{NLO} {Higgs} boson production via
  vector-boson fusion matched with shower in {\POWHEG}'',} \textit{ JHEP}
  \textbf{ 02} (2010) 037,
  \href{http://dx.doi.org/10.1007/JHEP02(2010)037}{\doi{10.1007/JHEP02(2010)037}},
\href{http://www.arXiv.org/abs/0911.5299}{\texttt{arXiv:0911.5299}}.

\bibitem{Luisoni:2013kna}
\hrefCMSnoop {}{G.~Luisoni, P.~Nason, C.~Oleari, and F.~Tramontano,
  ``{$\PH\PW^{\pm}$/$\PH\PZ$} + 0 and 1 jet at {NLO} with the {\POWHEG
  \textsc{box}} interfaced to \textsc{GoSam} and their merging within
  \textsc{MiNLO}'',} \textit{ JHEP} \textbf{ 10} (2013) 083,
  \href{http://dx.doi.org/10.1007/JHEP10(2013)083}{\doi{10.1007/JHEP10(2013)083}},
\href{http://www.arXiv.org/abs/1306.2542}{\texttt{arXiv:1306.2542}}.

\bibitem{Hartanto:2015uka}
\hrefCMSnoop {}{H.~B. Hartanto, B.~Jager, L.~Reina, and D.~Wackeroth, ``{Higgs}
  boson production in association with top quarks in the {\POWHEG}
  \textsc{box}'',} \textit{ Phys. Rev. D} \textbf{ 91} (2015) 094003,
  \href{http://dx.doi.org/10.1103/PhysRevD.91.094003}{\doi{10.1103/PhysRevD.91.094003}},
\href{http://www.arXiv.org/abs/1501.04498}{\texttt{arXiv:1501.04498}}.

\bibitem{POWHEG3}
\hrefCMSnoop {}{S.~Alioli, P.~Nason, C.~Oleari, and E.~Re, ``{A general
  framework for implementing NLO calculations in shower Monte Carlo programs:
  the \POWHEG \textsc{box}}'',} \textit{ JHEP} \textbf{ 06} (2010) 043,
  \href{http://dx.doi.org/10.1007/JHEP06(2010)043}{\doi{10.1007/JHEP06(2010)043}},
\href{http://www.arXiv.org/abs/1002.2581}{\texttt{arXiv:1002.2581}}.

\bibitem{Bolognesi:2012mm}
S.~Bolognesi\hrefCMSnoop {}{ { et~al.}, ``On the spin and parity of a
  single-produced resonance at the {LHC}'',} \textit{ Phys. Rev. D} \textbf{
  86} (2012) 095031,
  \href{http://dx.doi.org/10.1103/PhysRevD.86.095031}{\doi{10.1103/PhysRevD.86.095031}},
\href{http://www.arXiv.org/abs/1208.4018}{\texttt{arXiv:1208.4018}}.

\bibitem{Sjostrand:2014zea}
T.~Sj{\"o}strand\hrefCMSnoop {}{ { et~al.}, ``An introduction to {\PYTHIA}
  8.2'',} \textit{ Comput. Phys. Commun.} \textbf{ 191} (2015) 159,
  \href{http://dx.doi.org/10.1016/j.cpc.2015.01.024}{\doi{10.1016/j.cpc.2015.01.024}},
\href{http://www.arXiv.org/abs/1410.3012}{\texttt{arXiv:1410.3012}}.

\bibitem{Alwall_2014}
J.~Alwall\hrefCMSnoop {}{ { et~al.}, ``{The automated computation of tree-level
  and next-to-leading order differential cross sections, and their matching to
  parton shower simulations}'',} \textit{ JHEP} \textbf{ 07} (2014) 079,
  \href{http://dx.doi.org/10.1007/JHEP07(2014)079}{\doi{10.1007/JHEP07(2014)079}},
  \href{http://www.arXiv.org/abs/1405.0301}{\texttt{arXiv:1405.0301}}.

\bibitem{Gleisberg:2008ta}
T.~Gleisberg\hrefCMSnoop {}{ { et~al.}, ``{Event generation with \SHERPA
  1.1}'',} \textit{ JHEP} \textbf{ 02} (2009) 007,
  \href{http://dx.doi.org/10.1088/1126-6708/2009/02/007}{\doi{10.1088/1126-6708/2009/02/007}},
  \href{http://www.arXiv.org/abs/0811.4622}{\texttt{arXiv:0811.4622}}.

\bibitem{Sherpa:2019gpd}
\hrefCMSnoop {}{{Sherpa} Collaboration, ``{Event Generation with \SHERPA
  2.2}'',} \textit{ SciPost Phys.} \textbf{ 7} (2019) 034,
  \href{http://dx.doi.org/10.21468/SciPostPhys.7.3.034}{\doi{10.21468/SciPostPhys.7.3.034}},
  \href{http://www.arXiv.org/abs/1905.09127}{\texttt{arXiv:1905.09127}}.

\bibitem{Gleisberg:2008fv}
\hrefCMSnoop {}{T.~Gleisberg and S.~Hoeche, ``{\textsc{Comix}, a new matrix
  element generator}'',} \textit{ JHEP} \textbf{ 12} (2008) 039,
  \href{http://dx.doi.org/10.1088/1126-6708/2008/12/039}{\doi{10.1088/1126-6708/2008/12/039}},
  \href{http://www.arXiv.org/abs/0808.3674}{\texttt{arXiv:0808.3674}}.

\bibitem{Alwall:2007fs}
J.~Alwall\hrefCMSnoop {}{ { et~al.}, ``Comparative study of various algorithms
  for the merging of parton showers and matrix elements in hadronic
  collisions'',} \textit{ Eur. Phys. J. C} \textbf{ 53} (2008) 473,
  \href{http://dx.doi.org/10.1140/epjc/s10052-007-0490-5}{\doi{10.1140/epjc/s10052-007-0490-5}},
\href{http://www.arXiv.org/abs/0706.2569}{\texttt{arXiv:0706.2569}}.

\bibitem{Hoeche:2012yf}
\hrefCMSnoop {}{S.~Hoeche, F.~Krauss, M.~Schonherr, and F.~Siegert, ``{QCD
  matrix elements + parton showers: The NLO case}'',} \textit{ JHEP} \textbf{
  04} (2013) 027,
  \href{http://dx.doi.org/10.1007/JHEP04(2013)027}{\doi{10.1007/JHEP04(2013)027}},
  \href{http://www.arXiv.org/abs/1207.5030}{\texttt{arXiv:1207.5030}}.

\bibitem{Frederix:2012ps}
\hrefCMSnoop {}{R.~Frederix and S.~Frixione, ``{Merging meets matching in
  \MCATNLO}'',} \textit{ JHEP} \textbf{ 12} (2012) 061,
  \href{http://dx.doi.org/10.1007/JHEP12(2012)061}{\doi{10.1007/JHEP12(2012)061}},
  \href{http://www.arXiv.org/abs/1209.6215}{\texttt{arXiv:1209.6215}}.

\bibitem{kallweit2015nlo}
S.~Kallweit\hrefCMSnoop {}{ { et~al.}, ``{NLO} electroweak automation and
  precise predictions for {\PW}+multijet production at the {LHC}'',} \textit{
  JHEP} \textbf{ 04} (2015) 012,
  \href{http://dx.doi.org/10.1007/JHEP04(2015)012}{\doi{10.1007/JHEP04(2015)012}},
  \href{http://www.arXiv.org/abs/1412.5157}{\texttt{arXiv:1412.5157}}.

\bibitem{Kallweit_2016}
S.~Kallweit\hrefCMSnoop {}{ { et~al.}, ``{NLO QCD+EW predictions for $\PV$+jets
  including off-shell vector-boson decays and multijet merging}'',} \textit{
  JHEP} \textbf{ 04} (2016) 021,
  \href{http://dx.doi.org/10.1007/JHEP04(2016)021}{\doi{10.1007/JHEP04(2016)021}},
  \href{http://www.arXiv.org/abs/1511.08692}{\texttt{arXiv:1511.08692}}.

\bibitem{kallweit2015nlo2}
S.~Kallweit\hrefCMSnoop {}{ { et~al.}, ``{NLO} {QCD}+{EW} automation and
  precise predictions for {\PV}+multijet production'',} in \textit{ {Proc. 50th
  Int. Conf. Rencontres de Moriond on QCD and High Energy Interactions}},
  p.~121.
\newblock 2015.
\newblock
  \href{http://www.arXiv.org/abs/1505.05704}{\texttt{arXiv:1505.05704}}.

\bibitem{Lindert_2017}
\hrefCMSnoop {}{J.~M. Lindert { et~al.}, ``{Precise predictions for $\PV$+jets
  dark matter backgrounds}'',} \textit{ Eur. Phys. J. C} \textbf{ 77} (2017)
  829,
  \href{http://dx.doi.org/10.1140/epjc/s10052-017-5389-1}{\doi{10.1140/epjc/s10052-017-5389-1}},
  \href{http://www.arXiv.org/abs/1705.04664}{\texttt{arXiv:1705.04664}}.

\bibitem{POWHEG1}
\hrefCMSnoop {}{P.~Nason, ``{A new method for combining NLO QCD with shower
  Monte Carlo algorithms}'',} \textit{ JHEP} \textbf{ 11} (2004) 040,
  \href{http://dx.doi.org/10.1088/1126-6708/2004/11/040}{\doi{10.1088/1126-6708/2004/11/040}},
\href{http://www.arXiv.org/abs/hep-ph/0409146}{\texttt{arXiv:hep-ph/0409146}}.

\bibitem{POWHEG2}
\hrefCMSnoop {}{S.~Frixione, P.~Nason, and C.~Oleari, ``{Matching NLO QCD
  computations with parton shower simulations: the POWHEG method}'',} \textit{
  JHEP} \textbf{ 11} (2007) 070,
  \href{http://dx.doi.org/10.1088/1126-6708/2007/11/070}{\doi{10.1088/1126-6708/2007/11/070}},
\href{http://www.arXiv.org/abs/0709.2092}{\texttt{arXiv:0709.2092}}.

\bibitem{Alioli:2011as}
\hrefCMSnoop {}{S.~Alioli, S.-O. Moch, and P.~Uwer, ``Hadronic top-quark
  pair-production with one jet and parton showering'',} \textit{ JHEP} \textbf{
  01} (2012) 137,
  \href{http://dx.doi.org/10.1007/JHEP01(2012)137}{\doi{10.1007/JHEP01(2012)137}},
\href{http://www.arXiv.org/abs/1110.5251}{\texttt{arXiv:1110.5251}}.

\bibitem{Alioli:2009je}
\hrefCMSnoop {}{S.~Alioli, P.~Nason, C.~Oleari, and E.~Re, ``{NLO single-top
  production matched with shower in \POWHEG: $s$- and $t$-channel
  contributions}'',} \textit{ JHEP} \textbf{ 09} (2009) 111,
  \href{http://dx.doi.org/10.1088/1126-6708/2009/09/111}{\doi{10.1088/1126-6708/2009/09/111}},
  \href{http://www.arXiv.org/abs/0907.4076}{\texttt{arXiv:0907.4076}}.
[Erratum: \DOI{10.1007/JHEP02(2010)011}].

\bibitem{Frederix:2012dh}
\hrefCMSnoop {}{R.~Frederix, E.~Re, and P.~Torrielli, ``Single-top $t$-channel
  hadroproduction in the four-flavour scheme with {\POWHEG} and {a\MCATNLO}'',}
  \textit{ JHEP} \textbf{ 09} (2012) 130,
  \href{http://dx.doi.org/10.1007/JHEP09(2012)130}{\doi{10.1007/JHEP09(2012)130}},
\href{http://www.arXiv.org/abs/1207.5391}{\texttt{arXiv:1207.5391}}.

\bibitem{Campbell:2010ff}
\hrefCMSnoop {}{J.~M. Campbell and R.~K. Ellis, ``{\MCFM} for the {Tevatron}
  and the {LHC}'',} \textit{ Nucl. Phys. Proc. Suppl.} \textbf{ 205} (2010) 10,
  \href{http://dx.doi.org/10.1016/j.nuclphysbps.2010.08.011}{\doi{10.1016/j.nuclphysbps.2010.08.011}},
\href{http://www.arXiv.org/abs/1007.3492}{\texttt{arXiv:1007.3492}}.

\bibitem{Sirunyan:2019dfx}
\hrefCMSnoop {}{{CMS Collaboration}, ``{Extraction and validation of a new set
  of CMS \PYTHIA{}8 tunes from underlying-event measurements}'',} \textit{ Eur.
  Phys. J. C} \textbf{ 80} (2020) 4,
  \href{http://dx.doi.org/10.1140/epjc/s10052-019-7499-4}{\doi{10.1140/epjc/s10052-019-7499-4}},
  \href{http://www.arXiv.org/abs/1903.12179}{\texttt{arXiv:1903.12179}}.

\bibitem{Ball_2017}
\hrefCMSnoop {}{{{NNPDF}} Collaboration, ``{Parton distributions from
  high-precision collider data}'',} \textit{ Eur. Phys. J. C} \textbf{ 77}
  (2017) 663,
  \href{http://dx.doi.org/10.1140/epjc/s10052-017-5199-5}{\doi{10.1140/epjc/s10052-017-5199-5}},
  \href{http://www.arXiv.org/abs/1706.00428}{\texttt{arXiv:1706.00428}}.

\bibitem{GEANT4}
\hrefCMSnoop {}{{GEANT4} Collaboration, ``{\GEANTfour}---a simulation
  toolkit'',} \textit{ Nucl. Instrum. Meth. A} \textbf{ 506} (2003) 250,
\href{http://dx.doi.org/10.1016/S0168-9002(03)01368-8}{\doi{10.1016/S0168-9002(03)01368-8}}.

\bibitem{geant4_2}
\hrefCMSnoop {}{J.~Allison { et~al.}, ``{\GEANTfour} developments and
  applications'',} \textit{ IEEE Trans. Nucl. Sci.} \textbf{ 53} (2006) 270,
  \href{http://dx.doi.org/10.1109/TNS.2006.869826}{\doi{10.1109/TNS.2006.869826}}.

\bibitem{cmsPF}
\hrefCMSnoop {}{{CMS Collaboration}, ``Particle-flow reconstruction and global
  event description with the {CMS} detector'',} \textit{ JINST} \textbf{ 12}
  (2017) P10003,
  \href{http://dx.doi.org/10.1088/1748-0221/12/10/p10003}{\doi{10.1088/1748-0221/12/10/p10003}},
  \href{http://www.arXiv.org/abs/1706.04965}{\texttt{arXiv:1706.04965}}.

\bibitem{CMS-TDR-15-02}
\href {http://cds.cern.ch/record/2020886}{{CMS Collaboration}, ``Technical
  proposal for the {Phase-II} upgrade of the {Compact Muon Solenoid}'',} CMS
  Technical Proposal CERN-LHCC-2015-010, CMS-TDR-15-02, 2015.

\bibitem{Cacciari:2008gp}
\hrefCMSnoop {}{M.~Cacciari, G.~P. Salam, and G.~Soyez, ``The anti-\kt jet
  clustering algorithm'',} \textit{ JHEP} \textbf{ 04} (2008) 063,
  \href{http://dx.doi.org/10.1088/1126-6708/2008/04/063}{\doi{10.1088/1126-6708/2008/04/063}},
  \href{http://www.arXiv.org/abs/0802.1189}{\texttt{arXiv:0802.1189}}.

\bibitem{Cacciari:2011ma}
\hrefCMSnoop {}{M.~Cacciari, G.~P. Salam, and G.~Soyez, ``{FastJet user
  manual}'',} \textit{ Eur. Phys. J. C} \textbf{ 72} (2012) 1896,
  \href{http://dx.doi.org/10.1140/epjc/s10052-012-1896-2}{\doi{10.1140/epjc/s10052-012-1896-2}},
\href{http://www.arXiv.org/abs/1111.6097}{\texttt{arXiv:1111.6097}}.

\bibitem{Sirunyan:2020foa}
\hrefCMSnoop {}{{CMS Collaboration}, ``{Pileup mitigation at CMS in 13\TeV
  data}'',} \textit{ JINST} \textbf{ 15} (2020) P09018,
  \href{http://dx.doi.org/10.1088/1748-0221/15/09/p09018}{\doi{10.1088/1748-0221/15/09/p09018}},
  \href{http://www.arXiv.org/abs/2003.00503}{\texttt{arXiv:2003.00503}}.

\bibitem{Bertolini:2014bba}
\hrefCMSnoop {}{D.~Bertolini, P.~Harris, M.~Low, and N.~Tran, ``Pileup per
  particle identification'',} \textit{ JHEP} \textbf{ 10} (2014) 059,
  \href{http://dx.doi.org/10.1007/JHEP10(2014)059}{\doi{10.1007/JHEP10(2014)059}},
\href{http://www.arXiv.org/abs/1407.6013}{\texttt{arXiv:1407.6013}}.

\bibitem{Khachatryan:2016kdb}
\hrefCMSnoop {}{{CMS Collaboration}, ``{Jet energy scale and resolution in the
  CMS experiment in $\Pp\Pp$ collisions at 8\TeV}'',} \textit{ JINST} \textbf{
  12} (2017) P02014,
  \href{http://dx.doi.org/10.1088/1748-0221/12/02/P02014}{\doi{10.1088/1748-0221/12/02/P02014}},
\href{http://www.arXiv.org/abs/1607.03663}{\texttt{arXiv:1607.03663}}.

\bibitem{CMS-PAS-JME-16-003}
\href {https://cds.cern.ch/record/2256875}{{CMS Collaboration}, ``{Jet
  algorithms performance in {13\TeV} data}'',} {CMS Physics Analysis Summary}
  CMS-PAS-JME-16-003, 2017.

\bibitem{CMS:2019ctu}
\hrefCMSnoop {}{{CMS Collaboration}, ``Performance of missing transverse
  momentum reconstruction in proton-proton collisions at {$\sqrt{s} = 13\TeV$}
  using the {CMS} detector'',} \textit{ JINST} \textbf{ 14} (2019) P07004,
  \href{http://dx.doi.org/10.1088/1748-0221/14/07/P07004}{\doi{10.1088/1748-0221/14/07/P07004}},
\href{http://www.arXiv.org/abs/1903.06078}{\texttt{arXiv:1903.06078}}.

\bibitem{Larkoski:2014wba}
\hrefCMSnoop {}{A.~J. Larkoski, S.~Marzani, G.~Soyez, and J.~Thaler, ``Soft
  drop'',} \textit{ JHEP} \textbf{ 05} (2014) 146,
  \href{http://dx.doi.org/10.1007/JHEP05(2014)146}{\doi{10.1007/JHEP05(2014)146}},
\href{http://www.arXiv.org/abs/1402.2657}{\texttt{arXiv:1402.2657}}.

\bibitem{CMS-DP-2025-010}
\href {https://cds.cern.ch/record/2929534}{{CMS Collaboration}, ``{Calibration
  of the top and \PW jet tagging efficiency in 13\TeV data collected by the CMS
  experiment in 2016--2018}'',} CMS Detector Performance Summary
  CMS-DP-2025-010, 2025.

\bibitem{CMS:2025kje}
\hrefCMSnoop {}{{CMS Collaboration}, ``{Performance of heavy-flavour jet
  identification in Lorentz-boosted topologies in proton-proton collisions at
  $\sqrt{s} = 13\TeV$}'',} \textit{ JINST} \textbf{ 20} (2025) P11006,
  \href{http://dx.doi.org/10.1088/1748-0221/20/11/P11006}{\doi{10.1088/1748-0221/20/11/P11006}},
  \href{http://www.arXiv.org/abs/2510.10228}{\texttt{arXiv:2510.10228}}.

\bibitem{Bols:2020bkb}
E.~Bols\hrefCMSnoop {}{ { et~al.}, ``Jet flavour classification using
  {DeepJet}'',} \textit{ JINST} \textbf{ 15} (2020) P12012,
  \href{http://dx.doi.org/10.1088/1748-0221/15/12/P12012}{\doi{10.1088/1748-0221/15/12/P12012}},
  \href{http://www.arXiv.org/abs/2008.10519}{\texttt{arXiv:2008.10519}}.

\bibitem{CMS-DP-2018-058}
\href {https://cds.cern.ch/record/2646773}{{CMS Collaboration}, ``{Performance
  of the DeepJet \PQb tagging algorithm using 41.9\fbinv of data from
  proton-proton collisions at 13\TeV with Phase 1 CMS detector}'',} {CMS
  Detector Performance Summary} CMS-DP-2018-058, 2018.

\bibitem{Khachatryan:2015hwa}
\hrefCMSnoop {}{{CMS Collaboration}, ``Performance of electron reconstruction
  and selection with the {CMS} detector in proton-proton collisions at
  {$\sqrt{s} = 8\TeV$}'',} \textit{ JINST} \textbf{ 10} (2015) P06005,
  \href{http://dx.doi.org/10.1088/1748-0221/10/06/P06005}{\doi{10.1088/1748-0221/10/06/P06005}},
\href{http://www.arXiv.org/abs/1502.02701}{\texttt{arXiv:1502.02701}}.

\bibitem{Rehermann:2010vq}
\hrefCMSnoop {}{K.~Rehermann and B.~Tweedie, ``Efficient identification of
  boosted semileptonic top quarks at the {LHC}'',} \textit{ JHEP} \textbf{ 03}
  (2011) 059,
  \href{http://dx.doi.org/10.1007/JHEP03(2011)059}{\doi{10.1007/JHEP03(2011)059}},
  \href{http://www.arXiv.org/abs/1007.2221}{\texttt{arXiv:1007.2221}}.

\bibitem{Li:2024htp}
C.~Li\hrefCMSnoop {}{ { et~al.}, ``Accelerating resonance searches via
  signature-oriented pre-training'',} 2024.
  \href{http://www.arXiv.org/abs/2405.12972}{\texttt{arXiv:2405.12972}}.

\bibitem{CMS:2024onh}
\hrefCMSnoop {}{{CMS Collaboration}, ``The {CMS} statistical analysis and
  combination tool: {\textsc{Combine}}'',} \textit{ Comput. Softw. Big Sci.}
  \textbf{ 8} (2024) 19,
  \href{http://dx.doi.org/10.1007/s41781-024-00121-4}{\doi{10.1007/s41781-024-00121-4}},
  \href{http://www.arXiv.org/abs/2404.06614}{\texttt{arXiv:2404.06614}}.

\bibitem{CMS:2025eyd}
\hrefCMSnoop {}{{CMS Collaboration}, ``{A method for correcting the
  substructure of multiprong jets using the Lund jet plane}'',} \textit{ JHEP}
  \textbf{ 11} (2025) 038,
  \href{http://dx.doi.org/10.1007/JHEP11(2025)038}{\doi{10.1007/JHEP11(2025)038}},
  \href{http://www.arXiv.org/abs/2507.07775}{\texttt{arXiv:2507.07775}}.

\bibitem{Dreyer:2018nbf}
\hrefCMSnoop {}{F.~A. Dreyer, G.~P. Salam, and G.~Soyez, ``The {Lund} jet
  plane'',} \textit{ JHEP} \textbf{ 12} (2018) 064,
  \href{http://dx.doi.org/10.1007/JHEP12(2018)064}{\doi{10.1007/JHEP12(2018)064}},
  \href{http://www.arXiv.org/abs/1807.04758}{\texttt{arXiv:1807.04758}}.

\bibitem{Catani:1993hr}
\hrefCMSnoop {}{S.~Catani, Y.~L. Dokshitzer, M.~H. Seymour, and B.~R. Webber,
  ``{Longitudinally invariant $K_t$ clustering algorithms for hadron hadron
  collisions}'',} \textit{ Nucl. Phys. B} \textbf{ 406} (1993) 187,
  \href{http://dx.doi.org/10.1016/0550-3213(93)90166-M}{\doi{10.1016/0550-3213(93)90166-M}}.

\bibitem{Ellis:1993tq}
\hrefCMSnoop {}{S.~D. Ellis and D.~E. Soper, ``{Successive combination jet
  algorithm for hadron collisions}'',} \textit{ Phys. Rev. D} \textbf{ 48}
  (1993) 3160,
  \href{http://dx.doi.org/10.1103/PhysRevD.48.3160}{\doi{10.1103/PhysRevD.48.3160}},
  \href{http://www.arXiv.org/abs/hep-ph/9305266}{\texttt{arXiv:hep-ph/9305266}}.

\bibitem{Krohn:2009th}
\hrefCMSnoop {}{D.~Krohn, J.~Thaler, and L.-T. Wang, ``Jet trimming'',}
  \textit{ JHEP} \textbf{ 02} (2010) 084,
  \href{http://dx.doi.org/10.1007/JHEP02(2010)084}{\doi{10.1007/JHEP02(2010)084}},
  \href{http://www.arXiv.org/abs/0912.1342}{\texttt{arXiv:0912.1342}}.

\bibitem{ftest}
\hrefCMSnoop {}{R.~A. Fisher, ``On the interpretation of $\chi^{2}$ from
  contingency tables, and the calculation of {P}'',} \textit{ J. R. Stat. Soc.}
  \textbf{ 85} (1922) 87,
  \href{http://dx.doi.org/10.2307/2340521}{\doi{10.2307/2340521}}.

\bibitem{CMS:2018zzl}
\hrefCMSnoop {}{{CMS Collaboration}, ``{Measurements of properties of the Higgs
  boson decaying to a \PW boson pair in pp collisions at $\sqrt{s}=
  13\TeV$}'',} \textit{ Phys. Lett. B} \textbf{ 791} (2019) 96,
  \href{http://dx.doi.org/10.1016/j.physletb.2018.12.073}{\doi{10.1016/j.physletb.2018.12.073}},
  \href{http://www.arXiv.org/abs/1806.05246}{\texttt{arXiv:1806.05246}}.

\bibitem{CMS-NOTE-2011-005}
\href {https://cds.cern.ch/record/1379837}{{ATLAS and CMS Collaborations, and
  LHC Higgs Combination Group}, ``Procedure for the {LHC} {Higgs} boson search
  combination in {Summer} 2011'',} Technical Report CMS-NOTE-2011-005,
  ATL-PHYS-PUB-2011-11, 2011.

\bibitem{CMS:2018mlc}
\hrefCMSnoop {}{{CMS Collaboration}, ``{Measurement of the inelastic
  proton-proton cross section at $\sqrt{s} = 13\TeV$}'',} \textit{ JHEP}
  \textbf{ 07} (2018) 161,
  \href{http://dx.doi.org/10.1007/JHEP07(2018)161}{\doi{10.1007/JHEP07(2018)161}},
  \href{http://www.arXiv.org/abs/1802.02613}{\texttt{arXiv:1802.02613}}.

\bibitem{Barlow:1993dm}
\hrefCMSnoop {}{R.~J. Barlow and C.~Beeston, ``{Fitting using finite Monte
  Carlo samples}'',} \textit{ Comput. Phys. Commun.} \textbf{ 77} (1993) 219,
  \href{http://dx.doi.org/10.1016/0010-4655(93)90005-W}{\doi{10.1016/0010-4655(93)90005-W}}.

\bibitem{Conway:2011in}
\hrefCMSnoop {}{J.~S. Conway, ``Incorporating nuisance parameters in
  likelihoods for multisource spectra'',} in \textit{ PHYSTAT 2011}, volume~1,
  p.~115.
\newblock 2011.
\newblock \href{http://www.arXiv.org/abs/1103.0354}{\texttt{arXiv:1103.0354}}.
\newblock
  \href{http://dx.doi.org/10.5170/CERN-2011-006.115}{\doi{10.5170/CERN-2011-006.115}}.

\bibitem{roofit}
\hrefCMSnoop {}{W.~Verkerke and D.~P. Kirkby, ``{The RooFit toolkit for data
  modeling}'',} in \textit{ {Proceedings of the 13th International Conference
  for Computing in High-Energy and Nuclear Physics (CHEP03)}}, L.~Lyons and
  M.~Karagoz, eds.
\newblock 2003.
\newblock
  \href{http://www.arXiv.org/abs/physics/0306116}{\texttt{arXiv:physics/0306116}}.

\bibitem{roostats}
L.~Moneta\hrefCMSnoop {}{ { et~al.}, ``{The RooStats Project}'',} in \textit{
  {Proceedings of 13th International Workshop on Advanced Computing and
  Analysis Techniques in Physics Research {\textemdash} PoS(ACAT2010)}},
  T.~Speer { et~al.}, eds., p.~057.
\newblock 2011.
\newblock \href{http://www.arXiv.org/abs/1009.1003}{\texttt{arXiv:1009.1003}}.
\newblock
  \href{http://dx.doi.org/10.22323/1.093.0057}{\doi{10.22323/1.093.0057}}.

\bibitem{CMS:2014fzn}
\hrefCMSnoop {}{{CMS Collaboration}, ``{Precise determination of the mass of
  the Higgs boson and tests of compatibility of its couplings with the standard
  model predictions using proton collisions at 7 and 8\TeV}'',} \textit{ Eur.
  Phys. J. C} \textbf{ 75} (2015) 212,
  \href{http://dx.doi.org/10.1140/epjc/s10052-015-3351-7}{\doi{10.1140/epjc/s10052-015-3351-7}},
  \href{http://www.arXiv.org/abs/1412.8662}{\texttt{arXiv:1412.8662}}.

\bibitem{Cowan:2010js}
\hrefCMSnoop {}{G.~Cowan, K.~Cranmer, E.~Gross, and O.~Vitells, ``Asymptotic
  formulae for likelihood-based tests of new physics'',} \textit{ Eur. Phys. J.
  C} \textbf{ 71} (2011) 1554,
  \href{http://dx.doi.org/10.1140/epjc/s10052-011-1554-0}{\doi{10.1140/epjc/s10052-011-1554-0}},
  \href{http://www.arXiv.org/abs/1007.1727}{\texttt{arXiv:1007.1727}}.
[Erratum: \DOI{10.1140/epjc/s10052-013-2501-z}].

\bibitem{CMS:2024ddc}
\hrefCMSnoop {}{{CMS Collaboration}, ``{Measurement of boosted Higgs bosons
  produced via vector boson fusion or gluon fusion in the {$\PH\to\bbbar$}
  decay mode using LHC proton-proton collision data at {$\sqrt{s} =
  13\TeV$}}'',} \textit{ JHEP} \textbf{ 12} (2024) 035,
  \href{http://dx.doi.org/10.1007/JHEP12(2024)035}{\doi{10.1007/JHEP12(2024)035}},
  \href{http://www.arXiv.org/abs/2407.08012}{\texttt{arXiv:2407.08012}}.

\end{thebibliography}\endgroup
\cmsinstitute{Yerevan Physics Institute, Yerevan, Armenia}
{\tolerance=6000
A.~Hayrapetyan, V.~Makarenko\cmsorcid{0000-0002-8406-8605}, A.~Tumasyan\cmsAuthorMark{1}\cmsorcid{0009-0000-0684-6742}
\par}
\cmsinstitute{Institut f\"{u}r Hochenergiephysik, Vienna, Austria}
{\tolerance=6000
W.~Adam\cmsorcid{0000-0001-9099-4341}, L.~Benato\cmsorcid{0000-0001-5135-7489}, T.~Bergauer\cmsorcid{0000-0002-5786-0293}, M.~Dragicevic\cmsorcid{0000-0003-1967-6783}, P.S.~Hussain\cmsorcid{0000-0002-4825-5278}, M.~Jeitler\cmsAuthorMark{2}\cmsorcid{0000-0002-5141-9560}, N.~Krammer\cmsorcid{0000-0002-0548-0985}, A.~Li\cmsorcid{0000-0002-4547-116X}, D.~Liko\cmsorcid{0000-0002-3380-473X}, M.~Matthewman, J.~Schieck\cmsAuthorMark{2}\cmsorcid{0000-0002-1058-8093}, R.~Sch\"{o}fbeck\cmsAuthorMark{2}\cmsorcid{0000-0002-2332-8784}, M.~Shooshtari\cmsorcid{0009-0004-8882-4887}, M.~Sonawane\cmsorcid{0000-0003-0510-7010}, N.~Van~Den~Bossche\cmsorcid{0000-0003-2973-4991}, W.~Waltenberger\cmsorcid{0000-0002-6215-7228}, C.-E.~Wulz\cmsAuthorMark{2}\cmsorcid{0000-0001-9226-5812}
% ...
\par}
\vskip\cmsinstskip
\dag:~Deceased\\
$^{1}$Also at Yerevan State University, Yerevan, Armenia\\
$^{2}$Also at TU Wien, Vienna, Austria\\
% ...
\end{sloppypar}
\end{document}